\documentclass[prd, tightenlines, twocolumn, twoside, secnumarabic, superscriptaddress, preprintnumbers, nofootinbib, notitlepage]{revtex4-1}
\usepackage{graphicx}
\usepackage{epstopdf}
\usepackage{amsmath}
\usepackage{amsfonts}
\usepackage{amssymb}
\usepackage{appendix}
\usepackage{comment}
\usepackage{color}
\usepackage{slashed}
\usepackage{caption}
\usepackage{subcaption}
\usepackage{setspace}
\usepackage{footnote}
\usepackage{multirow}
\usepackage{mathrsfs}
\usepackage{hyperref}
\usepackage{cleveref}
\usepackage{wrapfig}
\definecolor{niceblue}{rgb}{0.388235, 0.627451, 0.847059}
\definecolor{nicered}{rgb}{0.7,0.1,0.1}
\definecolor{nicegreen}{rgb}{0.1,0.5,0.1}
\hypersetup{colorlinks,citecolor= nicegreen,linkcolor= nicered}

\newcommand{\globes}{\texttt{GLoBES}}
\newcommand{\globesfit}{\texttt{GLoBESfit}}

\newcommand{\nn}{\nonumber}

\definecolor{JMBcomment}{RGB}{64, 214, 136}

\definecolor{ToDoColor}{RGB}{219, 157, 13}

\makeatletter
\def\l@subsubsection#1#2{}
\makeatother

\begin{document}

\singlespacing
\allowdisplaybreaks

\title{Sterile Neutrinos and the Global Reactor Antineutrino Dataset}

\author{Jeffrey M. Berryman}
\affiliation{Center for Neutrino Physics, Department of Physics, Virginia Tech, Blacksburg, VA 24061, USA}
\affiliation{Department of Physics and Astronomy, University of Kentucky, Lexington, KY 40506, USA}
\affiliation{Department of Physics, University of California, Berkeley, CA 94720, USA}

\author{Patrick Huber}
\affiliation{Center for Neutrino Physics, Department of Physics, Virginia Tech, Blacksburg, VA 24061, USA}


\date{\today}
\begin{abstract}
We present results from global fits to the available reactor antineutrino dataset, as of Fall 2019, to determine the global preference for a fourth, sterile neutrino. We have separately considered experiments that measure the integrated inverse-beta decay (IBD) rate from those that measure the energy spectrum of IBD events at one or more locations. The software used is the newly developed \globesfit \, tool set which is based on the publicly available \globes \, framework and will be released as open-source software.      
\end{abstract}

\maketitle

\begin{spacing}{1.0}
\tableofcontents
\end{spacing}


\section{Introduction}
\label{sec:Intro}
\setcounter{equation}{0}

Nuclear reactors have been workhorses for neutrino physics since its inception as an experimental science by Cowan and Reines~\cite{Cowan:1992xc}. A number of important measurements of fundamental neutrino\footnote{We consider ``neutrino'' to be synonymous with ``electron antineutrino'' in this work.} properties have been made using reactor neutrinos: we highlight the confirmation of solar neutrino flavor conversion by KamLAND~\cite{Eguchi:2002dm} and the measurement of $\theta_{13}$~\cite{An:2012eh,Ahn:2012nd,Abe:2011fz}. Neutrinos are not produced in the fission process itself but in the subsequent beta decays of neutron-rich fission fragments. The first reactor neutrino experiments already had to deal with the question of how to predict the neutrino flux from these complex sources~\cite{Carter:1959qm}; interestingly, the uncertainty quoted was 5-10\%. The neutrino flux question has taken on a central role with reactor antineutrino anomaly (RAA)~\cite{Mention:2011rk}: an initial reactor flux re-evaluation~\cite{Mueller:2011nm}, later confirmed by one of the present authors~\cite{Huber:2011wv}, lead to an increase in predicted reactor neutrino event rates and thus to an observed deficit of reactor neutrinos.

An exciting, albeit speculative, explanation would be provided by oscillations into a sterile neutrino with $\Delta m^2\simeq 1\,\mathrm{eV}^2$. Unsurprisingly, this hypothesis has triggered significant interest~\cite{Abazajian:2012ys}, resulting in a number of new reactor neutrino experiments aimed at finding an actual oscillation signature. At the same time, this interest also resulted in a significant effort, both theoretically and experimentally, to improve reactor neutrino flux predictions; see, for instance, Ref.~\cite{Hayes:2016qnu} and references therein.

The interpretation of reactor neutrino data in terms of sterile neutrino oscillations is relatively straightforward and free of major tension, even when combined with other available electron neutrino disappearance data; see, e.g., Ref.~\cite{Dentler:2018sju,Gariazzo:2017fdh}. However, when an attempt is made to explain also the anomalous results in the $\bar\nu_\mu\rightarrow\bar\nu_e$ and $\nu_\mu\rightarrow\nu_e$ channels as observed by LSND~\cite{Aguilar:2001ty} and MiniBooNE~\cite{Aguilar-Arevalo:2013pmq} significant tension arises: for a sterile neutrino to be able to mediate the transition between two different active flavors, the sterile neutrino must mix with \emph{both} of these flavors. As a result, there is a well-defined relationship between the amplitude of $\nu_\mu\rightarrow\nu_e$ oscillations and the expected effect in the disappearance channels $\nu_\mu\rightarrow\nu_\mu$ and $\nu_e\rightarrow\nu_e$. The size of the RAA is consistent with sterile neutrino interpretation of LSND and MiniBooNE, but no evidence is found for commensurate disappearance in the $\nu_\mu\rightarrow\nu_\mu$ channel. This issue has persisted for some time and there is debate about the statistical significance of this tension, for a recent review see Ref.~\cite{Diaz:2019fwt}. Nonetheless, the lack of $\nu_\mu$ disappearance presents a considerable deficiency in the sterile neutrino hypothesis. Here, we remain agnostic with respect to LSND and MiniBooNE and focus entirely on the electron neutrino sector.

Given the significant ongoing experimental effort in reactor neutrino experiments, combined with the potential for a major discovery of new physics, we will discuss in detail the existing reactor antineutrino data. We will compare the data with the Huber-Mueller flux model, as well as with two state-of-the-art flux predictions, one based on the summation approach~\cite{Estienne:2019ujo} and one based on the conversion approach~\cite{Hayen:2019eop}. We will critically examine how these differing flux models affect the evidence for the sterile neutrino hypothesis. Moreover, we present an open-source software framework called \globesfit\, based on the established \globes~\cite{Huber:2004ka,Huber:2007ji} software suite, which will allow users to fit their preferred model to the data. We also hope that this open framework can be adopted by the experimental community to share their results.
 
A major result of this work is that we find if we restrict the analysis to data taken after 2010 -- i.e., the modern $\theta_{13}$ reactor experiments and their short-baseline counterparts -- then the reactor rate anomaly seems to be enhanced. This restriction should not be misconstrued as a lack of trust in the older experiments, but it does reflect the fact that data curation and release practices were less refined in the past. As a result, much less detailed information is available for properly including older experiments into a global fit.

Best practices in data sharing are crucial for this field to make progress, since we have previously shown~\cite{Berryman:2019hme} that no single, existing reactor experiment can definitively test the currently preferred parameter space. Therefore, global fits will play an important role in neutrino physics for the foreseeable future, reinforcing the value of an open fitting framework.

This manuscript is arranged as follows. In Sec.~\ref{sec:Fluxes}, we discuss the predictions of the fluxes of antineutrinos at nuclear reactors that we use, including how systematic uncertainties on these predictions are incorporated in our analyses. In Sec.~\ref{sec:RateExps}, we discuss the total IBD rate experiments that we have included in our fits, and in Sec.~\ref{sec:RateResults}, we present our combined analyses of various subsets of these data. Similarly, in Sec.~\ref{sec:SpectrumExp}, we discuss the IBD spectrum experiments that we have included in our analyses, and in Sec.~\ref{sec:SpecAnalysis} we present combined analyses of these experiments. We address the effect of the infamous 5 MeV bump~\cite{RENO:2015ksa,An:2015nua,Abe:2015rcp} on sterile neutrino searches in Sec.~\ref{sec:Bump}, and offer concluding remarks in Sec.~\ref{sec:Conclusions}.

In addition to communicating our findings, one of the intentions of this manuscript is to provide documentation for the tools that we have developed. Pursuant to this, we have included several appendices that deal with technical aspects of \globesfit. In Appendix \ref{app:OscEngines}, we provide an account of how oscillations involving a sterile neutrino have been included in \globes, and in Appendix \ref{app:TheCode}, we provide a general overview of the files that constitute \globesfit, including their content and some aspects of their functionality. Lastly, we provide extensive data tables and supplemental figures in Appendix \ref{app:SuppData}; these data are important for our analyses, but would have been intrusive to have included in the main text.


\section{Reactor Antineutrino Flux Models}
\label{sec:Fluxes}
\setcounter{equation}{0}

Predictions of the spectra of antineutrinos from reactors have played a central role from the inception of neutrino physics as experimental science~\cite{PhysRev.113.280}. More recently, the RAA \cite{Mention:2011rk} has put a spotlight on flux models; for a recent review on flux models and how they are produced, see Ref.~\cite{Hayes:2016qnu}. For the discussion here, we note the following the salient features. Neutrinos are produced not by the fission process itself, but by the beta decays of about 800 fission fragment isotopes, corresponding to about 10,000 beta decay branches. A large fraction of these beta decays are of (unique and non-unique) forbidden type, which implies a significant dependence on details of nuclear structure for both the emitted electron and antineutrino spectra. We will use the following three flux models:
\begin{enumerate}
  \item Huber-Mueller (HM): This flux
    model~\cite{Mueller:2011nm,Huber:2011wv} has become the {\it de facto} standard in the field and is based on the conversion of integrated beta spectrum measurements performed in the 1980s at the Institut Laue-Langevin~\cite{VonFeilitzsch:1982jw,Schreckenbach:1985ep,Hahn:1989zr}. This flux model has relatively little dependence on nuclear databases and employs the allowed approximation, i.e., all beta decays are treated as allowed decays.
    \item \emph{Ab initio}: This flux model~\cite{Estienne:2019ujo} uses fission fragment yields and information on individual beta decays from nuclear databases to directly compute the neutrino flux. It also employs the allowed approximation.
      \item Hayen-Kostensalo-Severijns-Suhonen (HKSS): This flux model~\cite{Hayen:2019eop} shares many of the features of HM, that is, it is based on a conversion of the integrated beta spectra. Importantly, it is the first attempt to include effects from forbidden decays in a systematic fashion.
  \end{enumerate}
 
\begin{figure*}[t]
\centering

\begin{subfigure}[t]{0.45\textwidth}
\centering
\includegraphics[width=\linewidth]{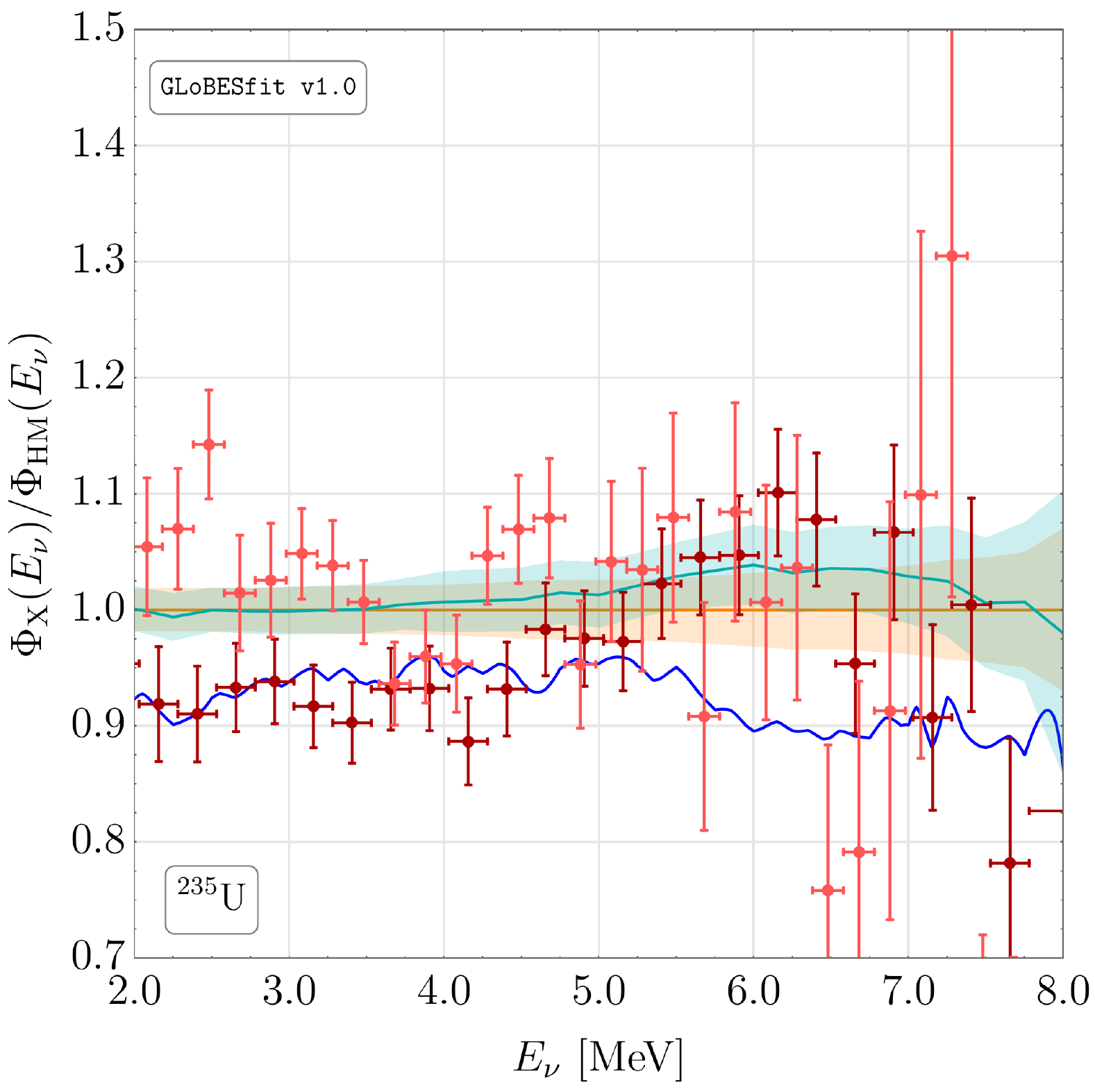}
        \caption{}\label{fig:fig_a}
\end{subfigure}
\begin{subfigure}[t]{0.45\textwidth}
\centering
\includegraphics[width=\linewidth]{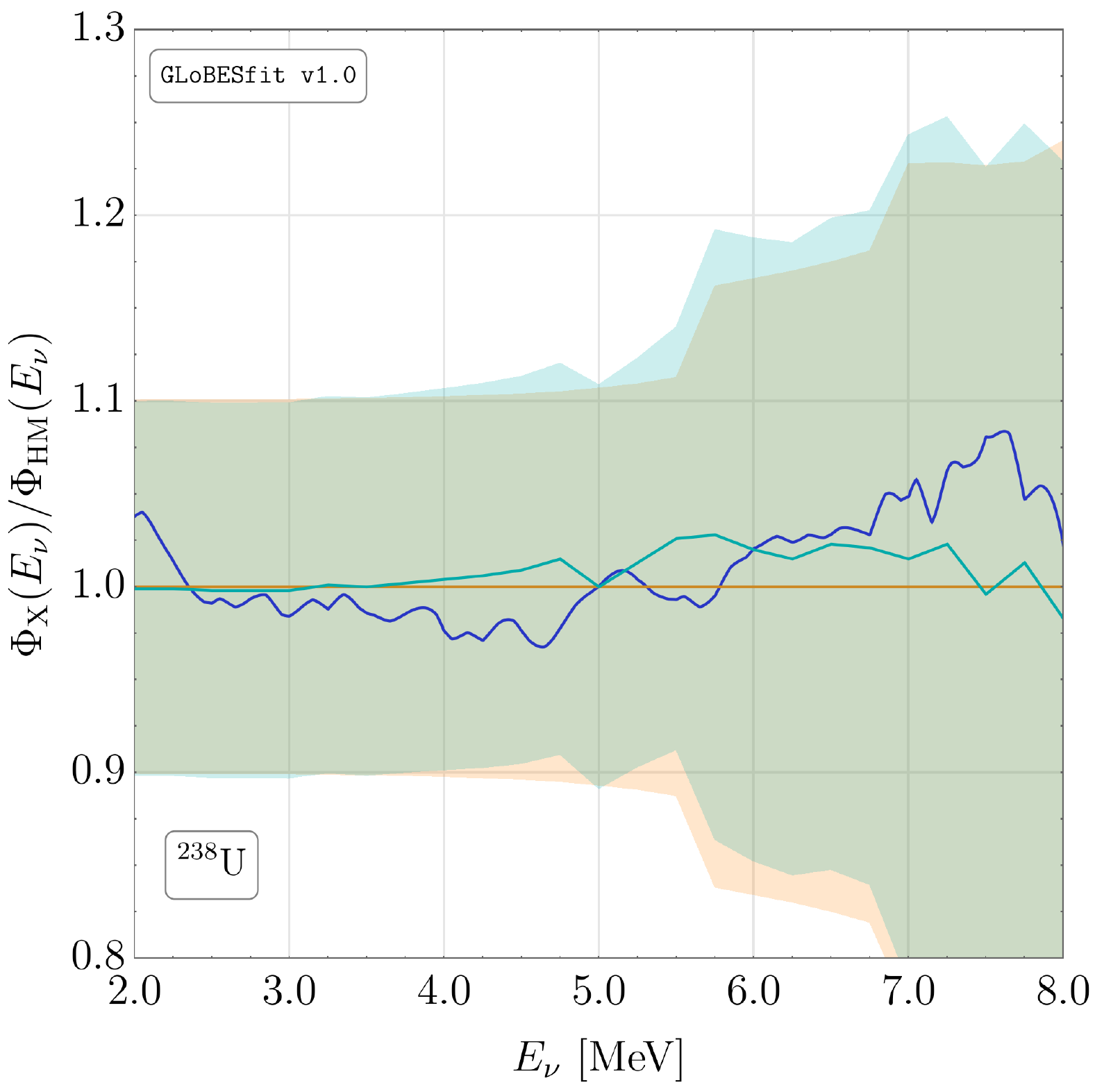}
\caption{}\label{fig:fig_b}
\end{subfigure}

\medskip

\begin{subfigure}[t]{0.45\textwidth}
\centering
\vspace{0pt}
\includegraphics[width=\linewidth]{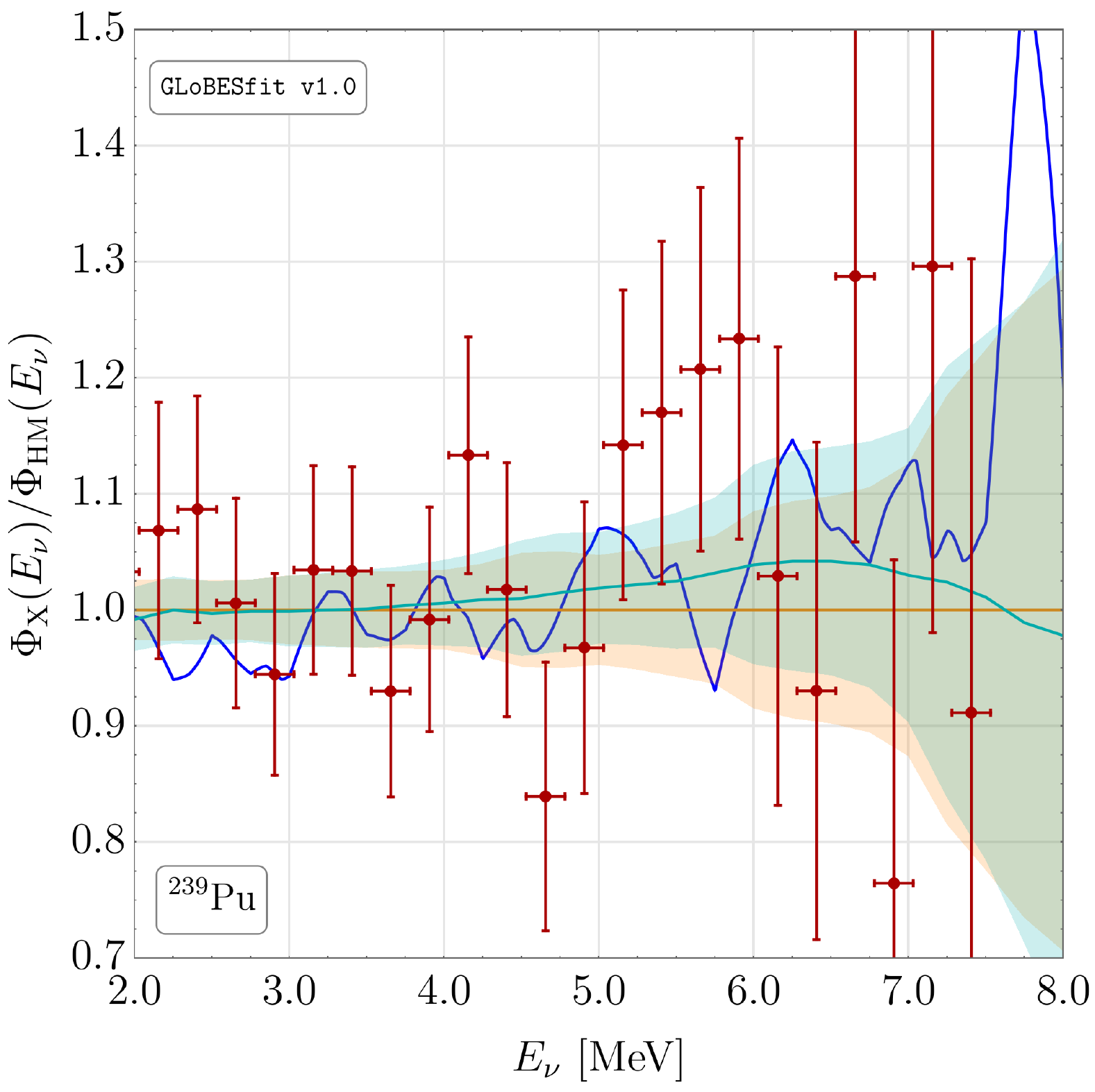}
\caption{}\label{fig:fig_c}
\end{subfigure}
\begin{subfigure}[t]{0.45\textwidth}
\centering
\vspace{0pt}
\includegraphics[width=\linewidth]{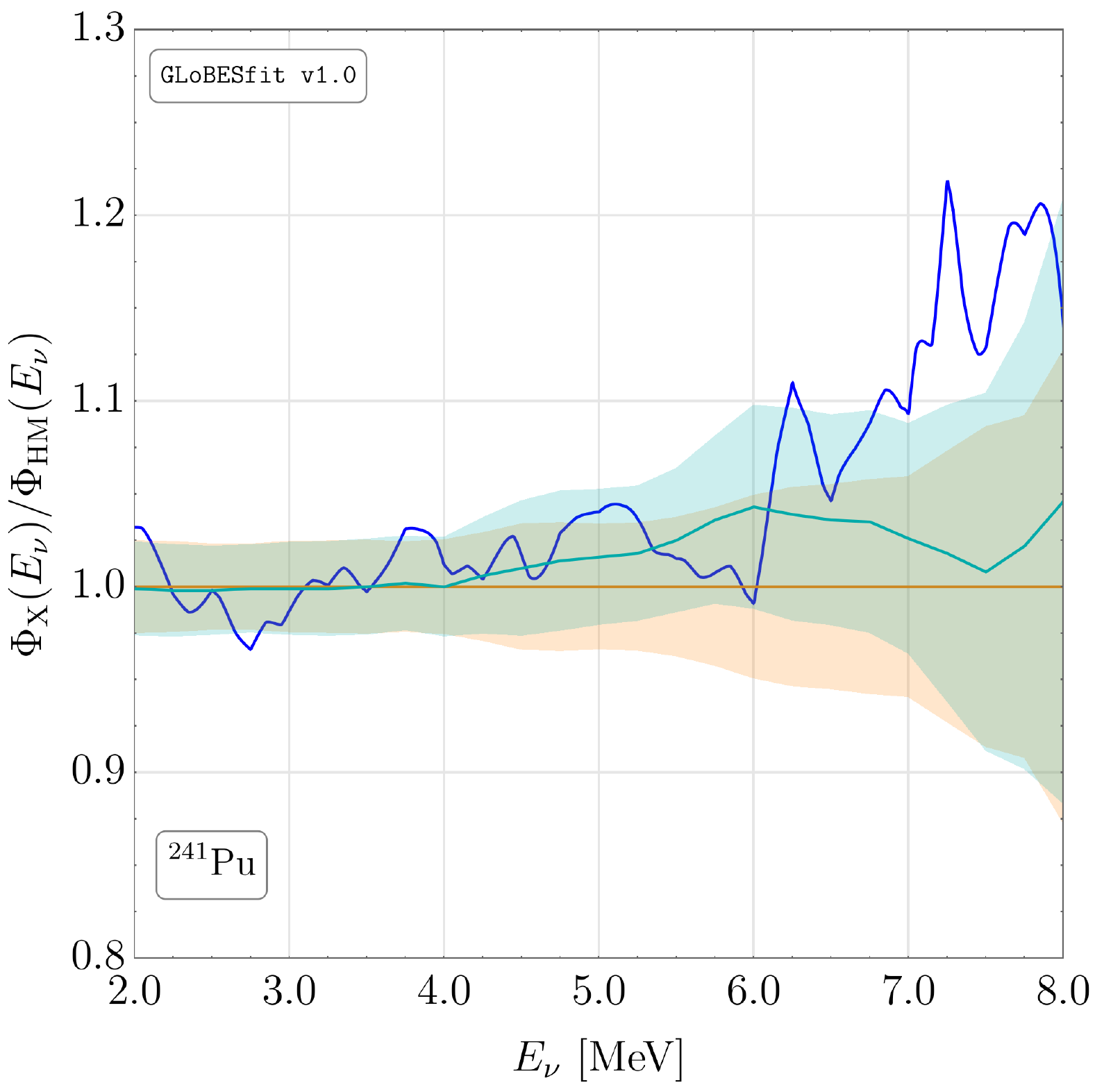}
\caption{}\label{fig:fig_d}
\end{subfigure}
\caption{The ratios of the three antineutrino flux models we consider relative to the traditional Huber-Mueller (HM) flux predictions for (a) $^{235}$U, (b) $^{238}$U, (c) $^{239}$Pu and (d) $^{241}$Pu. The orange curves are for the HM flux predictions \cite{Mueller:2011nm,Huber:2011wv} and are, of course, equal to one everywhere. The blue curves are for the \emph{ab initio} fluxes presented in Ref.~\cite{Estienne:2019ujo}. The dark cyan curves are for the updated conversion method calculation presented in Ref.~\cite{Hayen:2019eop}, which we call HKSS. The shaded regions represent the corresponding uncertainties, being equal to the square root of the diagonal elements of the global covariance matrix. Note that the \emph{ab initio} fluxes have no corresponding uncertainties; see text for details. The data points are based on antineutrino measurements performed at reactors, Daya Bay~\cite{Adey:2019ywk} (red) and PROSPECT~\cite{Ashenfelter:2018jrx} (pink).}
\end{figure*}

The ratios for the \emph{ab initio} and HKSS flux predictions relative to the HM predictions are shown in Figs.~\ref{fig:fig_a}-\ref{fig:fig_d} for, respectively, $^{235}$U, $^{238}$U, $^{239}$Pu and $^{241}$Pu. In each panel, the orange line at $1.0$ represents the HM predictions; the blue curve represents the \emph{ab initio} predictions; and the dark cyan curves represent the HKSS predictions. The associated bands represent the ($1\sigma$) systematic uncertainties on these predictions; we discuss how these are calculated below. For comparison, we also show the \emph{measured} neutrino spectrum from Daya Bay~\cite{Adey:2019ywk} (red) and PROSPECT~\cite{Ashenfelter:2018jrx} (pink). Daya Bay can separate the contributions from $^{235}$U and $^{239}$Pu by comparing data taken at different points in the fuel cycle. Meanwhile, PROSPECT is deployed at HFIR, which is fueled with highly-enriched uranium meaning nearly all fissions are of $^{235}$U. Visual inspection indicates that all three flux models and the neutrino data agree for $^{239}$Pu within the large error bars. However, the {\it ab initio} flux model predicts a lower flux for $^{235}$U than either HM or HKSS, which seems to agree somewhat better with data. Comparison of the data with any of three flux models reveals a bump in the region around 5-6 MeV. In later sections, we will perform a more careful analysis; impressions from this intial inspection will essentially be confirmed.

In all these calculations, one should formally be accounting for nonequilibrium corrections to the antineutrino flux; these are approximated by Mueller, et al., in Ref.~\cite{Mueller:2011nm} for $^{235}$U, $^{239}$Pu and $^{241}$Pu. However, the collaborations often do not publish information about their burn-up, so it is unclear how large the nonequilibrium correction should be. Even taking a nonequilibrium correction corresponding to 450 days of irradiation -- the largest such correction presented in Ref.~\cite{Mueller:2011nm} -- the difference between this the absence of nonequilibrium correction is typically overwhelmed by both experimental and other theoretical uncertainties. Therefore, we do not expect these to meaningfully impact the final results; we therefore elect to exclude them from the present analyses.


The experimental quantity of interest in all cases is the convolution of the neutrino flux with the inverse beta decay (IBD) cross section $\sigma_{\rm IBD}$ and the electron-type survival probability $P_{ee}$ (see Eq.~\eqref{eq:DefinePee} in Appendix \ref{app:OscEngines}). We define the flux weighted cross sections $\sigma_X^i$ as follows:
\begin{equation}
\label{eq:EventRates}
\sigma_{\rm X}^i =  \int_{E_{\rm min}}^{E_{\rm max}} d E_\nu \, \Phi^i_{\rm X}(E_\nu) \sigma_{\rm IBD} (E_\nu) P_{ee} (E_\nu).
\end{equation} 
This quantity provides the IBD rate for antineutrinos with \emph{true} energy in the range $\left[ E_{\rm min},\, E_{\rm max}\right]$ calculated using flux model X = HM, AI, HKSS for isotope $i = 235, \, 238, \, 239,\, 241$. The total IBD rate in this energy window is then the sum of the $\sigma_{\rm X}^i$, weighted by the appropriate \emph{effective fuel fraction} $f_i$, i.e., the fraction of fissions due to each isotope during the operation of the experiment. In this work, we use the IBD cross section published in Ref.~\cite{Vogel:1999zy} -- particularly, the results up to and including $\mathcal{O}(1/M)$ as given in their Eqs.~(14) and (15). The uncertainty on this cross section is negligible compared to the uncertainties on the fluxes.

We consider two types of reactor antineutrino measurements in this work. In Secs.~\ref{sec:RateExps} and \ref{sec:RateResults}, we consider experiments that measure the total IBD rate, but not necessarily the specific energy of a given event. For these, we will take $\left[ E_{\rm min},\, E_{\rm max}\right] = \left[ 1.8, \, 8.0   \right]$ MeV. On the other hand, in Secs.~\ref{sec:SpectrumExp} and \ref{sec:SpecAnalysis}, we consider measurements of the IBD spectrum over various energy bins, each with its own $\left[ E_{\rm min},\,   E_{\rm max}\right]$. Our treatment of the flux systematics at total IBD rate experiments is common to all such experiments; the flux systematics at spectrum experiments must be handled on a case-by-case basis.

We use the following generic procedure to incorporate systematic
uncertainties stemming from flux predictions into our analyses.

\begin{enumerate}
\item We use information presented in the literature to assign a covariance matrix to a set of flux predictions.
\item We then generate random variations of these fluxes using this covariance matrix to numerically calculate $\sigma_{\rm X}^i$ in Eq.~\eqref{eq:EventRates}, assuming $P_{ee}(E_\nu) = 1$, for some specified $\left[ E_{\rm min},\, E_{\rm max}\right]$.
\item For total rate analyses, we take $\left[ E_{\rm min},\, E_{\rm max}\right] = \left[ 1.8, \, 8.0 \right]$ MeV and calculate the $\sigma_{\rm X}^i$ and their correlations. These results will be applied to all total IBD rate measurements simultaneously, as will be described in Sec.~\ref{sec:RateResults}.
\item For spectral analyses, we calculate fractional deviations in the individual bins, including correlations. Usually, we will be interested in the ratio of two such spectra. The resulting covariance matrices are then combined with other experimental systematics; see Secs.~\ref{sec:SpectrumExp} and \ref{sec:SpecAnalysis} for more details.
  
\end{enumerate}

When we numerically integrate over the predicted fluxes, we do so by logarithmically interpolating/extrapolating on the published flux models.

\subsubsection*{HM Fluxes}

For the HM fluxes we use the central values published in Tables VII-IX of Ref.~\cite{Huber:2011wv} for the fluxes from $^{235}$U, $^{239}$Pu and $^{241}$Pu, respectively, whereas we use the values from the ``$N_{\overline{\nu}}$, 12 h'' column of Table III of Ref.~\cite{Mueller:2011nm} for $^{238}$U. Regarding errors, we use the following prescription. For $^{235}$U, $^{239}$Pu and $^{241}$Pu, we take the (percent) errors listed in the columns headed ``stat.'' and ``bias err." from Tables VII-IX of Ref.~\cite{Huber:2011wv}, respectively, to be fully uncorrelated between bins and these three isotopes; and the columns headed ``$\overline{Z}$,'' ``WM'' and ``norm.'' are added in quadrature and taken to be fully correlated between bins and these three isotopes. We assume all errors to be Gaussian; asymmetric errors are replaced by a Gaussian with width given by the geometric mean of the upper and lower uncertainties. For $^{238}$U, we combine the errors in the ``Nuclear databases'' column of Table II in Ref.~\cite{Mueller:2011nm}-- the correlations of which are shown in Table I of the same reference -- along with the ``Missing info.'' column of the same -- which we take to be fully correlated. These uncertainties yield the orange bands in Figs.~\ref{fig:fig_a}-\ref{fig:fig_d}.

With the information described above,  we obtain the following IBD yields for the HM fluxes:
\begin{align*}
\sigma_{\rm HM}^{235} = & \left(6.60 \pm 0.14 \right) \times 10^{-43} \text{ cm$^2$/fission } \\
\sigma_{\rm HM}^{238} = & \left(10.00 \pm 1.12 \right) \times 10^{-43} \text{ cm$^2$/fission }\\
\sigma_{\rm HM}^{239} = & \left( 4.33 \pm 0.11 \right) \times 10^{-43} \text{ cm$^2$/fission }\\
\sigma_{\rm HM}^{241} = & \left( 6.01 \pm 0.13 \right) \times 10^{-43}\text{ cm$^2$/fission }
\end{align*}
The uncertainties on the HM fluxes are described by the following
correlation matrix:
\begin{equation}
\rho_{\rm HM} = \left( \begin{array}{cccc} 1 & 0 & 0.943 & 0.971 \\
0 & 1 & 0 & 0 \\
0.943 & 0 & 1 & 0.928 \\
0.971 & 0 & 0.928 & 1 \end{array} \right);
\end{equation}
the ordering here is \{$^{235}$U, $^{238}$U, $^{239}$Pu, $^{241}$Pu\}. We reiterate that we are ignoring uncertainties from the IBD cross section; the dominant theoretical uncertainty is that on the antineutrino flux.

\subsubsection*{\emph{Ab Initio} Fluxes}

The values of the antineutrino fluxes can be found in the Supplementary Material to Ref.~\cite{Estienne:2019ujo}.\footnote{We thank Muriel Fallot for providing us with these fluxes in machine readable format~\cite{MFallot}.} These predictions, however, are not published with systematic uncertainties; consequently, we show no blue shaded regions in Figs.~\ref{fig:fig_a}-\ref{fig:fig_d}.
 
 In our analyses, we will always apply the systematic errors associated to the HM fluxes to the \emph{ab initio} fluxes, as well. This is likely an optimistic assignment; though precise uncertainties have not been calculated, the true systematic uncertainty on these predictions is likely of order $\gtrsim 5\%$. However, we will argue below that this treatment is sufficient for the conclusions of this work. For now, we simply point out that this treatment results in a more aggressive assignment of statistical significances than a more realistic error budget would produce.

We calculate the following isotopic IBD yields:
\begin{align*}
\sigma_{\rm AI}^{235} = & \left(6.17 \pm 0.13 \right) \times 10^{-43} \text{ cm$^2$/fission } \\
\sigma_{\rm AI}^{238} = & \left(9.94 \pm 1.09 \right) \times 10^{-43} \text{ cm$^2$/fission }\\
\sigma_{\rm AI}^{239} = & \left( 4.32 \pm 0.11 \right) \times 10^{-43} \text{ cm$^2$/fission }\\
\sigma_{\rm AI}^{241} = & \left( 6.10 \pm 0.13 \right) \times 10^{-43}\text{ cm$^2$/fission }
\end{align*}
While the isotopic fluxes for $^{238}$U, $^{239}$Pu and $^{241}$Pu here are consistent with those calculated for the HM fluxes, the $^{235}$U flux is substantially lessened. This is apparent from Fig.~\ref{fig:fig_a}: the \emph{ab initio} $^{235}$U flux is about $\sim5-10\%$ less than the corresponding HM prediction for most energies. The corresponding correlation matrix is given by
\begin{equation}
\rho_{\rm AI} = \left( \begin{array}{cccc} 1 & 0 & 0.941 & 0.970 \\
0 & 1 & 0 & 0 \\
0.941 & 0 & 1 & 0.923 \\
0.970 & 0 & 0.923 & 1 \end{array} \right).
\end{equation}

\subsubsection*{HKSS Fluxes}

The fluxes are presented in Tables VI-IX of Ref.~\cite{Hayen:2019eop}; specifically, we use the results in the columns headed ``$\delta N$ Num.," which represent the (percent) rescaling introduced by including first-forbidden decays to the reactor antineutrino flux. Systematics are handled by splicing together a subset of the errors published in Ref.~\cite{Hayen:2019eop} with those for the HM fluxes from Ref.~\cite{Mueller:2011nm,Huber:2011wv}. Specifically, we use all the components of the HM uncertainty budget discussed above, and we add to this the uncertainties under the columns headed ``$g_A$'' and ``Param.'' in Tables VI-IX of Ref.~\cite{Hayen:2019eop}; these latter components are each taken to be totally correlated between bins and isotopes. These uncertainties yield the dark cyan bands in Figs.~\ref{fig:fig_a}-\ref{fig:fig_d}.

The procedure outlined above gives us the following isotopic IBD yields:
\begin{align*}
\sigma_{\rm HKSS}^{235} = & \left(6.67 \pm 0.15 \right) \times 10^{-43} \text{ cm$^2$/fission } \\
\sigma_{\rm HKSS}^{238} = & \left(10.08 \pm 1.14 \right) \times 10^{-43} \text{ cm$^2$/fission }\\
\sigma_{\rm HKSS}^{239} = & \left( 4.37 \pm 0.12 \right) \times 10^{-43} \text{ cm$^2$/fission }\\
\sigma_{\rm HKSS}^{241} = & \left( 6.06 \pm 0.14 \right) \times 10^{-43}\text{ cm$^2$/fission }
\end{align*}
Note that these have increased by $\sim1\%$ relative to the nominal HM predictions. The uncertainties have also grown, stemming from uncertainty related to $g_A$ and the parametrization procedure used in Ref.~\cite{Hayen:2019eop}, but the change is quite modest --- including first-forbidden contributions has not resulted in a grossly inflated error budget.
These results are described by the following correlation matrix:
\begin{equation}
\rho_{\rm HKSS} = \left( \begin{array}{cccc} 1 & 0.057 & 0.949 & 0.972 \\
0.057 & 1 & 0.052 & 0.059 \\
0.949 & 0.052 & 1 & 0.934 \\
0.972 & 0.059 & 0.934 & 1 \end{array} \right).
\end{equation}
Unlike for the HM and \emph{ab initio} flux predictions, the flux from $^{238}$U is now correlated with the three other isotopes via the axial coupling $g_A$.


\section{Rate Experiments}
\label{sec:RateExps}
\setcounter{equation}{0}

In this section, we describe the experiments that have measured the IBD rate that enter into our analysis and how these are defined in \globes. We are ultimately interested in the ratio of the measured IBD rate relative to one of (1) the HM predictions, (2) the \emph{ab initio} predictions, or (3) the HKSS predictions. These are compared against the experimentally determined ratios relative to the same flux prediction. Many of the details regarding specific implementations of these experiments are the same between them; we relegate discussion of these to Appendix \ref{app:TheCode}.

\subsection{Short-Baseline Experiments}
\setcounter{equation}{0}

\subsubsection*{Bugey-4}

The detector width is taken to be 3.0 m and the core is assumed to be point-like. The center-to-center distance between the core and the detector is 15 m. The fuel fractions published by the collaboration \cite{Declais:1994ma} are
\[
\left( f_{235}, \, f_{238}, \, f_{239}, \, f_{241} \right) = \left( 0.614, \, 0.074, \, 0.274, \, 0.038 \right).
\]
The experimental resolution is not published by the collaboration. In the absence of further input, we take the resolution to be $6\%/\sqrt{E\rm{ [MeV]}}$, which is the stated resolution for Bugey-3. The published total IBD yield is $\sigma = 5.752 \times 10^{-43}$ cm$^{2}$/fission \cite{Declais:1994ma}; this leads us to $R_{\rm HM} = 0.941$, $R_{\rm AI} = 0.979$ and $R_{\rm HKSS} = 0.933$. Following Refs.~\cite{Giunti:2016elf,Gariazzo:2017fdh}, the experimental uncertainty is taken to be $1.4\%$, which is entirely correlated with Rovno 91.

\subsubsection*{Rovno 91}

Rovno 91 used the same detector as Bugey-4; the baseline, however, is different -- namely, 18 m. The fuel fractions published by the collaboration \cite{Kuvshinnikov:1990ry} are\footnote{Refs.~\cite{Giunti:2016elf,Gariazzo:2017fdh} use the following values in their analyses:
\[ \left( f_{235}, \, f_{238}, \, f_{239}, \, f_{241} \right) = \left(  0.606 , \, 0.074 , \, 0.277 , \, 0.043  \right). \] We use the values from the collaboration, instead.}
\[
\left( f_{235}, \, f_{238}, \, f_{239}, \, f_{241} \right) = \left( 0.614, \, 0.074, \, 0.274, \, 0.038 \right).
\]
The resolution is not published by the collaboration. In the absence of further input, we take the resolution to be $6\%/\sqrt{E\rm{ [MeV]}}$, as for Bugey-4. The absolute IBD cross section measured at Rovno 91 is $\sigma = 5.85 \times 10^{-43}$ cm$^2$/fission \cite{Kuvshinnikov:1990ry}. We find $R_{\rm HM} = 0.939$, $R_{\rm AI} = 0.983$ and $R_{\rm HKSS} = 0.931$. The total experimental uncertainty is $2.8\%$, of which $1.4\%$ is correlated with Bugey-4. 

\begin{figure}[t]
\includegraphics[width=\linewidth]{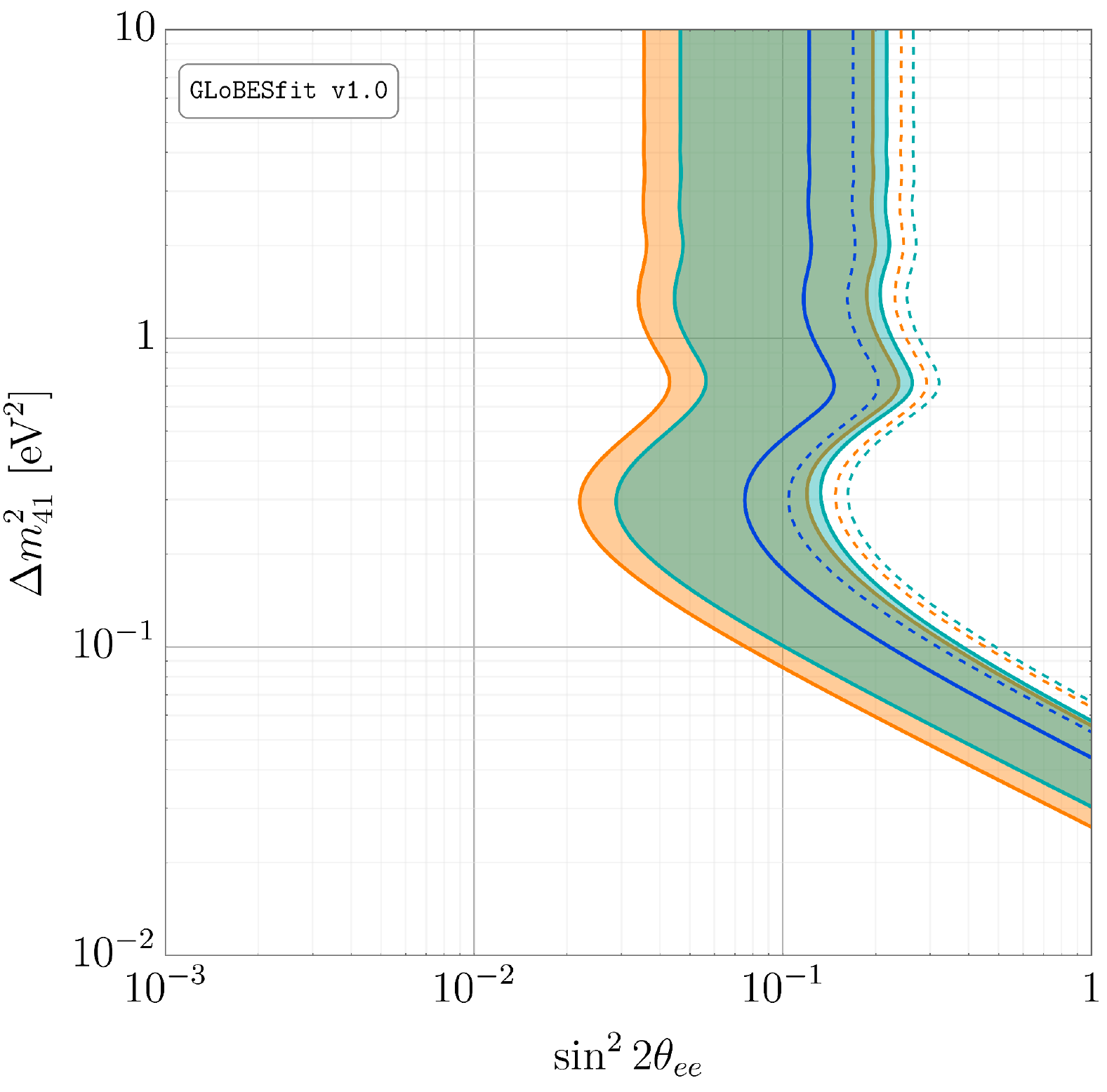}
\caption{The 68.3\% (solid) and 95\% (dashed) C.L. contours from our analyses of Bugey-4 and Rovno 91. We show results for the HM (orange), \emph{ab initio} (blue) and HKSS (dark cyan) reactor antineutrino flux models.}
\label{fig:Block0}
\end{figure}

\subsubsection*{Bugey-3}

The collaboration reports IBD rate measurements at 15 m, 40 m and 95 m. The fuel fractions published by the collaboration \cite{Declais:1994su} are:
\[
\left( f_{235}, \, f_{238}, \, f_{239}, \, f_{241} \right) = \left( 0.614, \, 0.074, \, 0.274, \, 0.038 \right),
\]
which we assume applies to all three measurements. The collaboration claims an energy resolution of $6\%/\sqrt{E\rm{ [MeV]}}$.

The collaboration only publishes the ratios of their measurements with respect to Refs.~\cite{Schreckenbach:1985ep,Hahn:1989zr}; we instead derive IBD cross sections from their stated event rates and experimental specifics and find $\sigma = \{ 5.75, \, 5.79, \, 5.33 \} \times 10^{-43}$ cm$^2$/fission. We find $R_{\rm HM} = \{ 0.941 , \, 0.947 , \, 0.872\}$, $R_{\rm AI} = \{0.979 , \, 0.985 , \, 0.907\}$; and $R_{\rm HKSS} = \{0.933 , \, 0.939 , \, 0.864\}$. The total uncertainties are, respectively, \{4.2\%, 4.3\%, 15.2\%\}, of which 4.0\% is correlated between each of these three experiments.

\begin{figure}[t]
\includegraphics[width=\linewidth]{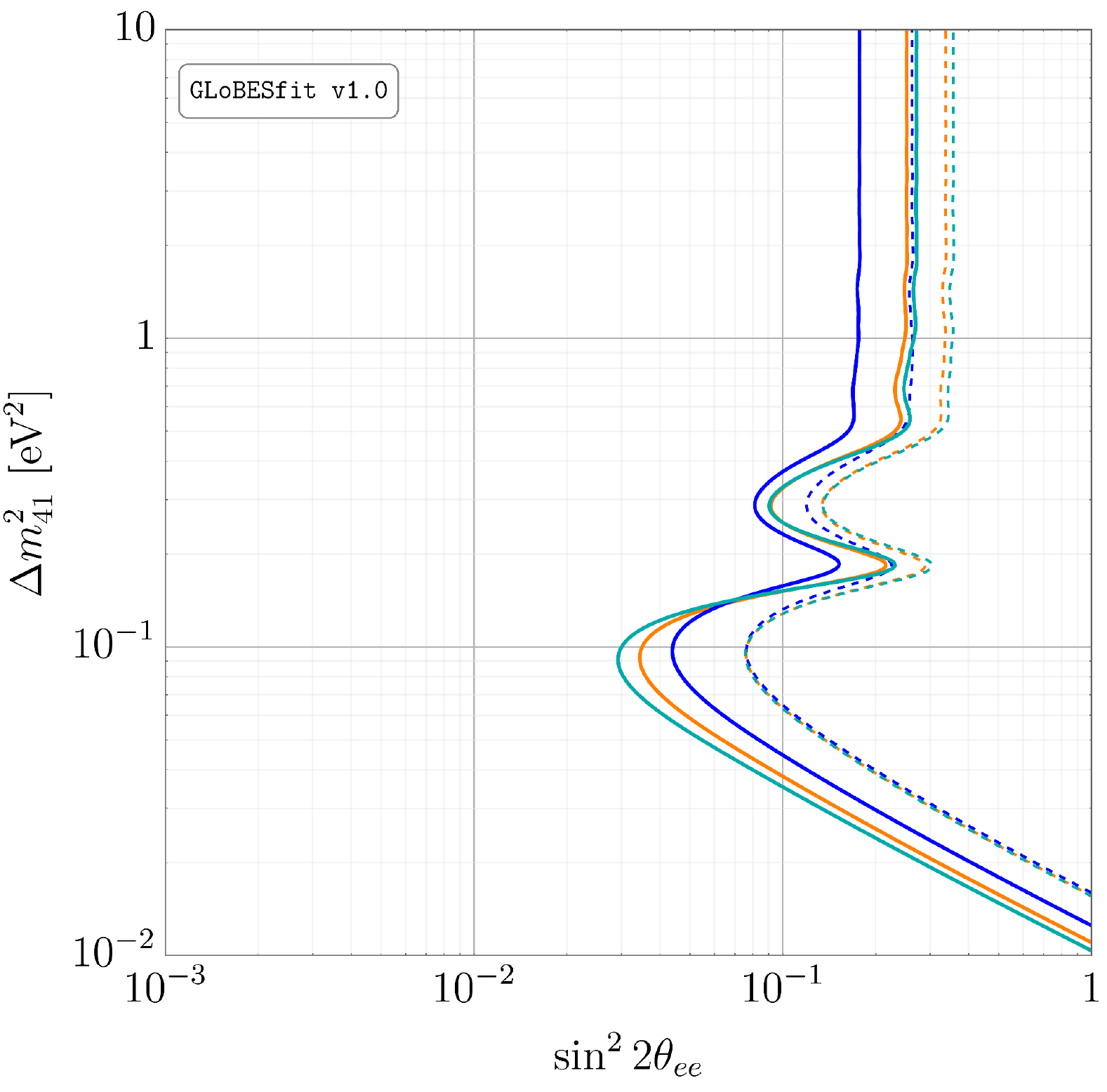}
\caption{The 68.3\% (solid) and 95\% (dashed) C.L. contours from our analysis of Bugey-3. We show results for the HM (orange), \emph{ab initio} (blue) and HKSS (dark cyan) reactor antineutrino flux models.}
\label{fig:Block1}
\end{figure}

\subsubsection*{G\"osgen}

IBD rate measurements are reported \cite{Zacek:1986cu} at 37.9 m, 45.9 m and 64.7 m; the detector is taken to be 2.0 m wide, while the core is taken to be point-like. The effective fuel fractions for the three measurements are slightly different:
\begin{align*}
\left( f_{235}, \, f_{238}, \, f_{239}, \, f_{241} \right) & =  \left( 0.619, \, 0.067, \, 0.272, \, 0.042 \right)_{\rm 38 \, m} \\
& = \left( 0.584, \, 0.068, \, 0.298, \, 0.050 \right)_{\rm 46 \, m} \\
& = \left( 0.543, \, 0.070, \, 0.329, \, 0.058 \right)_{\rm 65 \, m}
\end{align*}
The collaboration claims an energy resolution $18\%/\sqrt{E\rm{ [MeV]}}$.

From Ref.~\cite{Zacek:1986cu}, we derive $\sigma = \{ 5.52, 5.53, 5.22 \} \times 10^{-43}$ cm$^2$/fission. This leads us to $R_{\rm HM} = \{0.972, \, 0.984 , \, 0.940 \}$; $R_{\rm AI} = \{1.013 , \, 1.024 , \, 0.975\}$; and $R_{\rm HKSS} = \{0.962 , \, 0.975 , \, 0.931\}$. The total uncertainties are, respectively, \{5.4\%, 5.4\%, 6.7\%\}, of which 2.0\% is correlated between each of these three experiments; there exists an additional 3.8\% correlation between each of these and the ILL experiment. 

\subsubsection*{ILL}

We mostly follow Ref.~\cite{Kwon:1981ua}; however, we also correct for the fact that Ref.~\cite{ILLupdate} reports that the original power of the ILL experiment was underreported by 9.5\%. The detector width is 2.0 m, the core is assumed to be point-like and the center-to-center distance is 8.76 m. The core consists of 93\% enriched $^{235}$U; our analysis assumes that this is the only relevant isotope. The energy resolution is taken to be $18\%/\sqrt{E\rm{ [MeV]}}$.

We determine from Refs.~\cite{Kwon:1981ua,ILLupdate} that $\sigma = 4.76 \times 10^{-43}$ cm$^2$/fission, implying the ratios $R_{\rm HM} = 0.824$, $R_{\rm AI} = 0.881$ and $R_{\rm HKSS} = 0.816$. The experimental uncertainty is 9.1\%, of which 3.8\% is correlated with each of the three G\"osgen measurements.

\begin{figure}[t]
\includegraphics[width=\linewidth]{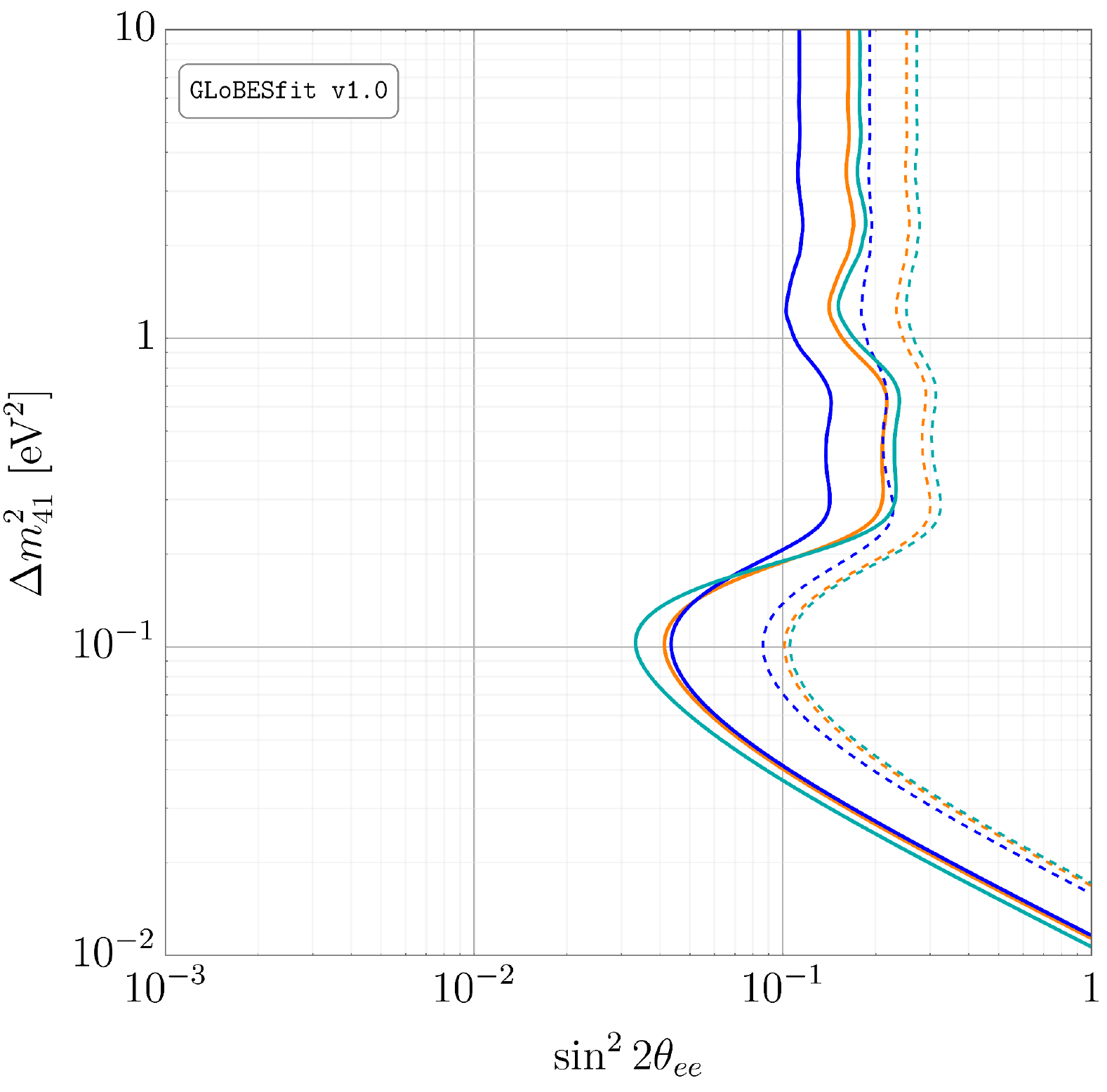}
\caption{The 68.3\% (solid) and 95\% (dashed) C.L. contours from our analysis of G\"osgen and ILL. We show results for the HM (orange), \emph{ab initio} (blue) and HKSS (dark cyan) reactor antineutrino flux models.}
\label{fig:Block2}
\end{figure}

\subsubsection*{Krasnoyarsk 87}

The detector is taken to have a width of 2.0 m while the core is point-like and comprised purely of $^{235}$U. The center-to-center distances at these experiments are 32.8 m and 92.3 m. The energy resolution is not specified in Ref.~\cite{Vidyakin:1987ue}; we assume it to be $20\%/\sqrt{E\rm{ [MeV]}}$. The collaboration explicitly publishes IBD yields for the two experiments: $\sigma = \{6.19, \, 6.30\} \times 10^{-43}$ cm$^2$/fission for \{33 m, 92 m\}. We calculate $R_{\rm HM} = \{0.936, \, 0.951\}$, $R_{\rm AI} = \{1.001, \, 1.018\}$ and $R_{\rm HKSS} = \{0.927, \, 0.942\}$. The respective errors are \{5.0\%, 20.4\%\}, of which 4.1\% is fully correlated between these experiments.

\begin{figure}[t]
\includegraphics[width=\linewidth]{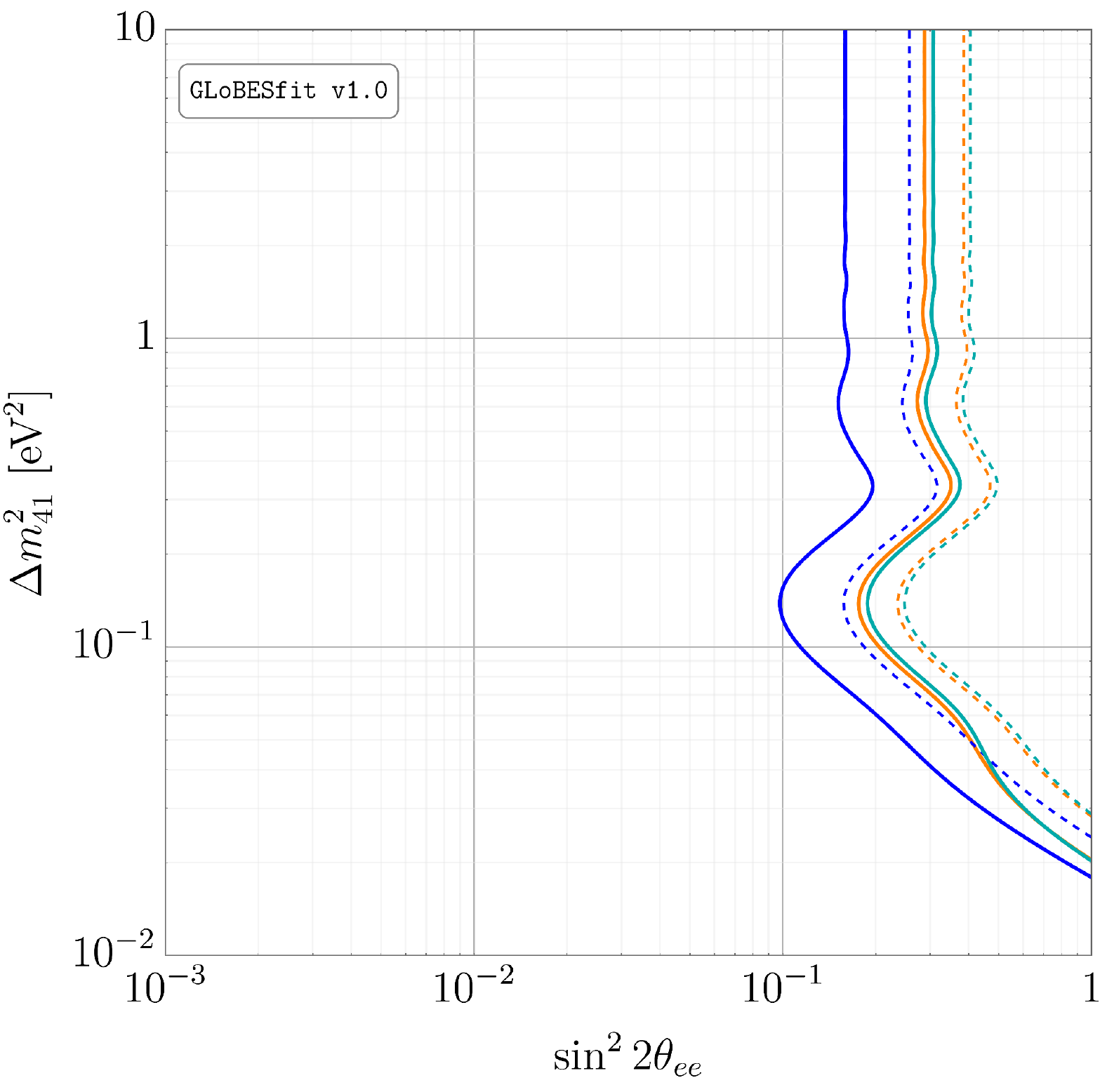}
\caption{The 68.3\% (solid) and 95\% (dashed) C.L. contours from our analysis of Krasnoyarsk 87. We show results for the HM (orange), \emph{ab initio} (blue) and HKSS (dark cyan) reactor antineutrino flux models.}
\label{fig:Block3}
\end{figure}

\subsubsection*{Krasnoyarsk 94}

We essentially repeat the analysis of Krasnoyarsk 87, except that the center-to-center distance is now taken to be 57.0 m. As before, the reactor is approximated to be purely $^{235}$U and the energy resolution is $20\%/\sqrt{E\rm{ [MeV]}}$. The collaboration publishes the IBD yield as $\sigma = 6.26 \times 10^{-43}$ cm$^2$/fission \cite{Vidyakin:1994ut}; this results in $R_{\rm HM} = 0.945$, $R_{\rm AI} = 1.011$ and $R_{\rm HKSS} = 0.936$. The uncertainty is 4.2\% and is uncorrelated with any other experiment.

\begin{figure}[t]
\includegraphics[width=\linewidth]{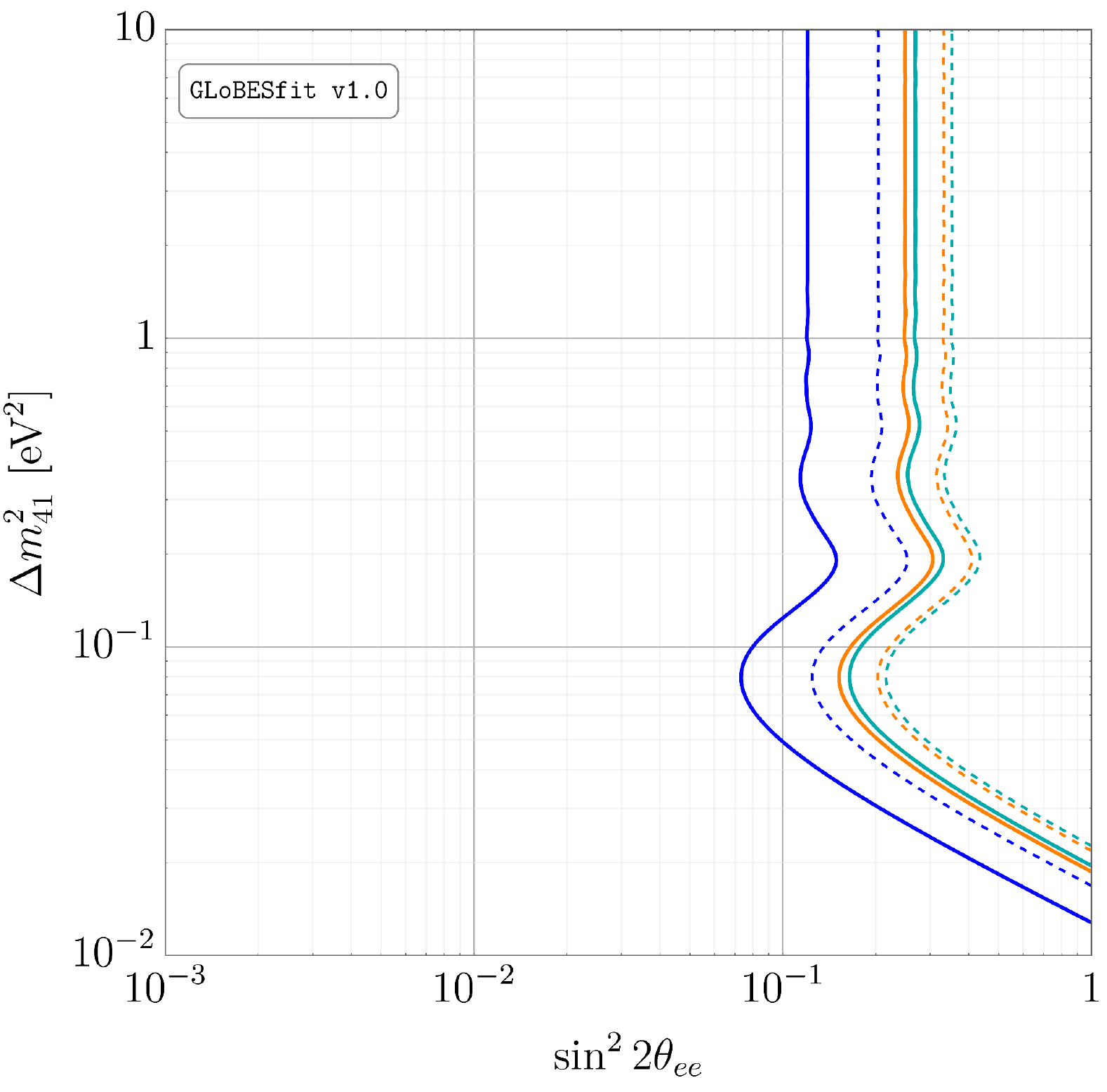}
\caption{The 68.3\% (solid) and 95\% (dashed) C.L. contours from our analysis of Krasnoyarsk 94. We show results for the HM (orange), \emph{ab initio} (blue) and HKSS (dark cyan) reactor antineutrino flux models.}
\label{fig:Block4}
\end{figure}

\subsubsection*{Krasnoyarsk 99}

We again re-use the Krasnoyarsk analysis with a center-to-center distance of 34.0 m and a core of pure $^{235}$U. The energy resolution is again taken to be $20\%/\sqrt{E\rm{ [MeV]}}$. The collaboration publishes the IBD yield as $\sigma = 6.39 \times 10^{-43}$ cm$^2$/fission \cite{Kozlov:1999ct}; this yields $R_{\rm HM} = 0.964$, $R_{\rm AI} = 1.032$ and $R_{\rm HKSS} = 0.956$. The corresponding uncertainty is 3.0\%, which is uncorrelated with any other experiment.

Ref.~\cite{Kozlov:1999ct} also includes measurements of charged- and neutral-current deuterium disintegration rates, but these are significantly less precise than the IBD rate measurement; we do not include these in our fits. 

\begin{figure}[t]
\includegraphics[width=\linewidth]{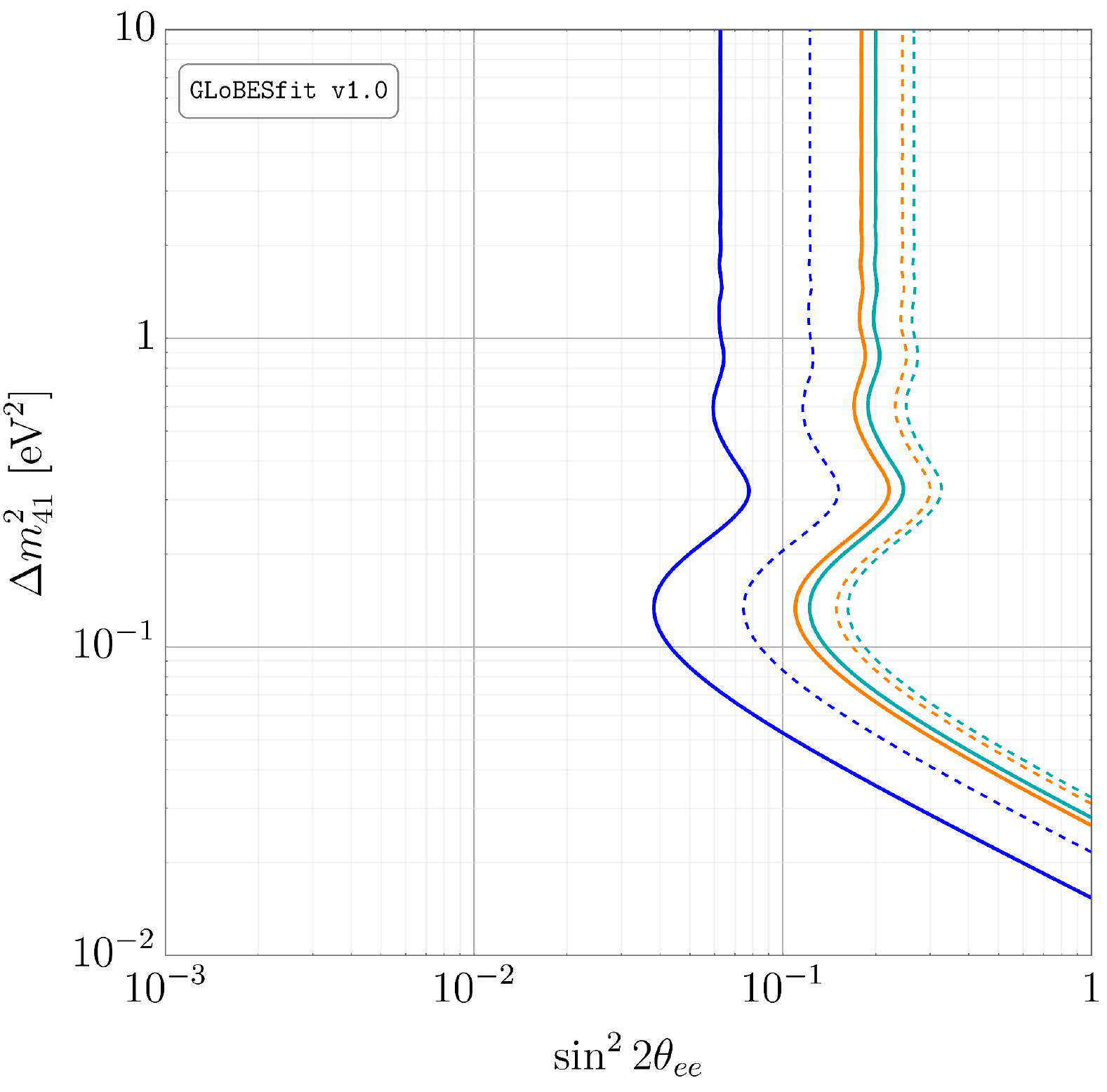}
\caption{The 68.3\% (solid) and 95\% (dashed) C.L. contours from our analysis of Krasnoyarsk 99. We show results for the HM (orange), \emph{ab initio} (blue) and HKSS (dark cyan) reactor antineutrino flux models.}
\label{fig:Block5}
\end{figure}

\subsubsection*{Savannah River}

The detector is taken to have a width of 2.0 m and the center-to-center distances are 18.2 m and 23.8 m in the two detector configurations with a core of essentially pure $^{235}$U. The energy resolution is taken to be $20\%/\sqrt{E\rm{ [MeV]}}$. The collaboration does not explicitly report their inferred IBD rate, but from the integrated rates in Table III  of Ref.~\cite{Greenwood:1996pb} and other experimental information, we derive $\sigma = \{ 6.07, 6.48 \} \times 10^{-43}$ cm$^2$/fission. These imply $R_{\rm HM} = \{0.917, \, 0.978\}$; $R_{\rm AI} = \{0.981, \, 1.047\}$; and $R_{\rm HKSS} = \{0.908, \, 0.969\}$. The respective uncertainties are \{2.8\%, 2.9\%\}, which are each uncorrelated with any other experiment, including each other.

\begin{figure}[t]
\includegraphics[width=\linewidth]{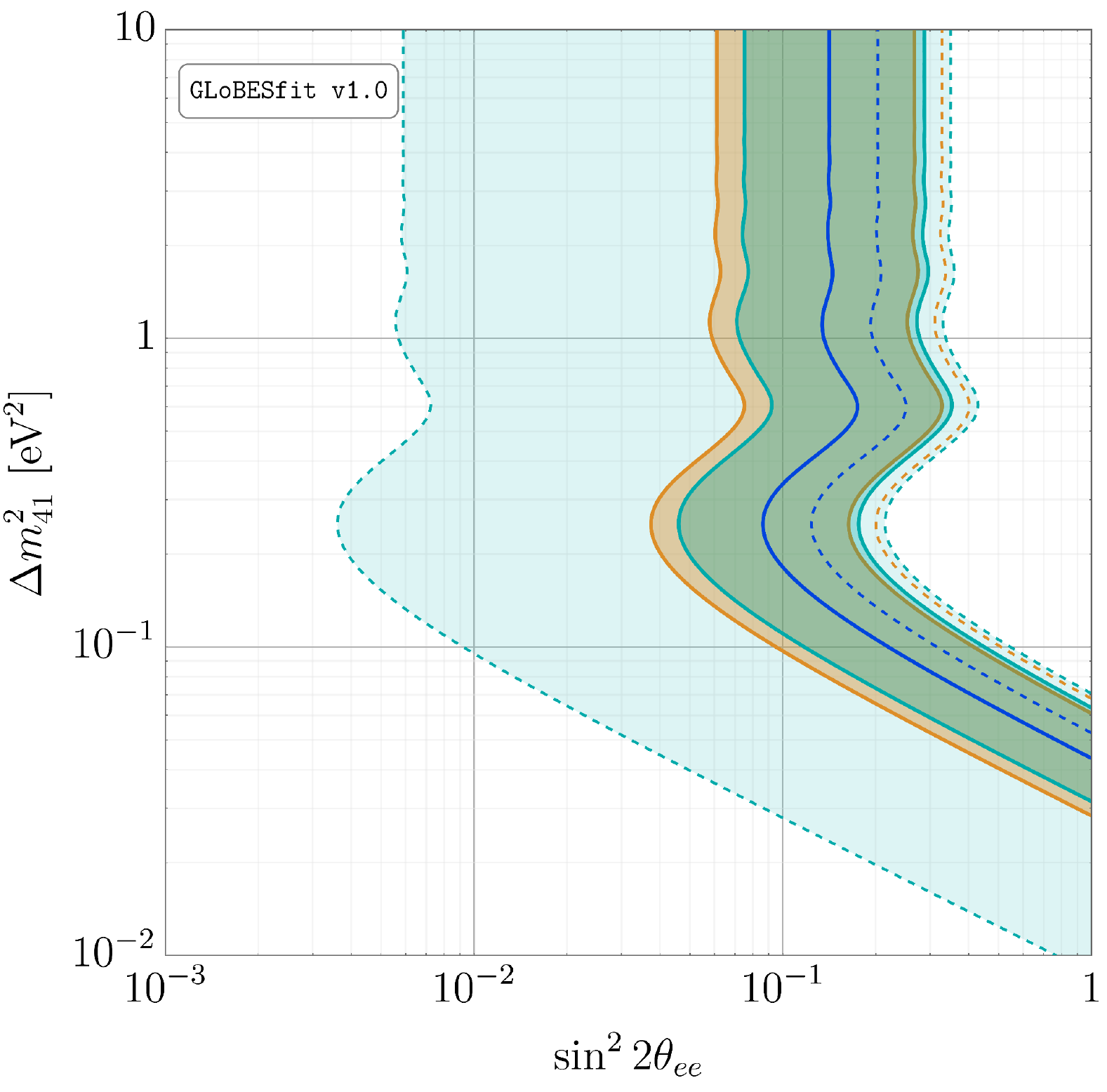}
\caption{The 68.3\% (solid) and 95\% (dashed) C.L. contours from our analysis of Savannah River (18 m). We show results for the HM (orange), \emph{ab initio} (blue) and HKSS (dark cyan) reactor antineutrino flux models.}
\label{fig:Block6}
\end{figure}

\begin{figure}[t]
\includegraphics[width=\linewidth]{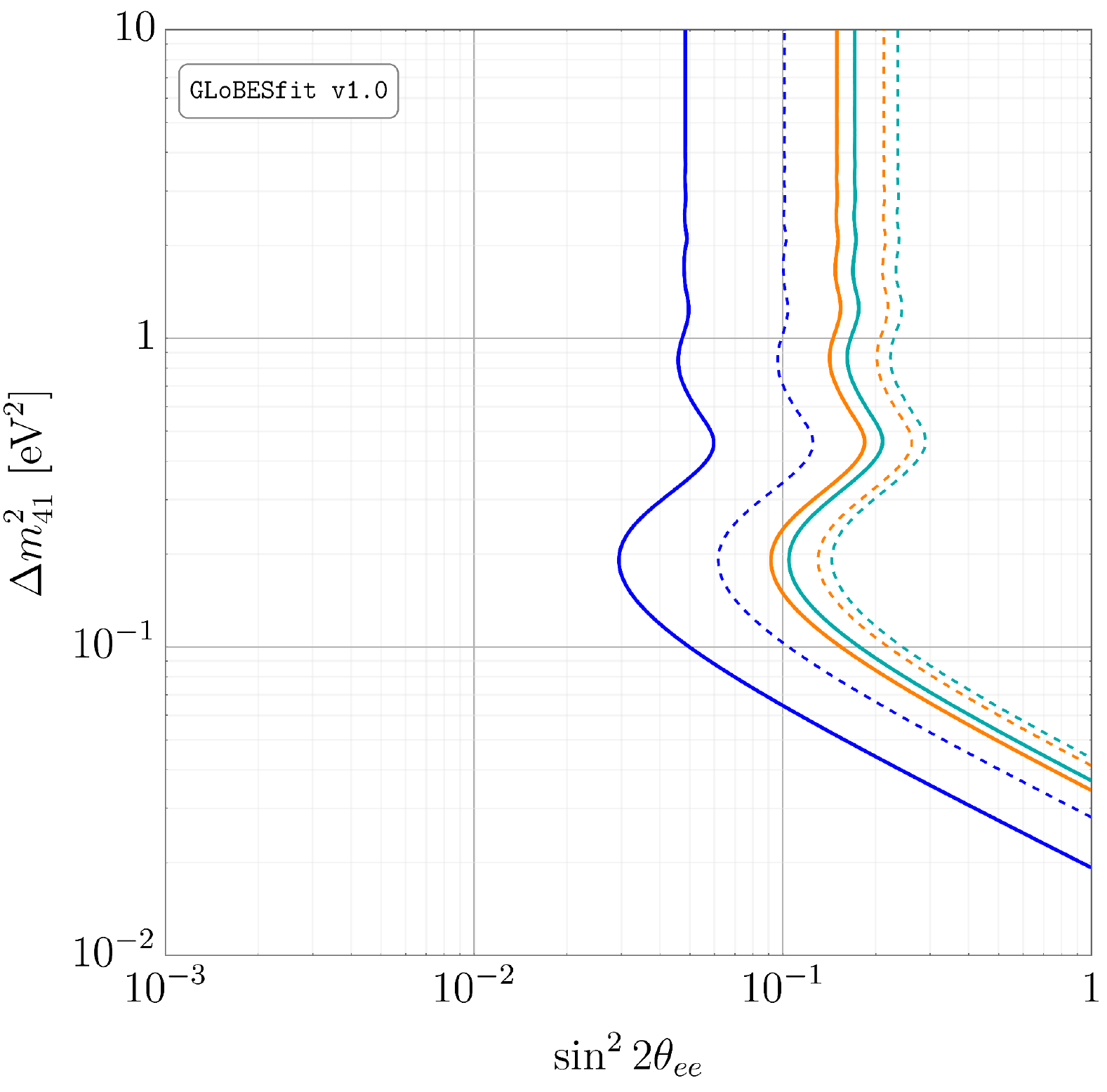}
\caption{The 68.3\% (solid) and 95\% (dashed) C.L. contours from our analysis of Savannah River (24 m). We show results for the HM (orange), \emph{ab initio} (blue) and HKSS (dark cyan) reactor antineutrino flux models.}
\label{fig:Block7}
\end{figure}

\subsubsection*{Rovno 88}

There are five measurements under the Rovno 88 umbrella \cite{Afonin:1988gx}. The first two (1I and 2I) are polyethylene detectors studded with $^3$He proportional counters, each located $\sim$18 m from the core. The latter three (1S, 2S and 3S) are liquid scintillation detectors; 1S and 3S are located 18 m from the core, while 2S is located 25 m from the core.\footnote{The 18 m experiments all formally have slightly different baselines --- this can be seen clearly in the values of $\sqrt{\langle R^2 \rangle}$ in Table II of Ref.~\cite{Afonin:1988gx}. However, we expect these differences to be small, so we ignore them.} Each detector has a width of 1.0 m, which is roughly consistent with Fig.~1 of Ref.~\cite{Afonin:1988gx}, and the resolution is taken to be $20\%/\sqrt{E\rm{ [MeV]}}$.

Each measurements takes different effective fuel fractions:
\begin{align*}
\left( f_{235}, \, f_{238}, \, f_{239}, \, f_{241} \right) & = \left( 0.607, \, 0.074, \, 0.277, \, 0.042 \right)_{\rm 1I} \\
& = \left( 0.603, \, 0.076, \, 0.276, \, 0.045 \right)_{\rm 2I} \\
& = \left( 0.606, \, 0.074, \, 0.277, \, 0.043 \right)_{\rm 1S} \\
& = \left( 0.557, \, 0.076, \, 0.313, \, 0.054 \right)_{\rm 2S} \\
& = \left( 0.606, \, 0.074, \, 0.274, \, 0.046 \right)_{\rm 3S}
\end{align*}

In Table III of Ref.~\cite{Afonin:1988gx}, the collaboration has provided their inferred IBD rate for each of these five experiments. However, in Table II of the same reference, they provide various experimental data that can, in principle, be used to rederive their results. Using the information in that table, Eq.~(8) and the updated energies per fission \cite{Kopeikin:2012zz}, we derive $\sigma = \{ 5.623, \, 6.023, \, 5.961, \, 6.231, \, 5.778 \} \times 10^{-43}$ cm$^2$/fission. From this, we calculate
\begin{eqnarray*}
& R_{\rm HM} = \{ 0.905, \, 0.969, \, 0.960, \, 1.018, \, 0.930 \}, & \\
& R_{\rm AI} = \{ 0.945, \, 1.011, \, 1.002, \, 1.059, \, 0.970 \}, & \\
& R_{\rm HKSS} = \{ 0.897, \, 0.959, \, 0.951, \, 1.008, \, 0.921 \}, &
\end{eqnarray*}
where the order here is \{1I, 2I, 1S, 2S, 3S\}. The respective errors are \{6.4\%, 6.4\%, 7.3\%, 7.3\%, 6.8\%\}; there exists a 3.1\% correlated error between 1I and 2I, as well as between each of 1S, 2S and 3S; moreover, there exists a further 2.2\% correlation between either of 1I and 2I and any of 1S, 2S and 3S.

\begin{figure}[t]
\includegraphics[width=\linewidth]{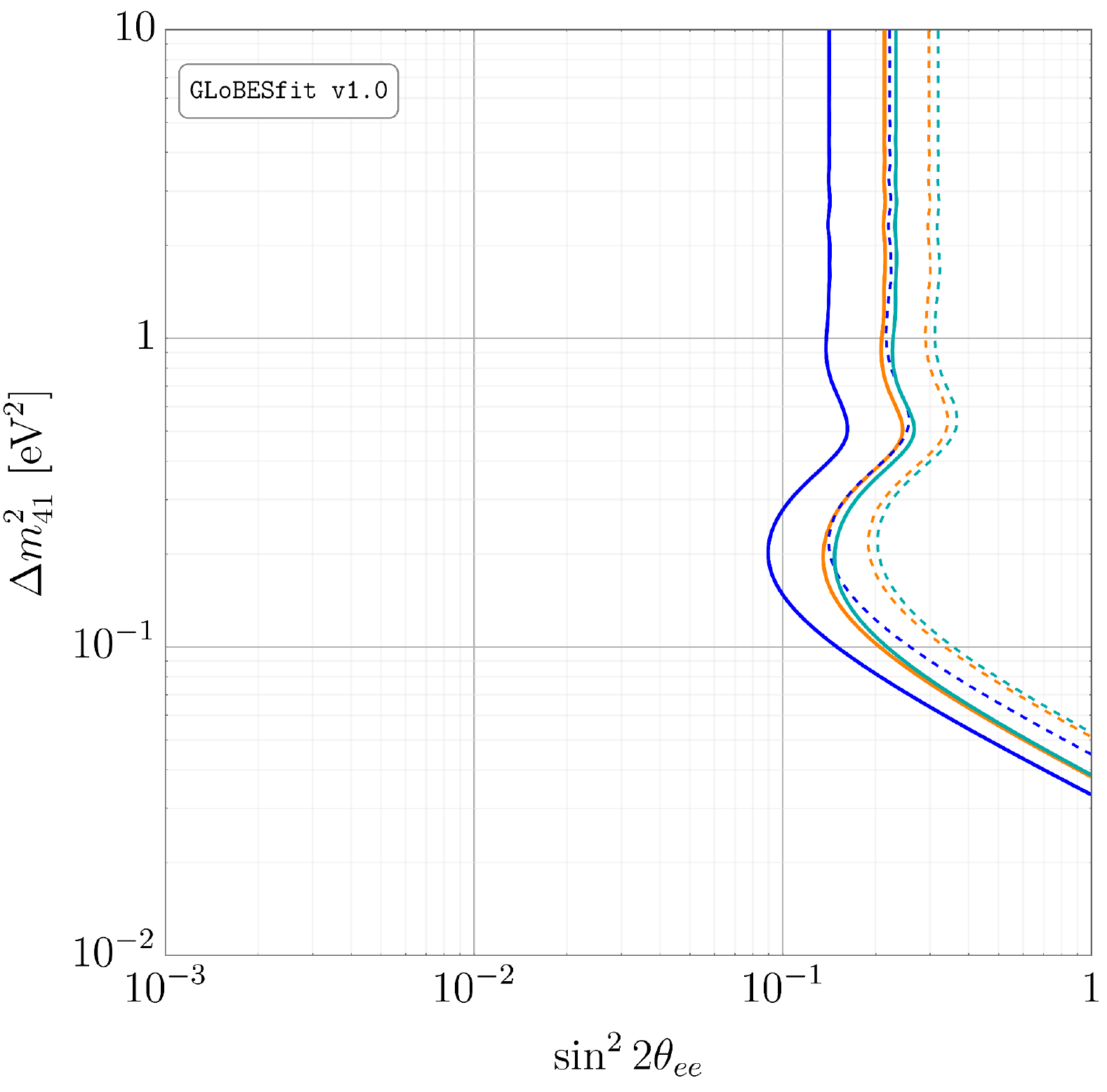}
\caption{The 68.3\% (solid) and 95\% (dashed) C.L. contours from our analysis of Rovno 88. We show results for the HM (orange), \emph{ab initio} (blue) and HKSS (dark cyan) reactor antineutrino flux models.}
\label{fig:Block8}
\end{figure}

\subsubsection*{Nucifer}

Given the short distance between the core and detector (7.21 m), one might worry that the finite extent of both the reactor core and of the detector may be important in determining the correct oscillation probabilities. However, given that Nucifer reports a total rate measurement, and given the relatively poor energy resolution, we find that this effect is not quantitatively important here.

The composition of the core is \cite{Boireau:2015dda}
\[ \left( f_{235}, \, f_{238}, \, f_{239}, \, f_{241} \right) = \left( 0.926, \, 0.008, \, 0.061, \, 0.005 \right). \]
The core is dominated by $^{235}$U, but since the effective fuel fractions for all four isotopes have been provided, we take these into account. The collaboration claims an energy resolution of $20\%/\sqrt{E\rm{ [MeV]}}$. From their reported event rate and other experimental specifics, we estimate the IBD rate to be $\sigma = 6.847 \times 10^{-43}$ cm$^2$/fission. This leads to $R_{\rm HM} = 1.046$, $R_{\rm AI} = 1.115$ and $R_{\rm HKSS} = 1.036$. The experimental uncertainty is 10.7\% and is uncorrelated with any other experiment.

\begin{figure}[t]
\includegraphics[width=\linewidth]{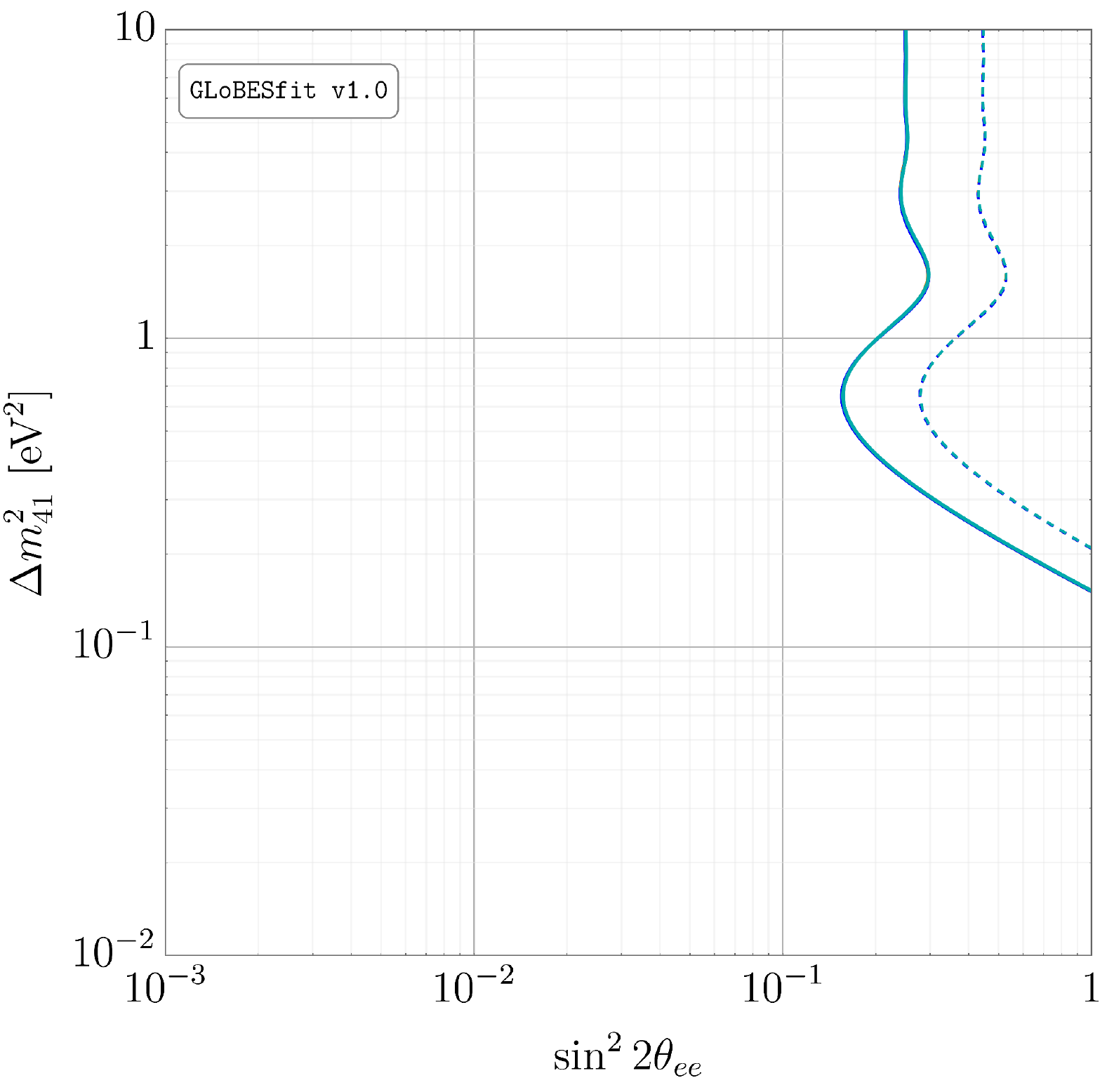}
\caption{The 68.3\% (solid) and 95\% (dashed) C.L. contours from our analysis of Nucifer. We show results for the HM (orange), \emph{ab initio} (blue) and HKSS (dark cyan) reactor antineutrino flux models.}
\label{fig:Block9}
\end{figure}


\subsection{Medium-Baseline Experiments -- Total Rates}

We now turn our attention to medium-baseline oscillation experiments. At these distances, $\mathcal{O}(10^2-10^3)$ m, two-flavor oscillations are insufficient; we must at least account for nonzero $\theta_{13}$ and $\Delta m_{31}^2$ (or the effective mass splitting $\Delta m_{ee}^2$). In practice, we include full four-neutrino oscillations, though we keep the relevant parameters from three-neutrino oscillations fixed at their best-fit values from Ref.~\cite{Esteban:2018azc}:
\begin{align*}
\Delta m_{21}^2 & = 7.39 \times 10^{-5} \text{ eV}^2; \\
\Delta m_{31}^2 & = 2.525 \times 10^{-3} \text{ eV}^2; \\
\sin^2 \theta_{12} & = 0.310; \\
\sin^2 \theta_{13} & = 0.02240.
\end{align*}
Note that $\sin^2 \theta_{23}$ does not factor into this analysis.

We define each detector width to be 4.0 m and take the reactors to be point-like; this helps high-frequency oscillations to average out smoothly, but otherwise does not impact our analysis. Since medium-baseline experiments are sourced by multiple reactors, our \globes \, definitions must account for the relative contributions of the reactors.

As mentioned, three-neutrino effects are important at these baselines; this can cause some ambiguity in the definition of the ratio $R$. In this work, we always assume that $R$ is defined with respect to \emph{the absence of oscillations}. We contrast this with the \emph{expected} deficit within the three-neutrino framework relative to the absence of oscillations, which we will call $R_{3\nu}$.

\subsubsection*{Palo Verde}

Palo Verde \cite{Boehm:2001ik} consists of three reactor cores; one is located 750 m from the detector and the other two are at 890 m. Table I of Ref.~\cite{Boehm:2001ik} presents the operating cycle over the course of the experiment. Assuming the total power of 11.63 GW$_{\rm th}$ is split evenly among the three reactors over the duration of the experiment and accounting for efficiency, the total exposure from the 750 m reactor is 372.6 GW$\cdot$yr, whereas the total exposure from both 890 m reactors is 706.0 GW$\cdot$yr.
The fuel fractions are not stated in Ref.~\cite{Boehm:2001ik}, so we take them from Refs.~\cite{Giunti:2016elf,Gariazzo:2017fdh}:
\[ \left( f_{235}, \, f_{238}, \, f_{239}, \, f_{241} \right) = \left( 0.600, \, 0.070, \, 0.270, \, 0.060 \right). \]
The resolution is also not stated in Ref.~\cite{Boehm:2001ik}, so we assume it to be $20\%/\sqrt{E\rm{ [MeV]}}$.

We use information in Ref.~\cite{Boehm:2001ik}, along with the total proton number in Ref.~\cite{Miller:2001qv},\footnote{We thank Giorgio Gratta for pointing us to this reference \cite{GGratta}.} to determine $\sigma = 6.036\times 10^{-43}$ cm$^2$/fission. From this, we calculate $R_{\rm HM} = 0.971$, $R_{\rm AI} = 1.014$ and $R_{\rm HKSS} = 0.963$. The uncertainty is taken to be 5.4\%, which is uncorrelated with any other experiment. For context, Ref.~\cite{Zhang:2013ela} reports a three-neutrino expectation of $R_{3\nu} = 0.967$.

\begin{figure}[t]
\includegraphics[width=\linewidth]{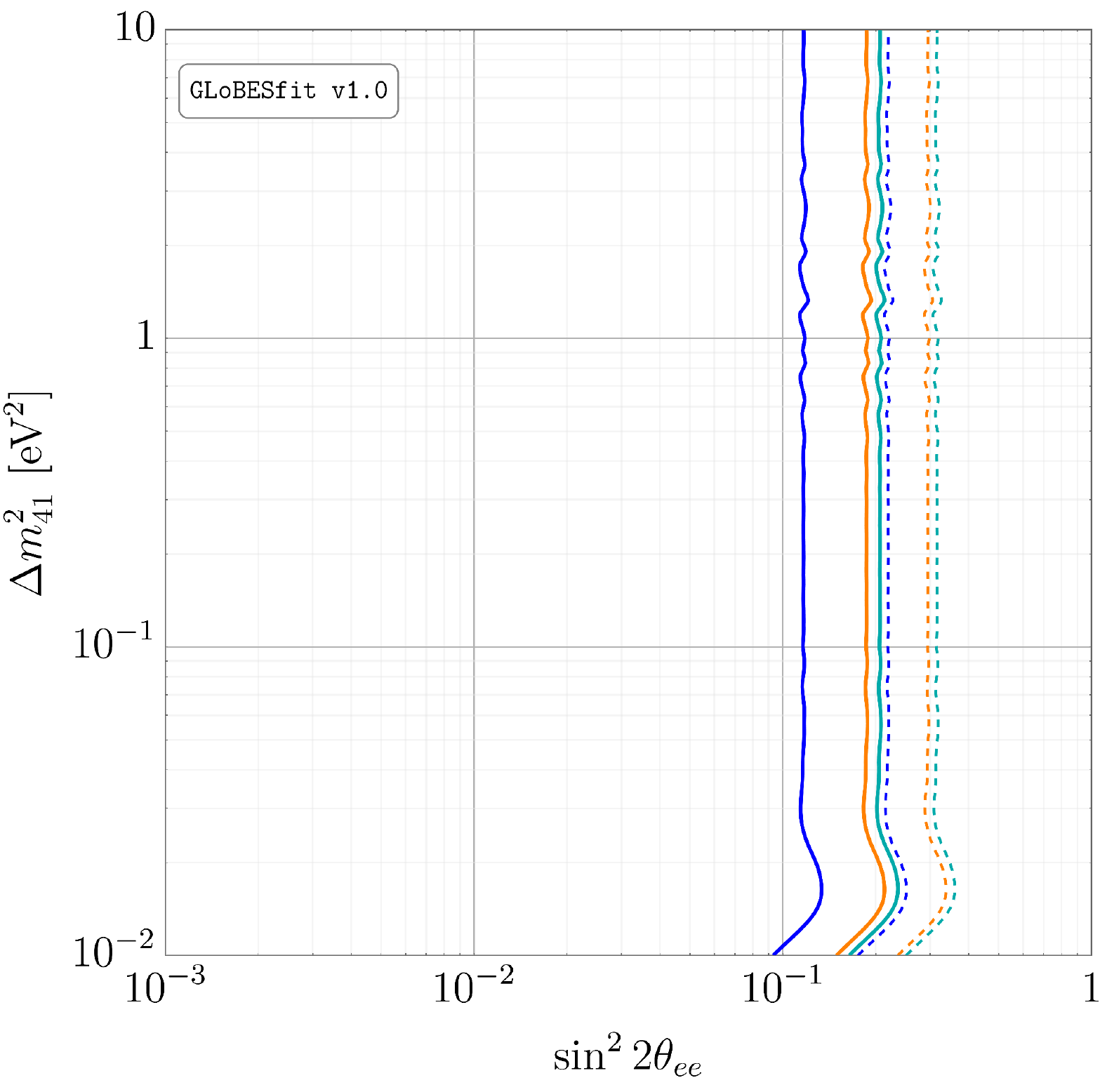}
\caption{The 68.3\% (solid) and 95\% (dashed) C.L. contours from our analysis of Palo Verde. We show results for the HM (orange), \emph{ab initio} (blue) and HKSS (dark cyan) reactor antineutrino flux models.}
\label{fig:Block10}
\end{figure}

\subsubsection*{Double Chooz}

\begin{figure}[t]
\includegraphics[width=\linewidth]{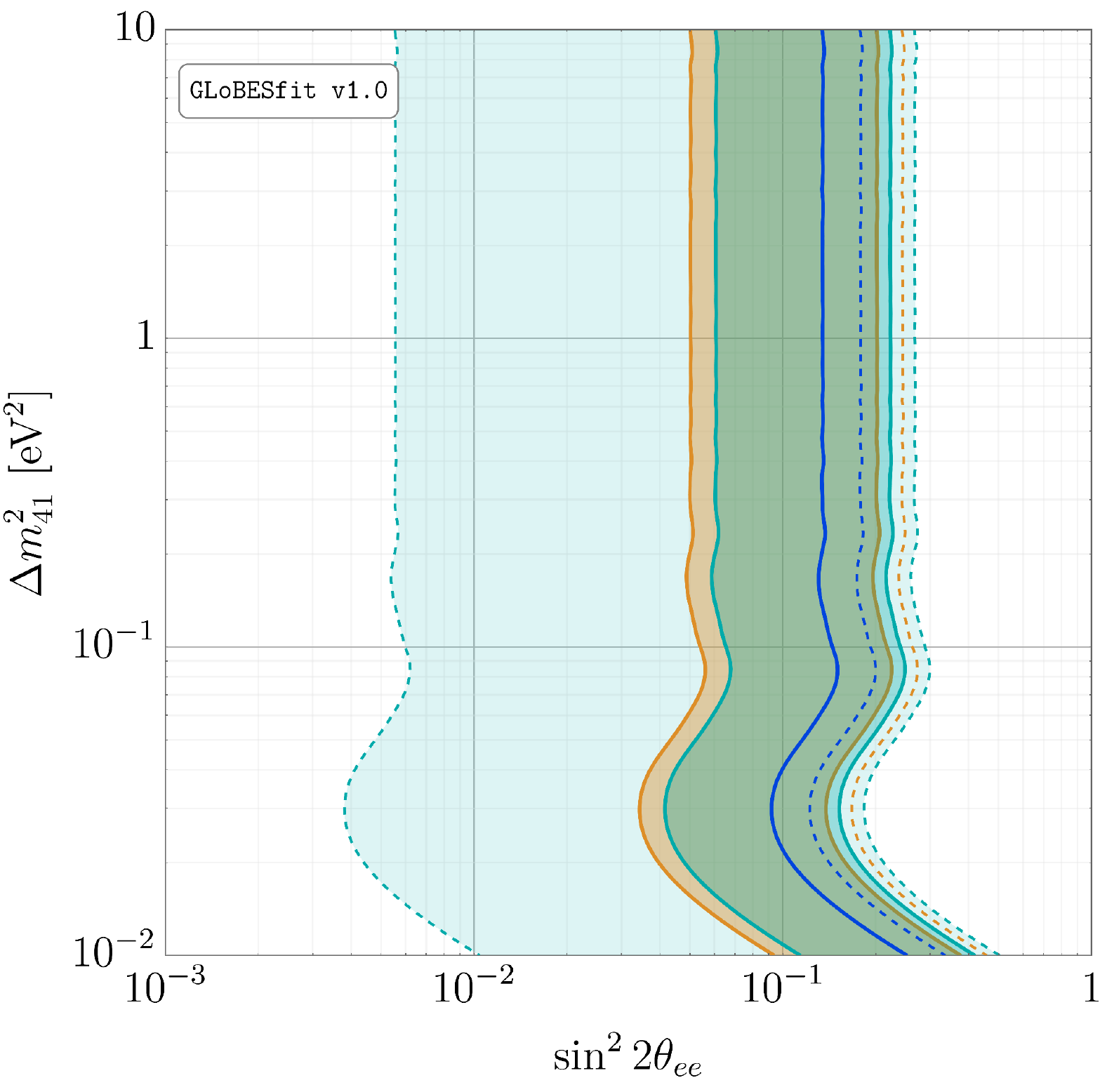}
\caption{The 68.3\% (solid) and 95\% (dashed) C.L. contours from our analysis of Double Chooz. We show results for the HM (orange), \emph{ab initio} (blue) and HKSS (dark cyan) reactor antineutrino flux models.}
\label{fig:Block11}
\end{figure}

Double Chooz \cite{Abe:2014bwa, DoubleChooz:2019qbj} consists of two reactors and two detectors; this measurement pertains to the near detector, from which the two reactors are 355 m and 469 m away. We assume that each reactor operates for the same amount of time over the course of the experiment, and that the average powers delivered during those times are the same. The effective fuel fractions are assumed to be the same for each reactor:
\[ \left( f_{235}, \, f_{238}, \, f_{239}, \, f_{241} \right) = \left( 0.520, \, 0.087, \, 0.333, \, 0.060 \right). \]
We take the energy resolution to be $8\%/\sqrt{E}$ \cite{Abe:2014bwa}.

Ref.~\cite{DoubleChooz:2019qbj} reports $\sigma = 5.71 \times 10^{-43}$ cm$^2$/fission. However, this result has been corrected for the effects of nonzero $\theta_{13}$. We calculate $R_{\rm HM} = 0.934$, $R_{\rm AI} = 0.969$ and $R_{\rm HKSS} = 0.926$. The corresponding uncertainty is 1.4\%, which is uncorrelated with any other experiment.

\subsubsection*{Chooz}

\begin{figure}[t]
\includegraphics[width=\linewidth]{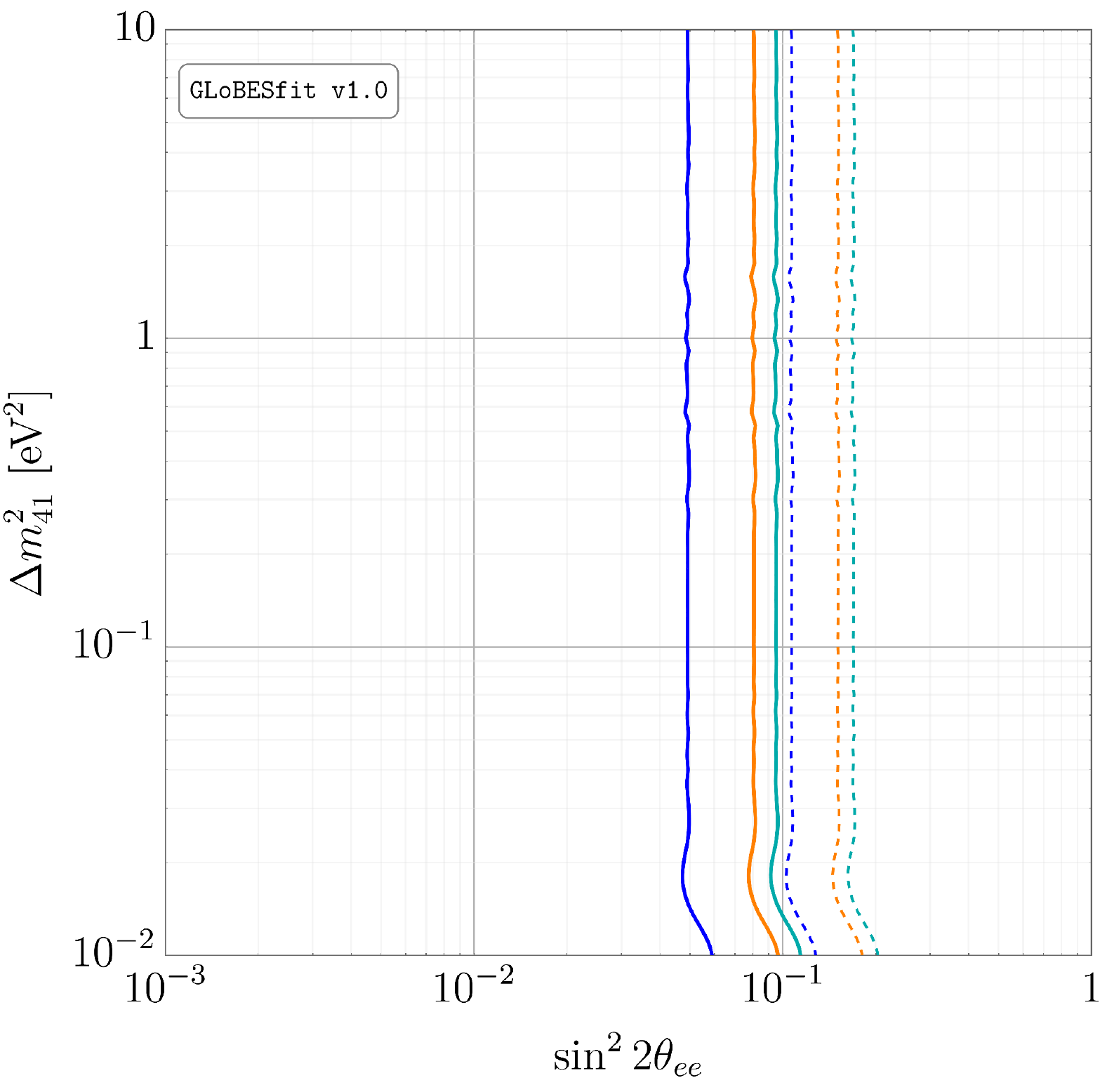}
\caption{The 68.3\% (solid) and 95\% (dashed) C.L. contours from our analysis of Chooz. We show results for the HM (orange), \emph{ab initio} (blue) and HKSS (dark cyan) reactor antineutrino flux models.}
\label{fig:Block12}
\end{figure}

Chooz \cite{Apollonio:2002gd} consists of two reactors located 998 m and 1115 m away from the detector. The two reactors have different run times and effective powers; the operation periods are broken down in Table 1 of Ref.~\cite{Apollonio:2002gd}. We assume that the reactors operate at the same average power during the period in which both are operational; this implies that reactor 1 (1115 m) delivered 12715.5 GW$\cdot$h over the experiment, while reactor 2 (998 m) delivered 8556.5 GW$\cdot$h. The effective fuel fractions are assumed to be the same for each reactor:
\[ \left( f_{235}, \, f_{238}, \, f_{239}, \, f_{241} \right) = \left( 0.496, \, 0.087, \, 0.351, \, 0.066 \right). \]
We take the energy resolution to be 0.5 MeV; Ref.~\cite{Apollonio:2002gd} claims a resolution of 0.33 MeV, but we use the more conservative estimate here. 

From Ref.~\cite{Apollonio:2002gd}, we derive $\sigma = 5.71 \times 10^{-43}$ cm$^2$/fission, giving us $R_{\rm HM} = 0.976$, $R_{\rm AI} = 1.013$ and $R_{\rm HKSS} = 0.968$. The experimental uncertainty is 3.2\%, which is uncorrelated with any other experiment. For context, Ref.~\cite{Zhang:2013ela} reports a three-neutrino expectation of $R_{3\nu} = 0.954$.


\subsection{Medium-Baseline Experiments -- Rate Evolution}

\subsubsection*{Daya Bay}

\begin{figure}[t]
\includegraphics[width=\linewidth]{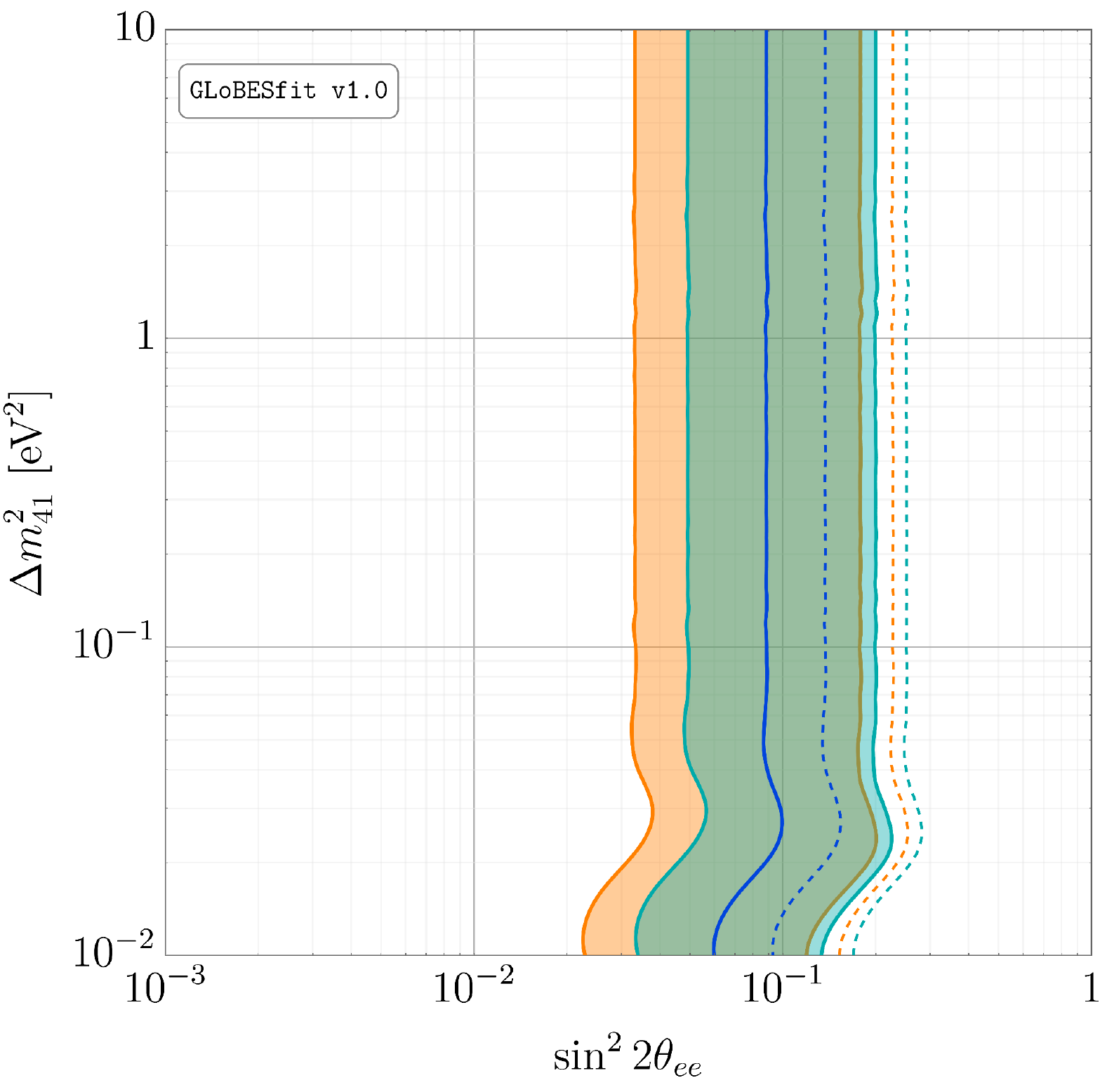}
\caption{The 68.3\% (solid) and 95\% (dashed) C.L. contours from our analysis of Daya Bay. We show results for the HM (orange), \emph{ab initio} (blue) and HKSS (dark cyan) reactor antineutrino flux models.}
\label{fig:Block13}
\end{figure}

The details about the experimental geometry are well documented in Refs.~\cite{An:2016srz,An:2016ses}. We tabulate relevant experimental specifics in Tables \ref{tab:DBlengths}-\ref{tab:DB_info} in Appendix \ref{app:SuppData}. The fuel evolution result that we use corresponds to the 1230-day data release; an analogous analysis for the 1958-day data appeared in Ref.~\cite{Adey:2019ywk}, which we look forward to including in the future.

AD8 did not operate for as long as AD1-3 in the 1230-day dataset; this is characterized by the exposure for detector $d$, defined as
\begin{equation}
\sum_r \left( P_r^{\rm 6AD} t^{\rm 6AD}_d + P_r^{\rm 8AD} t^{\rm 8AD}_d \right),
\end{equation}
where $P^{\rm 6AD, 8AD}_r$ represents the average power of reactor $r$ during either the 6AD or 8AD period, and $t^{\rm 6AD, 8AD}_d$ is the operating time of the detector during these periods. The powers are presented in Table \ref{tab:DBpowers}. For all detectors, $t^{\rm 8AD}_d = 920$ days; for ADs 1-3, $t^{\rm 6AD}_d = 182$ days, but AD8 has $t^{\rm 6AD}_d = 0$ days. Moreover, we account for the different total AD target masses and efficiencies, given in Table \ref{tab:DBfracs}.

The Daya Bay data are presented in bins of differing fuel fractions. We present the average fuel fractions of these bins and the corresponding IBD rate measurements in Table \ref{tab:DBfracs2}. These IBD rates have been inferred by the collaboration accounting for three-flavor oscillations. To construct our analyses, we correct these measurements by reinserting these effects, assuming the best-fit values presented in Ref.~\cite{An:2016ses}:
\begin{align*}
\sin^2 2\theta_{13} & = 0.0841 \\
\Delta m_{ee}^2 & = 2.50 \times 10^{-3} \text{ eV}^2
\end{align*}
Our calculations of the ratios $R_{\rm HM}$, $R_{\rm AI}$ and $R_{\rm HKSS}$ are also shown in Table \ref{tab:DBfracs2}. Our analyses employ the covariance matrices provided by the collaboration in the Supplementary Material to Ref.~\cite{An:2017osx}, appropriately converted from absolute IBD rate to ratio with respect to our IBD predictions. Ref.~\cite{Adey:2018qct} presents a total IBD rate measurement whose systematic uncertainty has been decreased by 29\% relative to prior measurements. Following the prescription of Ref.~\cite{Giunti:2019qlt}, we have decreased the systematic uncertainties from Ref.~\cite{An:2017osx} in our analyses by this amount.

\subsubsection*{RENO}

In Table \ref{tab:RENObaselines}, we show the distances between each RENO reactor and the near and far detectors; in Table \ref{tab:RENOpowers}, we show the average power of each reactor during the experiment;\footnote{We thank Soo-Bong Kim for providing this information \cite{SBKim}.} and in Table \ref{tab:RENOinfo}, we show other important experimental information, including the detector operating time and efficiency. These data will also be used in the RENO spectral analysis; here, we only need the information for the near detector.

We take the flux-weighted average baseline to be 433.1 m. Out definition of ``average baseline'' is different from that of RENO \cite{RENO:2018pwo}; we weight by the reactor flux at the detector, but RENO considers the average distance that a neutrino that has already interacted in the detector would have traveled. This is why we don't use the 410.6 m that the collaboration publishes. We take the energy resolution to be a constant 0.4 MeV.

As with Daya Bay, the RENO collaboration reports their total IBD yield for eight different sets of fuel fractions; we present these fuel fractions and the corresponding measured IBD rates, obtained by digitizing Figure 2 in Ref.~\cite{RENO:2018pwo}, in Table \ref{tab:RENOfracs}. As with Daya Bay, these IBD rates have been corrected for that three-flavor oscillations that have been removed in these measurements, assuming the best-fit values presented in Ref.~\cite{Bak:2018ydk}:
\begin{align*}
\sin^2 2\theta_{13} & = 0.0896 \\
\Delta m_{ee}^2 & = 2.68 \times 10^{-3} \text{ eV}^2
\end{align*}
In Table \ref{tab:RENO_info}, we show our estimates of the ratios using the HM, \emph{ab initio} and HKSS fluxes. The systematic uncertainty is given by the (quadrature) sum of the thermal power (0.5\%), fission fraction (0.7\%) and detection efficiency (1.93\%) uncertainties; this totals 2.1\% and is assumed to be totally correlated between data points.

\begin{figure}[t]
\includegraphics[width=\linewidth]{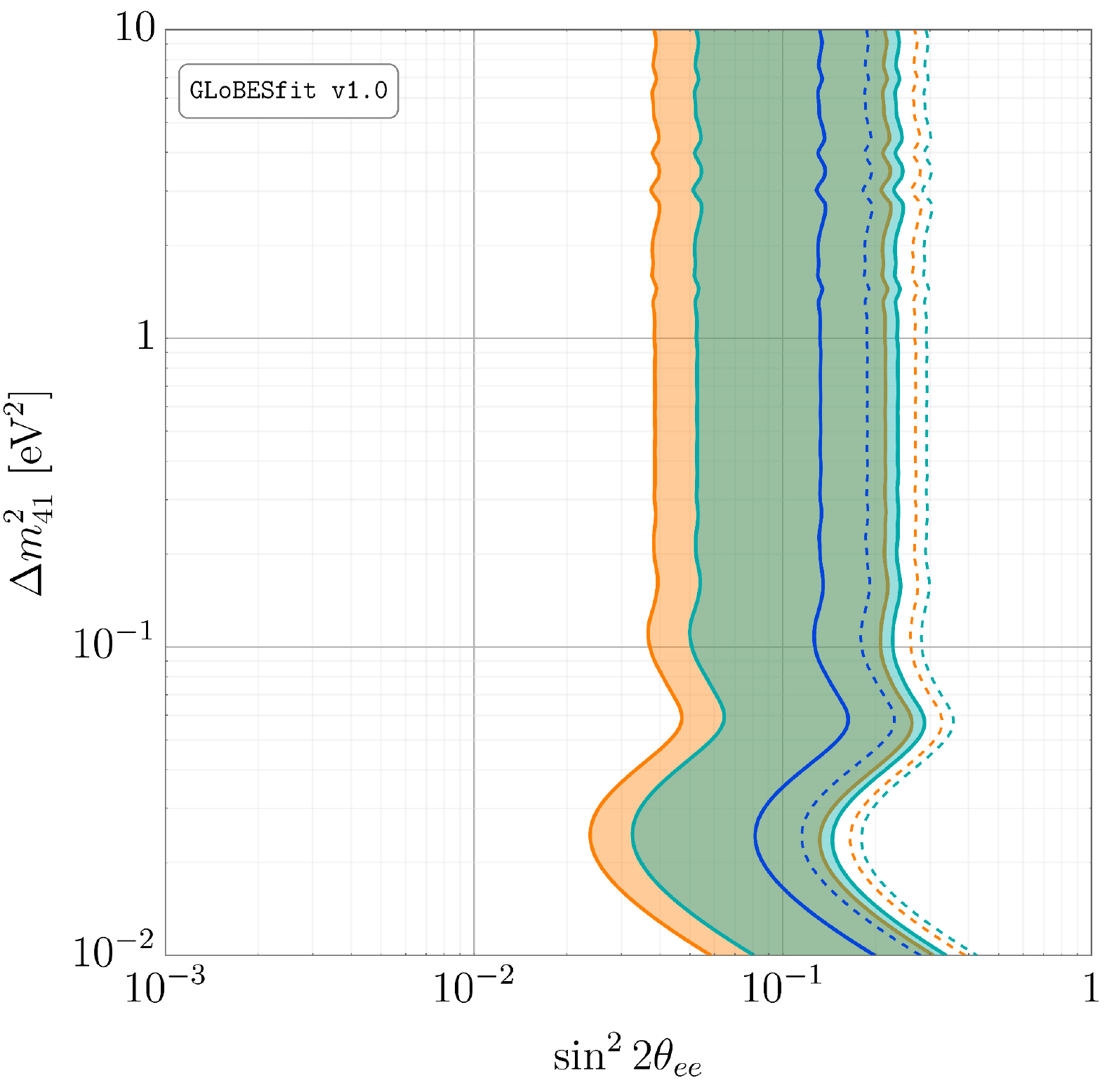}
\caption{The 68.3\% (solid) and 95\% (dashed) C.L. contours from our analysis of RENO. We show results for the HM (orange), \emph{ab initio} (blue) and HKSS (dark cyan) reactor antineutrino flux models.}
\label{fig:Block14}
\end{figure}


\section{Rate Analyses}
\label{sec:RateResults}
\setcounter{equation}{0}

In this section, we discuss how the experiments in the previous section constrain the mass and mixing of an additional sterile neutrino. This is one possible application of \globesfit \, -- the analysis techniques we describe can be modified for any one of a number of new-physics scenarios.

\subsection{Data -- A Combined View}

\begin{figure}
\includegraphics[width=\linewidth]{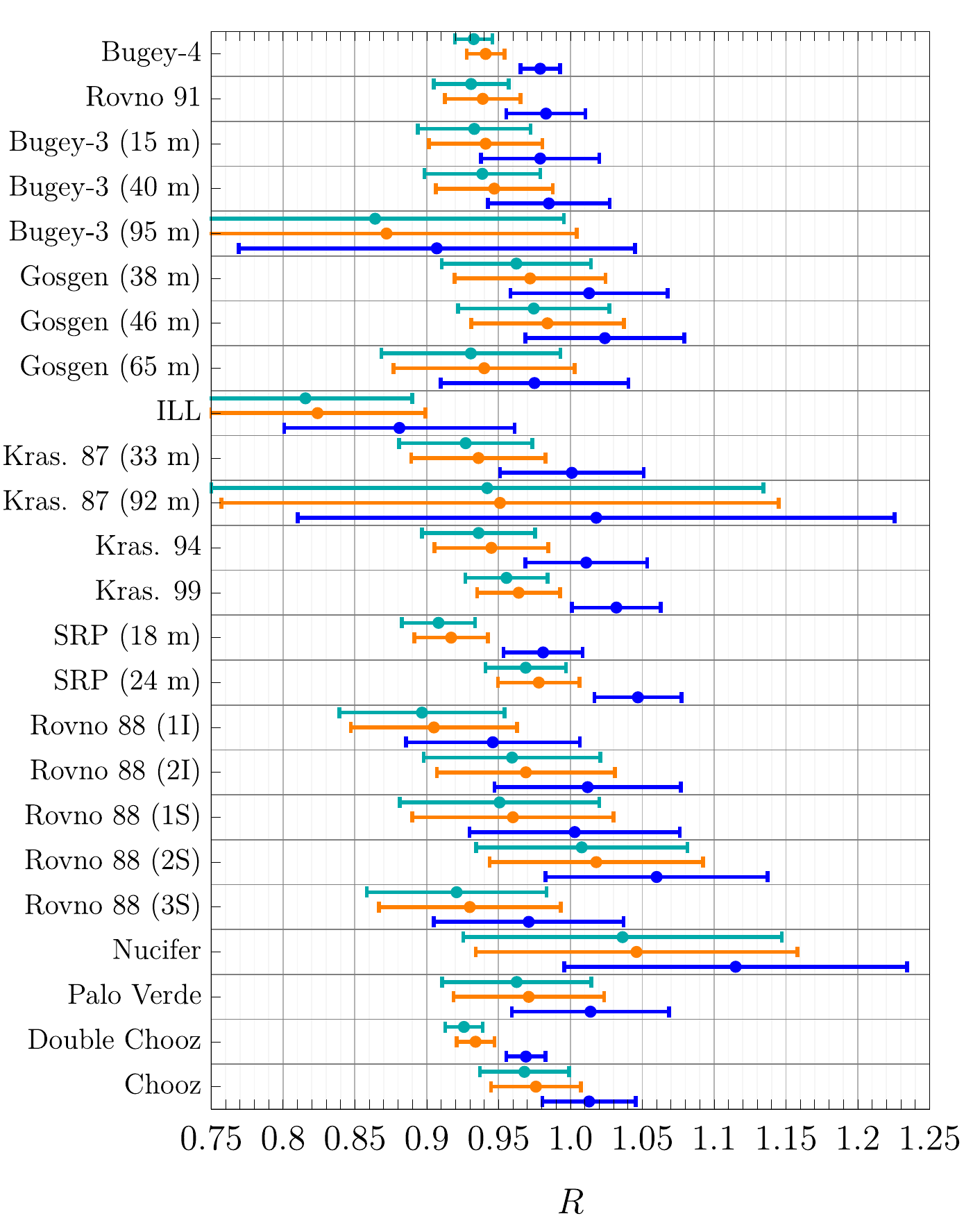}
\caption{The measurement-to-prediction ratio $R$ as measured by short- and medium-baseline IBD rate experiments for each flux model under consideration. Orange represents the HM fluxes; blue represents the \emph{ab initio} fluxes; and dark cyan represents the HKSS fluxes. Errors shown are experimental only.}
\label{fig:RateData}
\vspace{5mm}
\includegraphics[width=\linewidth]{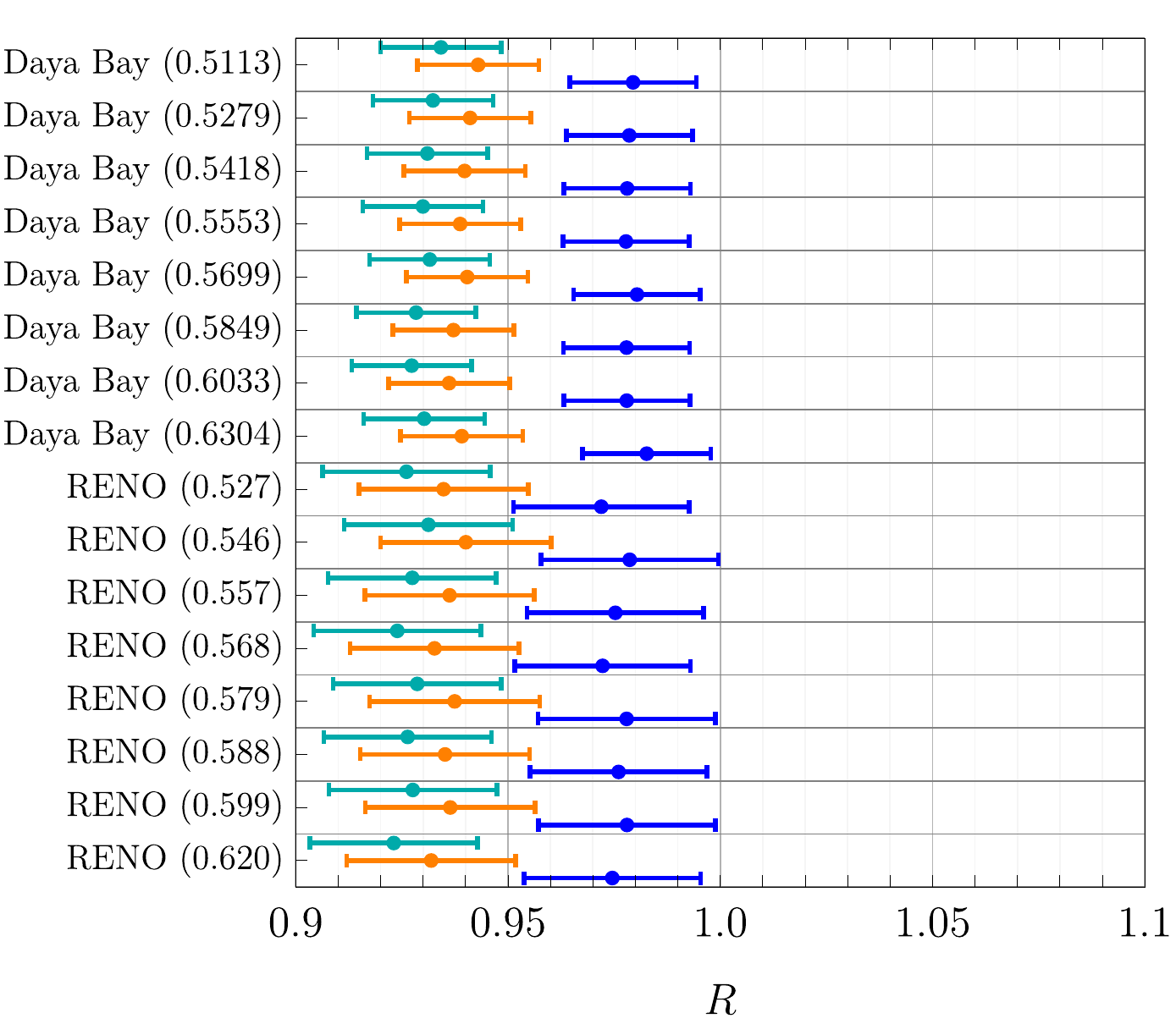}
\caption{Similar to Fig.~\ref{fig:RateData}, but for medium-baseline rate evolution experiments. The numbers in parentheses on the ordinate axis represent the effective $^{235}$U fission fraction for the particular data point.}
\label{fig:RateData2}
\end{figure}

We depict the data discussed above in Figs.~\ref{fig:RateData} and \ref{fig:RateData2}. The orange data points represent the ratio $R$ of measured IBD rates relative to the prediction calculated using the HM flux model. The blue data points are the same using the \emph{ab initio} fluxes; the dark cyan points employ the HKSS fluxes. The depicted error bands are experimental only; theoretical uncertainty is of order $\sim2.5\%$.

As discussed in Sec.~\ref{sec:Fluxes}, the reduction of the antineutrino flux from $^{235}$U, in particular, in the \emph{ab initio} fluxes results in a diminished predicted IBD rate, implying larger experiment-to-prediction ratios. On the other hand, the HKSS fluxes almost universally result in a higher predicted flux, resulting in a diminished ratio $R$.

For each group of correlated measurements, we show the 68.3\% and 95\% confidence level (C.L.) contours derived for each flux model with solid and dashed curves, respectively, in Figs.~\ref{fig:Block0}-\ref{fig:Block14}; the captions detail which experiments are depicted therein. Results using the HM fluxes are shown in orange; those using the \emph{ab initio} fluxes are show in blue; and those using the HKSS fluxes are shown in dark cyan.

In Tables \ref{tab:Stats_HM}, \ref{tab:Stats_SM} and \ref{tab:Stats_HKSS}, we tabulate relevant statistics for our analyses using the HM, \emph{ab initio} and HKSS fluxes, respectively. Specifically, we show the minimum value of the $\chi^2$ for each group of experiments, $\chi^2_{\rm min}$, as well as the $\chi^2$ in the limit of no mixing with a sterile neutrino, $\chi^2_{3\nu}$. The difference between these is used to determine the $p$-value for each group of experiments, assuming these are chi-squared distributed.

\begin{table}[!t]
\centering\begin{tabular}{|c||c|c|c|c|c|} \hline
Measurements & $\chi^2_{3\nu}$ & $\chi^2_{\rm min}$ &  $n_{\rm data}$ &  $p$ & $n\sigma$ \\ \hline \hline
Bugey-4/Rovno 91 & 4.6 & $\sim0$ & 2 & 0.10 & 1.6 \\ \hline
Bugey-3 & 1.8 & 0.29 & 3 & 0.48 & 0.71 \\ \hline
G\"osgen/ILL & 3.9 & 2.4 & 4 & 0.49 & 0.69 \\ \hline
Krasnoyarsk 87 & 1.4 & $\sim0$ & 2 & 0.50 & 0.68 \\ \hline
Krasnoyarsk 94 & 1.4 & 0.0 & 1 & 0.50 & 0.67 \\ \hline
Krasnoyarsk 99 & 0.97 & 0.0 & 1 & 0.62 & 0.5 \\ \hline
Savannah River, 18 m & 5.6 & 0.0 & 1 & $6.0 \times 10^{-2}$ & 1.9 \\ \hline
Savannah River, 24 m & 0.38 & 0.0 & 1 & 0.83 & 0.22  \\ \hline
Rovno 88 & 3.0 & 1.4 & 5 & 0.46 & 0.74 \\ \hline
Nucifer & 0.18 & 0.18 & 1& $\sim1$ & $\sim0$  \\ \hline \hline
Palo Verde & $\sim0$ & 0.0 & 1 & $\sim1$ & $\sim0$ \\ \hline
Double Chooz & 5.9 & 0.0 & 1 & $5.2 \times 10^{-2}$ & 1.9 \\ \hline
Chooz & 0.45 & 0.45 & 1 & $\sim1$ & $\sim0$ \\ \hline 
Daya Bay & 12.5 & 8.2 & 8 & 0.12 & 1.6 \\ \hline
RENO & 9.7 & 5.2 & 8 & 0.11 & 1.6 \\ \hline \hline
All SBL & 14.9 & 7.9 & 21 & $3.1 \times 10^{-2}$ & 2.2 \\ \hline
All MBL & 29.6 & 18.6 & 19 & $4.2 \times 10^{-3}$ & 2.9 \\ \hline \hline
All & 39.7 & 31.2 & 40 & $1.8 \times 10^{-2}$ & 2.5 \\ \hline
\end{tabular}
\caption{Values of $\chi^2$ for our analysis -- namely, the minimum value ($\chi^2_{\rm min}$), the value for $\sin^2 2\theta_{ee} = 0$ ($\chi^2_{3\nu}$) -- and the $p$-value at which three-neutrino mixing can be excluded for each experimental block for our analysis of the HM fluxes. The number of data points in each block is denoted $n_{\rm data}$. We convert the $p$-value to the number of $\sigma$ in the last column.}
\label{tab:Stats_HM}
\end{table}

\begin{table}[h]
\centering\begin{tabular}{|c||c|c|c|c|c|}\hline
Measurement(s) & $\chi^2_{3\nu}$ & $\chi^2_{\rm min}$ &  $n_{\rm data}$ &  $p$ & $n\sigma$ \\ \hline \hline
Bugey-4/Rovno 91 & 0.59 & $\sim0$ & 2 & 0.74 & 0.33 \\ \hline
Bugey-3 & 0.50 & 0.27 & 3 & 0.89 & 0.14 \\ \hline
G\"osgen/ILL & 2.7 & 2.7 & 4 & 0.96 & 0.05 \\ \hline
Krasnoyarsk 87 & $\sim0$ & $\sim0$ & 2 & $\sim1$ & $\sim0$ \\ \hline
Krasnoyarsk 94 & 0.06 & 0.05 & 1 & $\sim1$ & $\sim0$ \\ \hline
Krasnoyarsk 99 & 0.77 & 0.76 & 1 & $\sim1$ & $\sim0$ \\ \hline
Savannah River, 18 m & 0.29 & 0.0 & 1 & 0.86 & 0.17 \\ \hline
Savannah River, 24 m & 1.7 & 1.7 & 1 & $\sim1$ & $\sim0$  \\ \hline
Rovno 88 & 1.9 & 1.8 & 5 & 0.95 & 0.07 \\ \hline
Nucifer & 0.18 & 0.18 & 1& $\sim1$ & $\sim0$  \\ \hline \hline
Palo Verde & 0.46 & 0.46 & 1 & $\sim1$ & $\sim0$ \\ \hline
Double Chooz & 1.2 & $\sim0$ & 1 & 0.55 & 0.60 \\ \hline
Chooz & 2.6 & 2.6 & 1 & $\sim1$ & $\sim0$ \\ \hline 
Daya Bay & 4.2 & 4.2 & 8 & $\sim1$ & $\sim0$ \\ \hline
RENO & 5.9 & 5.5 & 8 & 0.80 & 0.25 \\ \hline \hline
All SBL & 11.2 & 9.9 & 21 & 0.52 & 0.65 \\ \hline 
All MBL & 19.5 & 17.2 & 19 & 0.30 & 1.20 \\ \hline \hline
All & 30.5 & 29.3 & 40 & 0.57 & 0.56 \\ \hline
\end{tabular}
\caption{The analog of Table \ref{tab:Stats_HM} for our analysis of \emph{ab initio} fluxes.}
\label{tab:Stats_SM}
\end{table}

\begin{table}[h]
\centering\begin{tabular}{|c||c|c|c|c|c|}\hline
Measurements & $\chi^2_{3\nu}$ & $\chi^2_{\rm min}$ &  $n_{\rm data}$ &  $p$ & $n\sigma$ \\ \hline \hline
Bugey-4/Rovno 91 & 5.2 & 0.0 & 2 & $7.6 \times 10^{-2}$ & 1.8 \\ \hline
Bugey-3 & 2.1 & 0.26 & 3 & 0.40 & 0.84 \\ \hline
G\"osgen/ILL & 4.1 & 2.2 & 4 & 0.38 & 0.87 \\ \hline
Krasnoyarsk 87 & 1.7 & $\sim0$ & 2 & 0.42 & 0.81 \\ \hline
Krasnoyarsk 94 & 1.8 & 0.0 & 1 & 0.41 & 0.82 \\ \hline
Krasnoyarsk 99 & 1.4 & 0.0 & 1 & 0.50 & 0.67 \\ \hline
Savannah River, 18 m & 6.4 & 0.0 & 1 & $4.1\times10^{-2}$ & 2.0 \\ \hline
Savannah River, 24 m & 0.70 & 0.0 & 1 & 0.70 & 0.38  \\ \hline
Rovno 88 & 3.3 & 1.3 & 5 & 0.37 & 0.90 \\ \hline
Nucifer & 0.18 & 0.18 & 1& $\sim1$ & $\sim0$  \\ \hline \hline
Palo Verde & 0.04 & 0.0 & 1 & 0.98 & 0.03 \\ \hline
Double Chooz & 6.4 & 0.0 & 1 & $3.9\times 10^{-2}$ & 2.1 \\ \hline
Chooz & 0.20 & 0.20 & 1 & $\sim1$ & $\sim0$ \\ \hline 
Daya Bay & 13.4 & 8.2 & 8 & $7.5\times 10^{-2}$ & 1.8 \\ \hline
RENO & 10.3 & 5.2 & 8 & $7.6\times 10^{-2}$ & 1.8 \\ \hline \hline
All SBL & 15.5 & 8.0 & 21 & $2.4 \times 10^{-2}$ & 2.3 \\ \hline 
All MBL & 30.2 & 18.6 & 19 & $3.0 \times 10^{-3}$ & 3.0 \\ \hline \hline
All & 40.3 & 31.1 & 40 & $1.0 \times 10^{-3}$ & 2.6 \\ \hline
\end{tabular}
\caption{The analog of Tables \ref{tab:Stats_HM} and \ref{tab:Stats_SM} for our analysis of HKSS fluxes.}
\label{tab:Stats_HKSS}
\end{table}

\subsection{Combining Experiments}

To analyze all of these experiments simultaneously, we combine the uncertainties on the integrated rate experiments into one covariance matrix, the elements of which are given by correlated and uncorrelated uncertainties described in Sec.~\ref{sec:RateExps}. We use separate covariance matrices for Daya Bay and RENO, which have been provided by the respective collaborations. We incorporate systematic uncertainties stemming from the flux predictions via four nuisance parameters (one for each of $^{235}$U, $^{238}$U, $^{239}$Pu and $^{241}$Pu), which we bundle into a vector $\vec{\xi}$. These flux predictions are correlated; we account for these correlations using the covariance matrices discussed in Sec.~\ref{sec:Fluxes}.

The chi-squared function we employ has the following form:
\begin{align}
\chi^2 = & \, (\vec{R}_{\rm exp} - \vec{R}_{\rm pred})^T \cdot \left(V_{\rm exp}\right)^{-1} \cdot (\vec{R}_{\rm exp} - \vec{R}_{\rm pred}) \nonumber \\
& + \vec{\xi}^T \cdot \left(V_{\rm th}\right)^{-1} \cdot \vec{\xi},
\end{align}
where $\vec{R}_{\rm exp}$ is the vector of experimental ratios and $\vec{R}_{\rm pred} = \vec{R}_{\rm pred}(\sin^2 2\theta_{ee}, \Delta m_{41}^2, \vec{\xi} )$ is the vector of predicted ratios. Our treatment of theoretical uncertainties has been previously described in Sec.~\ref{sec:Fluxes}; the theoretical covariance matrix $V_{\rm th}$ accounts for these in our calculations. Furthermore, $V_{\rm exp}$ is the covariance matrix describing experimental uncertainties. We minimize over $\vec\xi$ for each point in $\sin^2 2\theta_{ee}$--$\Delta m_{41}^2$ parameter space.

\begin{figure}
\includegraphics[width=\linewidth]{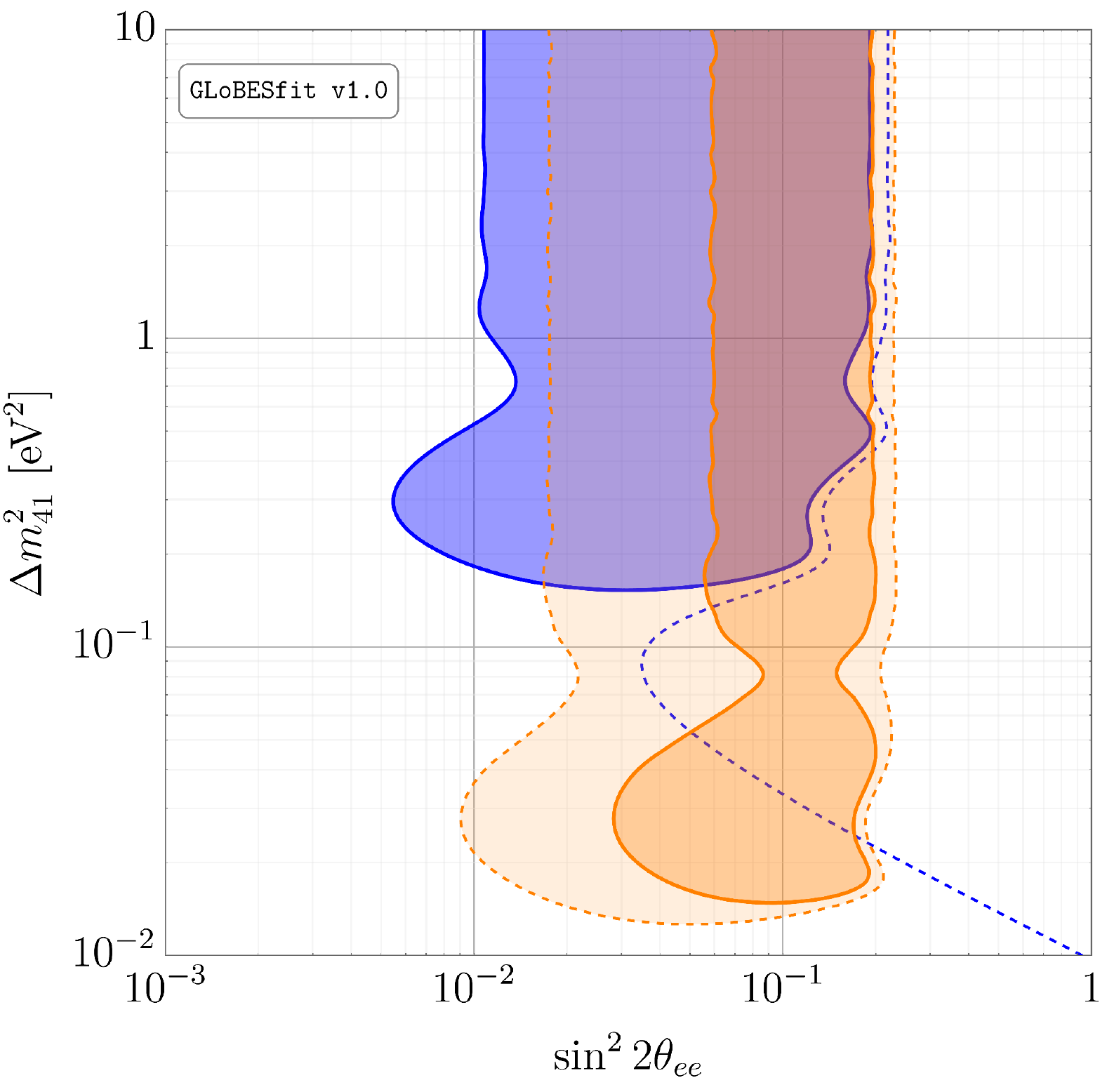}
\caption{The 95\% (solid) and 99\% (dashed) C.L. contours from our analyses of SBL (blue) and MBL (orange) rate experiments using the HM flux predictions.}
\label{fig:SBL_MBL_HM}
\end{figure}

\begin{figure}
\includegraphics[width=\linewidth]{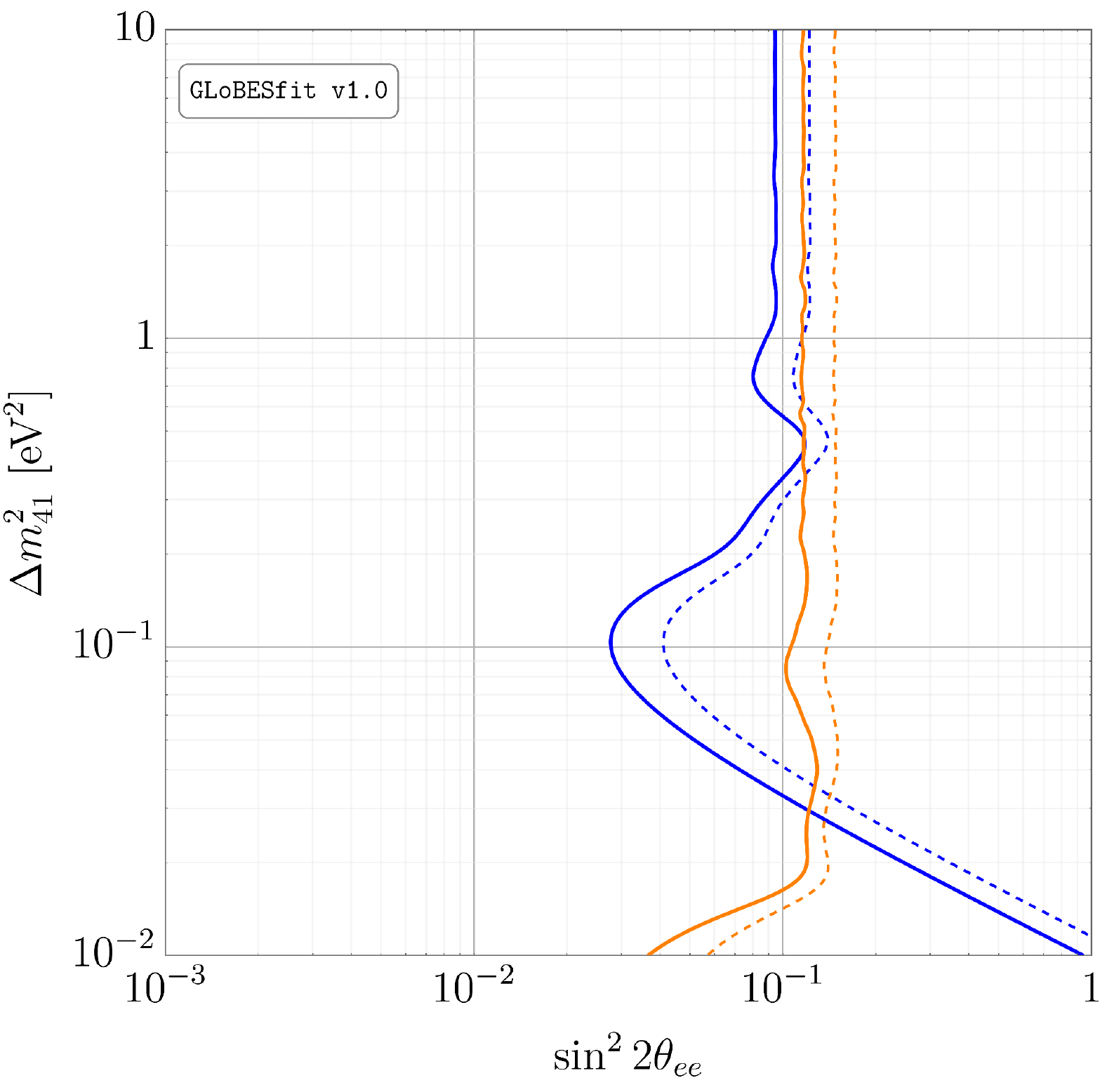}
\caption{The 95\% (solid) and 99\% (dashed) C.L. contours from our analyses of SBL (blue) and MBL (orange) rate experiments using the \emph{ab initio} flux predictions.}
\label{fig:SBL_MBL_SM}
\end{figure}

\begin{figure}
\includegraphics[width=\linewidth]{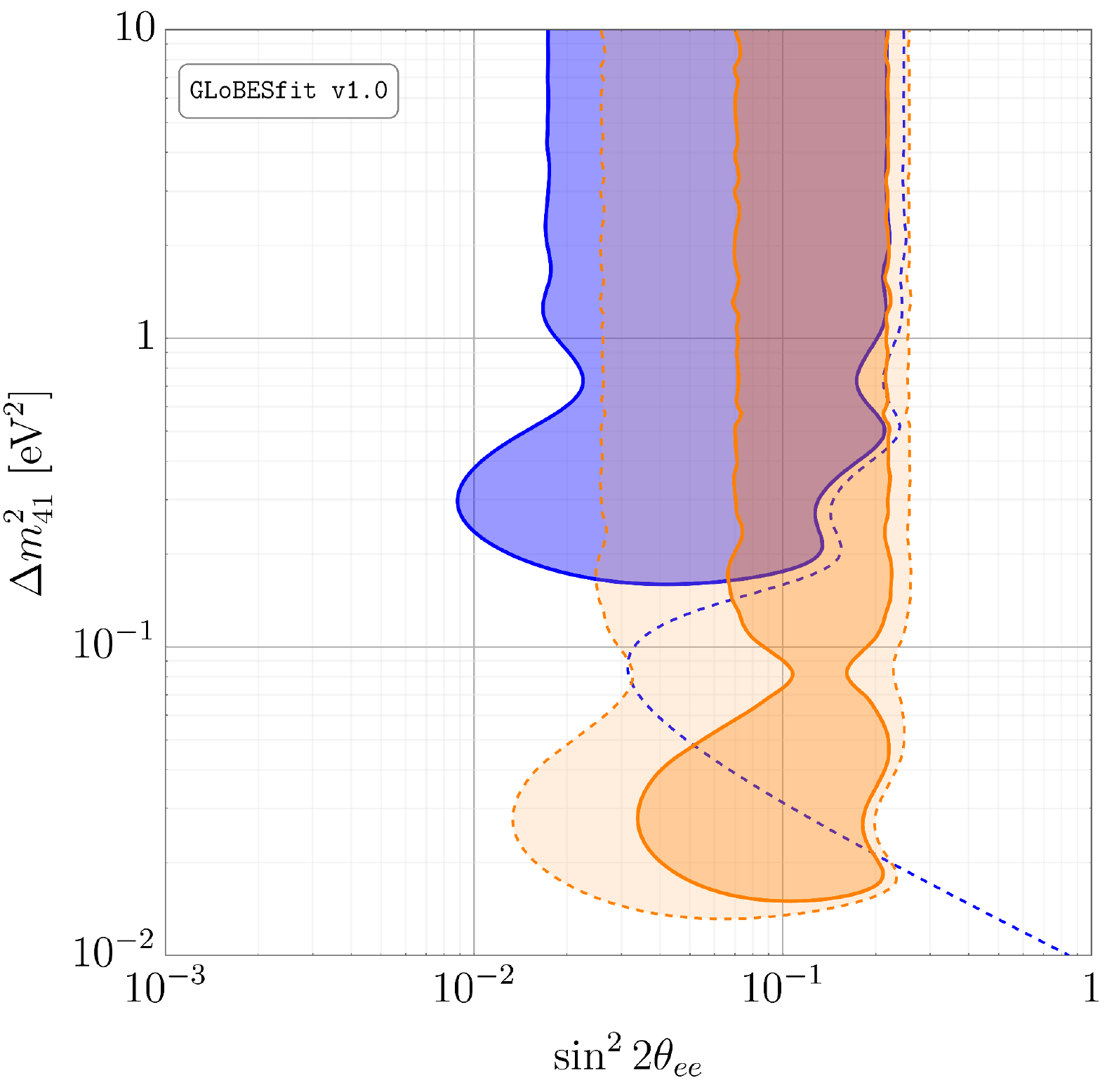}
\caption{The 95\% (solid) and 99\% (dashed) C.L. contours from our analyses of SBL (blue) and MBL (orange) rate experiments using the HKSS flux predictions.}
\label{fig:SBL_MBL_HKSS}
\end{figure}

In Figs.~\ref{fig:SBL_MBL_HM}, \ref{fig:SBL_MBL_SM} and \ref{fig:SBL_MBL_HKSS}, we separately show the regions preferred by short-baseline (SBL) and medium-baseline (MBL) experiments using the HM, \emph{ab initio} and HKSS fluxes, respectively. In each figure, we show SBL experiments in blue and MBL experiments in orange. Solid contours represent 95\% C.L., whereas dashed contours represent 99\% C.L. Relevant statistics are compiled in Tables \ref{tab:Stats_HM}-\ref{tab:Stats_HKSS}.

In each case, the MBL experiments prefer a smaller value of $\Delta m_{41}^2$ than the SBL experiments, as one would expect. In the region above $\Delta m_{41}^2 \gtrsim 1$ eV$^2$, we see that the MBL experiments are generally consistent with the region preferred by the SBL experiments. Both SBL and MBL experiments separately indicate $\gtrsim2\sigma$ evidence for a sterile neutrino for the HM and HKSS flux models, though the latter is stronger than the former. Conversely, the \emph{ab initio} fluxes are much more indifferent to the existence of a sterile neutrino; the 95\% C.L. curves do not close for either SBL or MBL experiments, as can be seen from Fig.~\ref{fig:SBL_MBL_SM}.

\begin{figure}
\includegraphics[width=\linewidth]{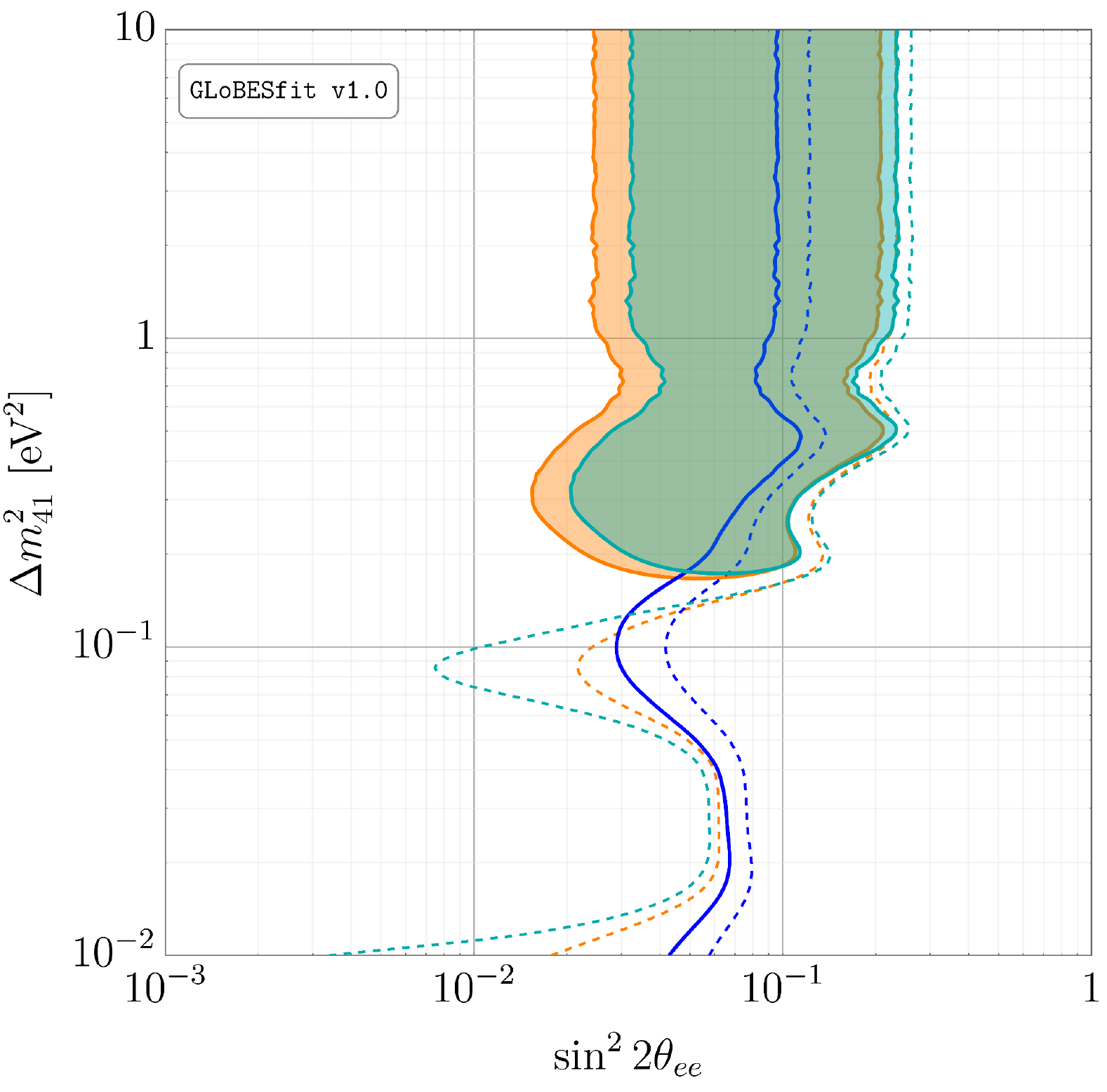}
\caption{The 95\% (solid) and 99\% (dashed) C.L. contours from our analyses of all IBD rate experiments using the HM (orange), \emph{ab initio} (blue) and HKSS (dark cyan) flux models.}
\label{fig:RateCombo}
\end{figure}

In Fig.~\ref{fig:RateCombo}, we show results from combining all SBL and MBL reactor experiments for all flux models we have considered. The solid and dashed curves depict the 95\% and 99\% C.L. regions, respectively. As before, the HM fluxes are shown in orange; the \emph{ab initio} fluxes are show in blue; and the HKSS fluxes are shown in dark cyan.

\begin{figure}
\includegraphics[width=\linewidth]{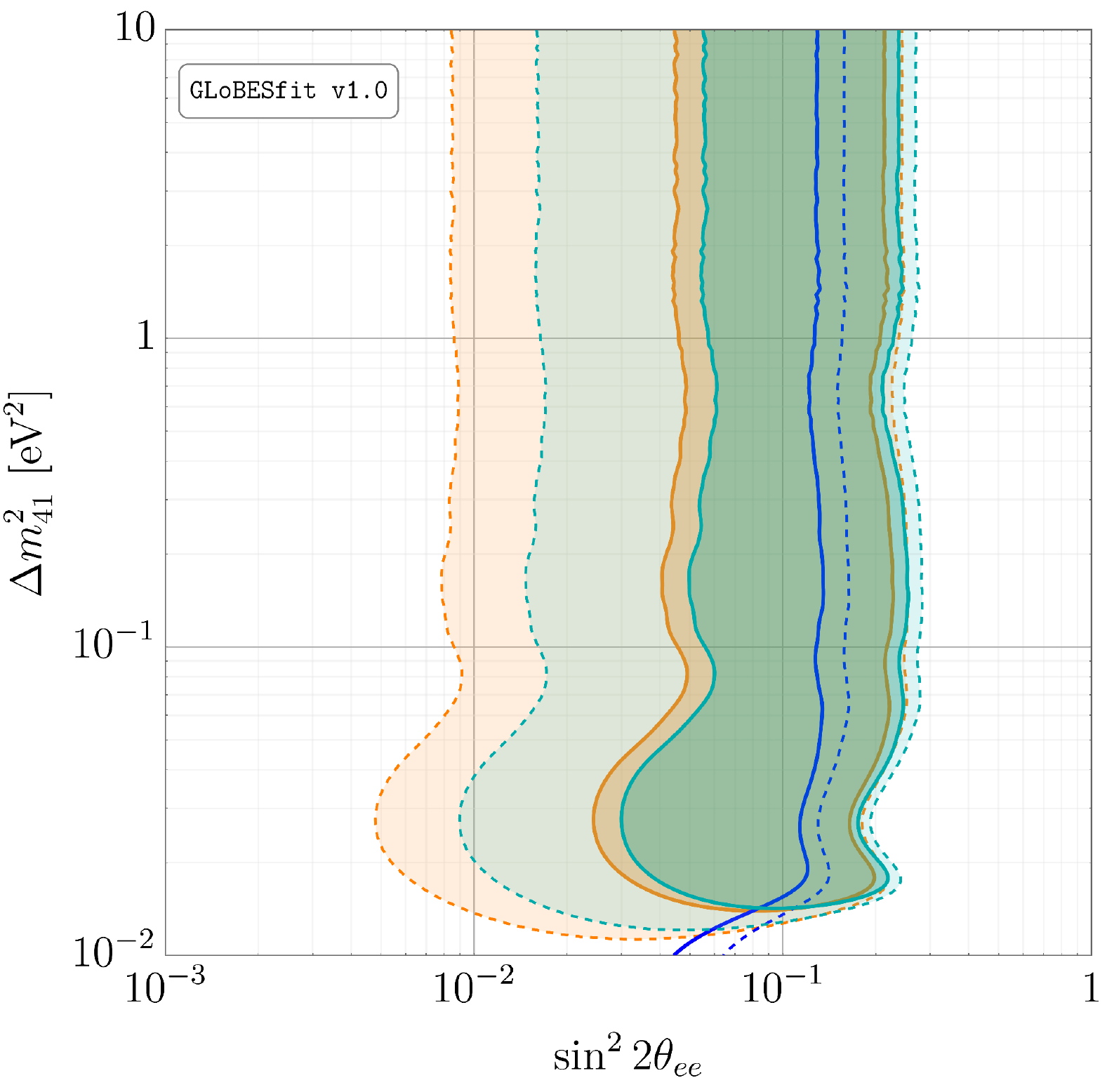}
\caption{The 95\% (solid) and 99\% (dashed) C.L. contours from our analyses of IBD rate experiments from the 2010s using the HM (orange), \emph{ab initio} (blue) and HKSS (dark cyan) flux models.}
\label{fig:Modern1}
\end{figure}

Using the \emph{ab initio} fluxes decreases the global preference for a sterile neutrino relative to the HM fluxes: the evidence decreases from $2.5\sigma$ to $0.56\sigma$. The reason for this, as we have mentioned, is that the overall flux normalization for the \emph{ab initio} flux prediction for $^{235}$U is $\sim10\%$ less than the corresponding HM prediction. Therefore, the \emph{ab initio} expected IBD rates are less than the HM rates, so the evidence for a sterile neutrino is diminished. Moreover, recall that assigning the HM uncertainties to the \emph{ab initio} flux predictions is optimistic; the true uncertainties associated with the calculation are assuredly larger than the $\sim2.5\%$ that we have ascribed. Consequently, this flux model is consistent with no sterile neutrino oscillation.

On the other hand, using the HKSS fluxes increases the global preference for a sterile neutrino: the evidence increases slightly from $2.5 \sigma$ to $2.6\sigma$. The effect of including first-forbidden decays is to increase the expected flux in the region around $E_\nu \sim 5-6$ MeV; the conversion process thus produces a higher overall antineutrino flux, implying a larger deficit than for the HM fluxes. Additionally, the inclusion of first-forbidden decays has not dramatically increased the uncertainty budget of this calculation, as had been speculated may happen~\cite{Hayes:2013wra}.

The observation of this diverging preference for a sterile neutrino is one of the central conclusions of this work. Clearly, at most one of these flux predictions is correct; determining which is paramount if the community wishes to determine how to best pursue this anomaly in the coming decade. It is critical to note that these predictions are ultimately derived from experimental measurements --- the conversion method requires reliable uranium and plutonium fission $\beta$ spectra, and the \emph{ab initio} method requires precisely known cumulative fission yields and $\beta$ decay strengths. Perhaps the most critical step to be taken in improving the antineutrino flux predictions is to improve the underlying data on which they are based.

It is important to stress that HKSS represents the first attempt to systematically include forbidden decays in flux predictions and there are a number of caveats one might raise with respect to the methods used. The shell model works best for spherical nuclei with near-magic numbers of nucleons, properties decidedly absent in many fission fragments. Moreover, the shape factors reported in Ref.~\cite{Hayen:2019eop} are larger than any observed shape factors. It is worthwhile to point out that for nuclei with non-unique forbidden decays for which the beta spectrum has been measured, the overall spectral shapes are close to the allowed one. A similar caveat applies to Ref.~\cite{Hayes:2013wra}: if the operator which causes by far the largest effect on the neutrino flux, $[\Sigma,r]^{1-}$, were to dominate a given beta decay branch, then a bimodal beta spectrum would result. However, this never has been observed. This may all be indicating that the concerns about forbidden decays in computing reactor neutrino fluxes may have been overstated.

\subsubsection*{Temporal Cut: Experiments from the 2010s}

Given the scarcity of experimental information that accompanies reports of older reactor experiments, one can reasonably ask how the evidence for the existence of a sterile neutrino changes if only recent experiments are included in our analysis. We introduce an \emph{ad hoc} temporal cut on these reactor data -- for concreteness, we only consider experiments from the 2010s, i.e., Nucifer, Double Chooz, Daya Bay and RENO -- and derive the constraints shown in Fig.~\ref{fig:Modern1}. As in previous figures, the 95\% (99\%) C.L. contour is shown in solid (dashed), and we simultaneously show results using the HM (orange), \emph{ab initio} (blue) and HKSS (dark cyan) flux predictions. Relevant statistics are shown in Table \ref{tab:Modern}.

Making this cut removes most SBL experiments, so the barycenters of the fits shift to smaller values of $\Delta m_{41}^2$. Imposing this cut increases the evidence for a sterile neutrino for all three flux models. For the HM and HKSS flux models, the evidence increases to $2.7\sigma$ and $2.8\sigma$, constituting modestly strong evidence. On the other hand, the significance for the \emph{ab initio} flux model rises to $0.81\sigma$.

To counteract the disproportionately high relevance of MBL experiments in this fit, we impose an additional \emph{ad hoc} prior requiring $\Delta m_{41}^2 > 0.1$ eV$^2$ and repeat the analysis. The resulting statistics are shown in Table \ref{tab:Modern2}. The evidence is weakened by imposing this prior, but the same basic trend emerges: the HM and HKSS flux models prefer a sterile neutrino at $\gtrsim2\sigma$, while the \emph{ab initio} fluxes show negligible preference.

\begin{table}[t]
\centering\begin{tabular}{|c||c|c|c|c|c|}\hline
Flux Model & $\chi^2_{3\nu}$ & $\chi^2_{\rm min}$ & d.o.f. & $p$ & $n\sigma$ \\ \hline \hline
HM & 24.8 & 14.7 & 14 & $6.4 \times 10^{-3}$ & 2.7 \\ \hline
\emph{Ab Initio} & 13.3 & 11.5 & 14 & 0.42 & 0.81 \\ \hline
HKSS & 25.6 & 14.8 & 14 & $4.6 \times 10^{-3}$ & 2.8 \\ \hline
\end{tabular}\caption{The relevant statistics from our scans over experiments from the 2010s, as described in the text.}
\label{tab:Modern}

\vspace{5mm}

\centering\begin{tabular}{|c||c|c|c|c|c|}\hline
Flux Model & $\chi^2_{3\nu}$ & $\chi^2_{\rm min}$ & d.o.f. & $p$ & $n\sigma$ \\ \hline \hline
HM & 24.8 & 16.3 & 14 & $1.4 \times 10^{-2}$ & 2.4 \\ \hline
\emph{Ab Initio} & 13.3 & 12.8 & 14 & 0.78 & 0.28 \\ \hline
HKSS & 25.6 & 16.2 & 14 & $1.7 \times 10^{-2}$ & 2.6 \\ \hline
\end{tabular}
\caption{Similar to Table \ref{tab:Modern}, except an \emph{ad hoc} prior has been imposed requiring $\Delta m_{41}^2 > 0.1$ eV$^2$.}
\label{tab:Modern2}
\end{table}

\subsection{Alternate Analysis: Rescaling the HM Fluxes}

We consider an alternative to the sterile-neutrino hypothesis: that the data can be explained by simply rescaling the HM fluxes. In particular, we only consider the two dominant fissile isotopes -- $^{235}$U and $^{239}$Pu -- and introduce a rescaling factor of each, respectively $r_{235}$ and $r_{239}$. Similar analyses have been performed in Refs.~\cite{An:2017osx,Giunti:2017yid,Gebre:2017vmm,RENO:2018pwo,Gariazzo:2018mwd,Giunti:2019qlt,Adey:2019ywk}. We scan over the $r_{235}$--$r_{239}$ plane in order to determine the extent to which the data prefer a rescaling of the HM fluxes and to determine if this is a more compelling explanation of reactor rate deficits than introducing a sterile neutrino. As in our sterile neutrino analyses, we treat systematics from $^{238}$U and $^{241}$Pu using nuisance parameters.

\begin{figure}
\includegraphics[width=\linewidth]{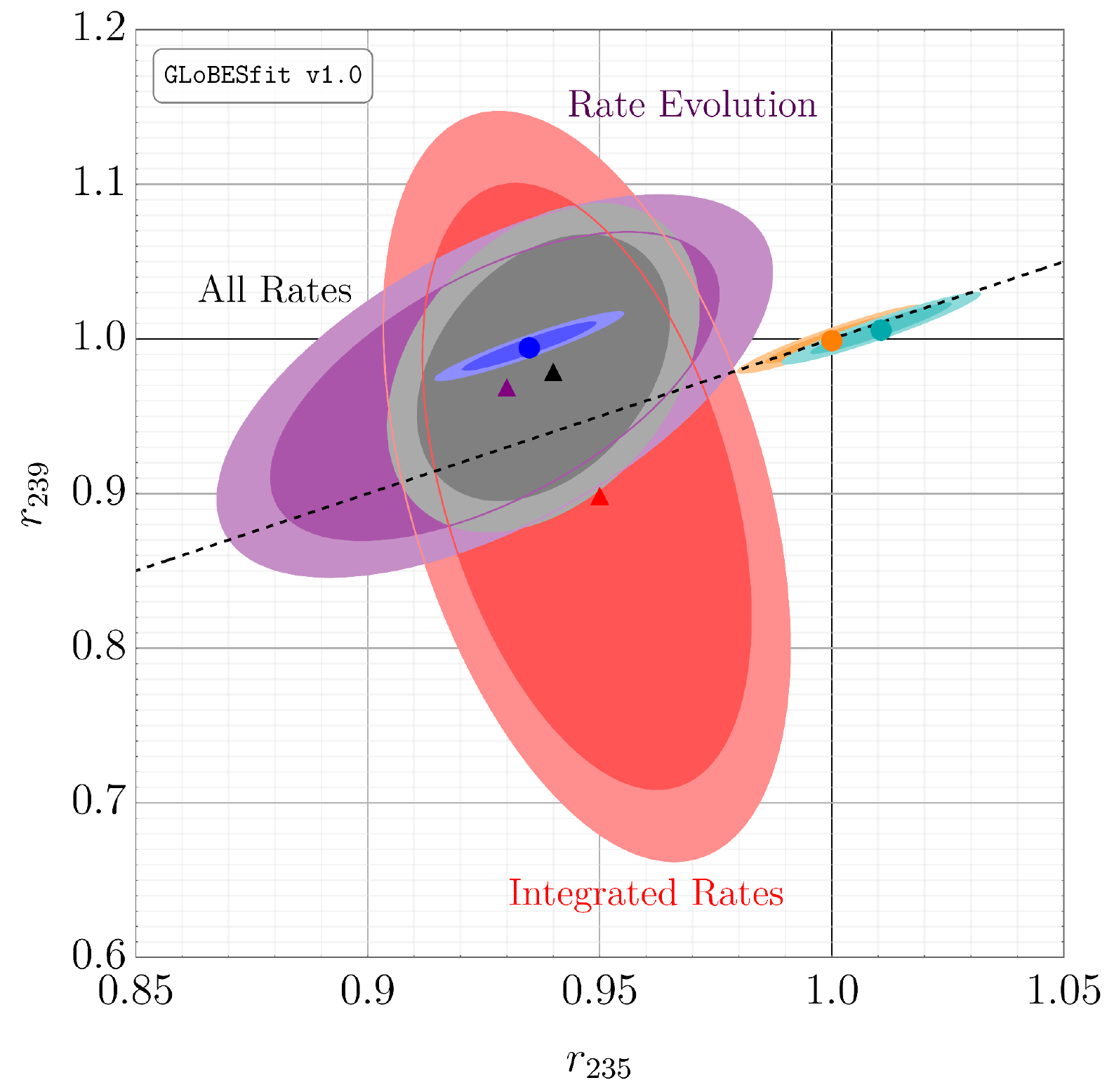}
\caption{The 95\% C.L. (dark) and 99\% C.L. (light) contours in $r_{235}$--$r_{239}$ plane for integrated rate (red), fuel evolution (purple) and all reactor experiments (black). The orange, blue star and cyan ellipses represent the expectations from the HM, \emph{ab initio} and HKSS flux models, respectively; $1\sigma$ ($2\sigma$) is shown in dark (light) shades. The black, dashed line represents the line along which $r_{235}=r_{239}$. The triangles represent the best-fit values for the three fits, and the circles show the central values for the flux models.}
\label{fig:Rescale2}
\includegraphics[width=\linewidth]{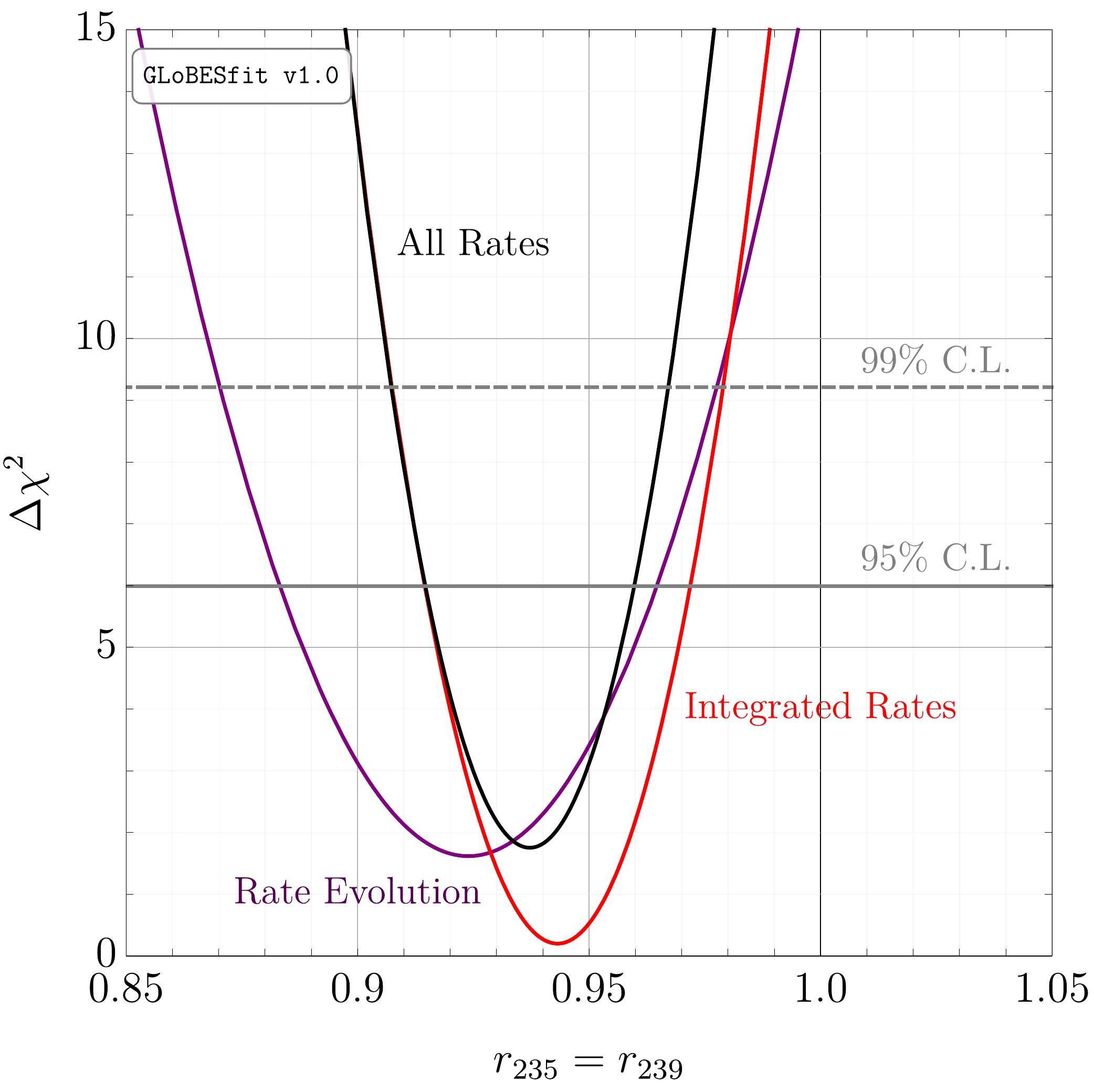}
\caption{The difference between $\chi^2$ calculated along the line $r_{235}=r_{239}$ and the minimum $\chi^2$ in the $r_{235}$--$r_{239}$ plane in Fig.~\ref{fig:Rescale2}. The red curve is for integrated rate experiments, the purple curve is for fuel evolution experiments and the black curve is for all reactor experiments. The solid (dashed) gray line represents the 95\% (99\%) C.L. limit for two degrees of freedom.}
\label{fig:Rescale3}
\end{figure}

In Fig.~\ref{fig:Rescale2}, we show contours of constant $\Delta \chi^2$ in the $r_{235}$--$r_{239}$ plane for different subsets of reactor experiments. The red regions refer to our analysis of SBL and MBL experiments that measure the total, time-integrated rate (namely, all of the above experiments except for Daya Bay and RENO); the purple regions refer to MBL experiments that measure the IBD rate as the effective fuel fractions change, i.e., Daya Bay and RENO; and the gray regions refer to a combined analysis of all experiments. Dark (light) shading represents the 95\% (99\%) C.L. region.

In Table \ref{tab:Rescale}, we present the minimum value of the $\chi^2$ ($\chi^2_{\rm min}$), its value at the point $r_{235} = r_{239} = 1$ ($\chi^2_{0}$) -- indicated by the orange circle in Fig.~\ref{fig:Rescale2} -- and both the corresponding $p$-value and $n\sigma$ at which the unscaled HM fluxes can be excluded for each dataset. We also show the $1\sigma$ ($2\sigma$) preferred region for the HM fluxes in dark (light) orange shading. While the fuel evolution dataset exhibits a moderate but clear preference ($3.7\sigma$), the integrated rate dataset presents $4.4\sigma$ evidence in favor of rescaling the HM flux predictions. Taken together, these data present $5.5\sigma$ evidence for this hypothesis, with the $^{235}$U flux, again, being particularly suspect.

\begin{table}
\centering\begin{tabular}{|c||c|c|c|c|c|}\hline
Analysis & $\chi^2_{\rm HM}$ & $\chi^2_{\rm min}$ & d.o.f. & $p$ & $n\sigma$  \\ \hline \hline
Fuel Evolution & 25.8 & 9.0 & 14 & $2.2\times 10^{-4}$ & 3.7 \\ \hline
Integrated Rate & 40.7 & 17.8 & 22 & $1.1 \times 10^{-6}$ & 4.4 \\ \hline
All & 62.6 & 28.0 & 38 & $3.1 \times 10^{-8}$ & 5.5 \\ \hline
\end{tabular}
\caption{Relevant statistics from our scans in the $r_{235}$--$r_{239}$ plane, as described in the text.}
\label{tab:Rescale}

\vspace{5mm}

\begin{tabular}{|c||c|c|c||c|c|c|}\hline
Analysis & $\chi^2_{\rm AI}$ & $p$ & $n\sigma$ & $\chi^2_{\rm HKSS}$ & $p$ & $n\sigma$ \\ \hline \hline
Fuel Evolution & 9.4 & 0.83 & 0.22 & 30.8 & $1.8 \times 10^{-5}$ & 4.3 \\ \hline
Integrated Rate & 19.2 & 0.49 & 0.69 & 49.6 & $1.2 \times 10^{-7}$ & 5.3 \\ \hline
All & 28.3 & 0.86 & 0.18 & 75.0 & $6.1 \times 10^{-11}$ & 6.5 \\ \hline
\end{tabular}
\caption{Relevant statistics from our comparison between the rescaled HM fluxes and the \emph{ab initio} and HKSS flux models.}
\label{tab:Rescale1}

\vspace{5mm}

\begin{tabular}{|c||c|c|c|c|}\hline
Analysis & $\Delta \chi^2_{\rm min, 1D}$ & $p$ & $n\sigma$ \\ \hline \hline
Fuel Evolution & 1.6 & 0.45 & 0.76 \\ \hline
Integrated Rate & 0.20 & 0.90 & 0.12 \\ \hline
All & 1.8 & 0.42 & 0.81 \\ \hline
\end{tabular}
\caption{Relevant statistics from our scan along the line $r_{235} = r_{239}$, as described in the text.}
\label{tab:Rescale2}
\end{table}

On the other hand, we can directly compare how the \emph{ab initio} and HKSS flux models compare to this rescaling of the HM fluxes. The blue and dark cyan circles, respectively, represent the relative sizes of the $^{235}$U and $^{239}$Pu isotopic IBD fluxes as compared to the HM prediction, calculated using our results from Sec.~\ref{sec:Fluxes}. As for the HM fluxes, the appropriately colored regions represent the $1\sigma$ (dark) and $2\sigma$ (light) regions for these flux models. These three models generally agree on the flux from $^{239}$Pu fissions, but disagree quite severely for $^{235}$U, with the \emph{ab initio} fluxes presenting a $\sim$6\% deficit and the HKSS fluxes presenting a $\sim$1\% enhancement with respect to HM. We quantify this in Table \ref{tab:Rescale1}, where we show how these alternate flux predictions compare to freely rescaling the $^{235}$U and $^{239}$Pu fluxes.

Unsurprisingly, we find that the \emph{ab initio} fluxes are consistent with the best-fit points from these analyses; the tension between them fails to surpass $1\sigma$ for any analysis we have performed. Conversely, the data strongly disfavor the HKSS fluxes: while the fuel evolution data exhibit $4.3\sigma$ tension, the integrated rate experiments present $5.3\sigma$ tension, and all experiments together present $6.5\sigma$ tension with this flux model. This again underscores the need to revisit the data that underpin these flux predictions.

An interesting subset of the rescaling hypothesis is to consider $r_{235}$ and $r_{239}$ being rescaled by the same factor. The line along which $r_{235} = r_{239}$ is shown in black dashing in Fig.~\ref{fig:Rescale2}. In Fig.~\ref{fig:Rescale3}, we show the values of $\Delta \chi^2$ \emph{relative to the minimum value in the entire $r_{235}$--$r_{239}$ plane} as a function of $r_{235} = r_{239}$. Shown are the respective curves for integrated rate (red), fuel evolution (purple) and all experiments (black). The solid (dashed) gray, horizontal line is the 95\% (99\%) C.L. exclusion limit, calculated for two degrees of freedom; we present this as a subspace of the $r_{235}$--$r_{239}$ plane of Fig.~\ref{fig:Rescale2}, and not necessarily as a separate hypothesis. 

In Table \ref{tab:Rescale2}, we show the minimum value of $\Delta \chi^2$ along this one-dimensional subspace, $\Delta \chi^2_{\rm 1D}$, the corresponding $p$ value and equivalent number of $\sigma$. The data are modestly consistent with a uniform rescaling of the $^{235}$U and $^{239}$Pu fluxes; a nonuniform rescaling of the $^{235}$U and $^{239}$Pu fluxes is preferred in an absolute sense, but not dramatically so. None of these analyses excludes a uniform rescaling of the HM fluxes by more than $1.1\sigma$.

We compare the sterile neutrino hypothesis and the rescaled-HM-fluxes hypothesis. For the former, we find $\chi^2_{\rm{min}}$/d.of. = 31.1/38, implying $p = 0.78$ (or $0.28\sigma$). For the latter, we find $\chi^2_{\rm{min}}$/d.o.f. = 28.0/38 and $p = 0.88$ ($0.15\sigma$). Clearly, the data mildly prefer for the fluxes to be independently rescaled over introducing a sterile neutrino. This is unsurprising: the the sterile neutrino hypothesis can be mapped onto rescaled-HM-fluxes hypothesis, assuming $\Delta m_{41}^2$ is sufficiently large that its corresponding oscillations average out at all experiments. Because the former is a subset of the latter, it cannot be more strongly preferred. Introducing the \emph{ad hoc} restriction that $\Delta m_{41}^2 \gtrsim 5$ eV$^2$ to ensure that oscillations would average out at all experiments, we find $\chi^2_{\rm{min}}$/d.o.f. = 32.3/39 and $p = 0.77$ ($0.30\sigma$).\footnote{We take the number of degrees of freedom to be 39 here. If $\Delta m_{41}^2$ is large enough for oscillations to average out, then its value cannot be measured and it is not a relevant degree of freedom.} 


\section{Spectrum Experiments}
\label{sec:SpectrumExp}
\setcounter{equation}{0}

In this section, we discuss the experiments that enter into our spectral analyses. Unlike the rate experiments, these are all taken to be mutually uncorrelated. Given the increase in the complexity of each measurement -- i.e., given that we are considering event spectra instead of integrated event rates -- the systematics become much richer than in the previous section(s). Consequently, we discuss the statistical analysis of each experiment separately in what follows.

For these analyses, we exclusively employ the \emph{ab initio} flux predictions. The reason for this is that, because we are interested in the ratios of spectra, one should expect that the choice of flux model does not affect the results of such a study. One could just as easily use the HM or HKSS fluxes; the result are essentially indistinguishable.

We have limited our analyses to considering Bugey-3 \cite{Declais:1994su}, DANSS \cite{Alekseev:2018efk}, Daya Bay \cite{Adey:2018zwh}, Double Chooz \cite{DoubleChooz:2019qbj}, NEOS \cite{Ko:2016owz} and RENO \cite{RENO:2018pwo}. While the short-baseline reactor experiments PROSPECT and STEREO have produced early results, these are not yet competitive with other experiments in the best-fit region around $\Delta m_{41}^2 \sim 1$ eV$^2$, so we do not include them at this time. Additional data releases are expected in the near future for both experiments,\footnote{While this manuscript was being prepared, Ref.~\cite{AlmazanMolina:2019qul} appeared on the preprint arXiv. We will include these data in future analyses, but we have not included them here.} as well as the SoLid experiment \cite{Ryder:2015sma}; we look forward to including all of these in future versions of \globesfit.


\subsection{Bugey-3}
\label{subsec:Bugey3}
 
\subsubsection*{Implementation in \globes}

Based on design specifications for a 900 MW$_e$ series reactor vessel described in Ref.~\cite{BugeySpecs}, we assume that the height of the active volume of the reactor core at Bugey-3 is 3.66 m tall with a radius of 1.5 m. We further assume that antineutrino production is uniform vertically\footnote{In reality this follows a $\cos z$ distribution with $z=0$ being in the center of the reactor, but since the reactor center and detector position are essentially at the same level, this has no practical impact on the baseline distribution.} and azimuthally, but that the radial distribution is given by a zeroth-order cylindrical Bessel function whose first zero lies at the edge of the core.

We only consider the antineutrino spectra measured at the 15 m and 40 m positions. The spectral measurement at 95 m is sufficiently imprecise that we do not expect to gain appreciable sensitivity to a sterile neutrino through its inclusion; this is borne out in Fig.~16 of Ref.~\cite{Declais:1994su}. Ref.~\cite{Declais:1994su} states that the detector modules used in the experiment are comprised of 98 separate $8.3 \times 8.3 \times 85.0$ cm$^3$ segments arranged in a $7 \times 14$ grid; the total dimensions of the modules are $58.1 \times 85.0 \times 116.2$ cm$^3$. The detector in the 15 m position is comprised of one such detector module, whereas the detector in the 40 m position is comprised of two of these, having total dimension $85.0 \times 116.2 \times 116.2$ cm$^3$. We find that our results are insensitive to the exact relative orientation of either detector and the core.

We precompute the oscillation probabilities at both positions using the geometries outlined above. The quantity of interest is the flux-averaged value of $\sin^2 (q L)$ at each experiment, where
\begin{equation}
q \equiv 1.267 \left(\frac{\Delta m_{41}^2}{\rm{eV}^2}\right) \left(\frac{\rm GeV}{E_\nu}\right)
\end{equation}
is independent of the experimental geometry. Note that this definition of $q$ assumes that lengths are given in km, which is the expectation in \texttt{GLoBES}. We define this average to be $F(q)$, and discuss it in more detail in the Appendix (see Eq.~\eqref{eq:define_F} and the surrounding discussion). We precompute $F(q)$ in intervals of 0.01 in $\log_{10} q$ over the range $q \in [ 10^0, 10^4 ]$.

The two positions are included with two separate AEDL experiments in \globesfit. We fix \texttt{@time}, \texttt{@power}, \texttt{@norm} and \texttt{\$target\_mass} to be 1.0 for both positions, and the spectrum is sampled in 25 bins on [2.8, 7.8] MeV in antineutrino energy, corresponding to [1.0, 6.0] MeV in positron energy. As for our rate analyses, we the energy resolution to be $6\%/\sqrt{E\rm{ [MeV]}}$. We set the values of \texttt{\$lengthtab} for each to the appropriate value of $1/ \sqrt{\langle L^{-2} \rangle}$ for each position; we calculate these to be 14.93 m and 39.96 m. We take the effective fission fractions to be
\begin{equation}
\left( f_{235}, \, f_{238}, \, f_{239}, \, f_{241} \right) = \left( 0.538, \, 0.078, \, 0.328, \, 0.056  \right) \nonumber
\end{equation}
for the duration of the experiment for both the 15 m and 40 m positions.

\subsubsection*{Statistical Analysis}

The upper panel of Fig.~15 of Ref.~\cite{Declais:1994su} shows the ratio of the spectrum measured at 40 m relative to that measured at 15 m. We have digitized this figure and the depicted statistical uncertainties for use in our analyses; we reproduce these data in Fig.~\ref{fig:Bugey3Spectrum} in Appendix~\ref{app:SuppData}. In addition to these statistical uncertainties, the collaboration also claims a 2.0\% systematic uncertainty on the relative normalization of the two spectra. We have also accounted for a 2.0\% energy scale uncertainty for each position and have assumed that these are totally uncorrelated. These are included in our analyses via the \globes \, function \texttt{glbShiftEnergyScale}; see the \globes \, manual for a description of this function \cite{Huber:2004ka,Huber:2007ji}.

We utilize a chi-squared function of the following form:
\begin{align}
\chi^2_{\rm Bugey-3} & = \left( \vec{S}_{\rm exp} - \vec{S}_{\rm pred} \right)^T \cdot \left( V_{\rm B} \right)^{-1} \cdot \left( \vec{S}_{\rm exp} - \vec{S}_{\rm pred} \right) \nn \\
& + \frac{\xi_{15}^2}{\sigma_{15}^2} + \frac{\xi_{40}^2}{\sigma_{40}^2},
\end{align}
where $S_{\rm exp}$ is the experimental spectrum and $S_{\rm pred}$ is the ratio predicted using \globes, assuming the existence of a sterile neutrino with mixing angle $\sin^2 2\theta_{ee}$ and mass-squared splitting $\Delta m_{41}^2$. The covariance matrix $V_B$ includes statistical uncertainties and the 2.0\% normalization uncertainty. The nuisance parameters $\xi_{15}$ and $\xi_{40}$ correspond to variations in the energy scale at 15 m and 40 m, respectively, with uncertainties of 2.0\% apiece. In our fits, we minimize over these nuisance parameters for each value of $\sin^2 2\theta_{ee}$ and $\Delta m_{41}^2$.

\begin{figure}[!t]
\includegraphics[width=\linewidth]{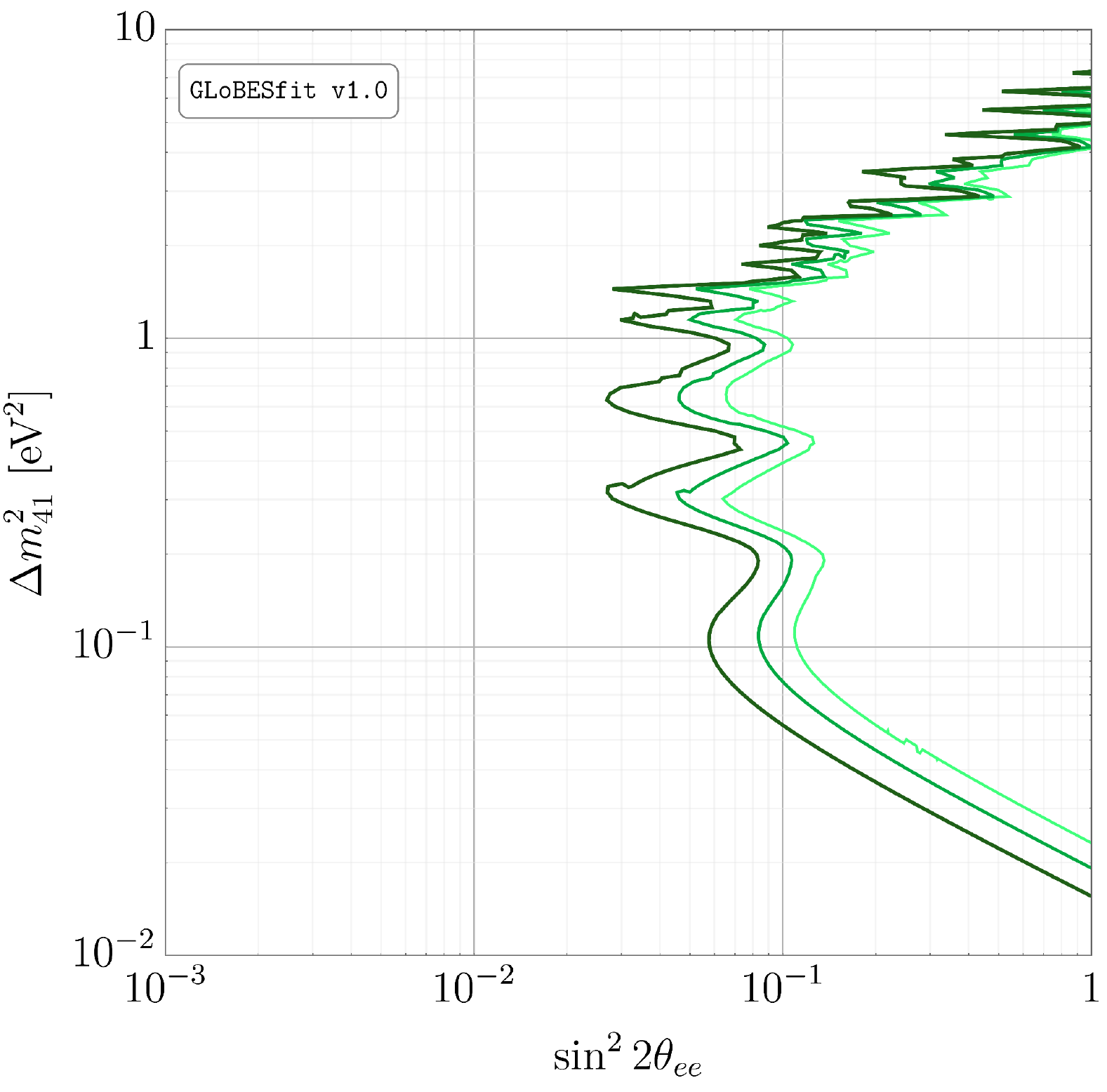}
\caption{The 95\% (dark green), 99\% (green) and 99.9\% (light green) C.L. curves for our sterile neutrino analysis of Bugey-3.}
\label{fig:Bugey}
\end{figure}

The results of our benchmark sterile neutrino analysis are shown in Fig.~\ref{fig:Bugey}. The dark green, green and dark green contours represent the 95\%, 99\% and 99.9\% C.L. contours, respectively. Relevant statistics are summarized in Table \ref{tab:SpecSummary}. Bugey-3 shows a modest preference for a sterile neutrino; we find that the evidence rises to the level of $1.4\sigma$.


\subsection{DANSS}
\label{subsec:DANSS}


\subsubsection*{Implementation in \globes}

The DANSS \cite{Alekseev:2018efk} reactor core is a 3.7-meter-tall cylinder with a diameter of 3.2 m. We assume production to be uniformly distributed, but that the radial distribution is given by the zeroth-order cylindrical Bessel function with zero at edge of core; we sample the vertical extent of the core following the burning profile in Ref.~\cite{DANSStalk}. The detector is a 1 m$^3$ cube located directly underneath the axis of the core; in our analyses, the centers of the core and the detector are taken to be either 10.7 m or 12.7 m apart, and we are ultimately interested in the ratio of the spectra obtained at these distances. We calculate the oscillation probabilities as described in Appendix ~\ref{app:OscEngines}. We calculate $F(q)$ (see Eq.~\eqref{eq:define_F}) on the range $q \in [ 10^1, 10^4 ]$ in intervals of 0.01 in $\log_{10}(q)$ for both the top ($10.7$ m) and bottom ($12.7$ m) positions.

We introduce separate experiments in \globesfit \, for the upper and lower positions at DANSS.\footnote{The collaboration also publishes the ratio of events in a middle position relative to the upper position \cite{Alekseev:2018efk}. Given that these are nontrivially correlated with the lower-to-upper ratio and that these are not likely to dramatically affect the final exclusion, we ignore these data.} Since we are interested in the \emph{ratio} of the spectra in the upper and lower positions and not in the \emph{absolute number} of events, we fix \texttt{@time}, \texttt{@power}, \texttt{@norm} and \texttt{\$target\_mass} to be 1.0 for both positions. The main difference, of course, is the values of \texttt{\$lengthtab}; this is given by the appropriate value of $1/\sqrt{\langle L^{-2} \rangle}$. These are determined to be 10.68 m and 12.69 m for the upper and lower positions, respectively, based on the previously described geometry. 

The spectrum is calculated for 24 bins on [2.8, 8.8] MeV in antineutrino energy, corresponding to [1.0, 7.0] MeV in positron energy. We estimate the energy resolution from Table 1 of Ref.~\cite{Alekseev:2018efk}; we fit the width $\sigma$ to a function of the form $a E + b \sqrt{E}$ and find $a = 0.0868, \, b = 0.006 \sqrt{\text{GeV}}$. We use the following fuel fractions:
\begin{equation}
\left( f_{235}, \, f_{238}, \, f_{239}, \, f_{241} \right) = \left( 0.56, \, 0.07, \, 0.31, \, 0.06  \right)
\end{equation}
These are the fuel fractions during the middle of a VVER-1000 reactor operating cycle as presented in Table 1 of Ref.~\cite{Kopeikin:2012zz}. 


\subsubsection*{Statistical Analysis}

Table 2 in Ref.~\cite{Alekseev:2018efk} tabulates the bottom/top spectral ratio, as well as the statistical errors on these ratios; we reproduce these data in Fig.~\ref{fig:DANSSSpectrum} in Appendix~\ref{app:SuppData}. The DANSS collaboration has not published a thorough analysis of their systematic uncertainties, but the impact of these have been estimated in the final result. Specifically, the following systematics have been included:
\begin{itemize}
\item The energy scale uncertainty is taken to be 2.0\%, and is fully correlated between the upper and lower configurations. This systematic is included via the \globes \, function \texttt{glbShiftEnergyScale}. A nuisance parameter is introduced corresponding to this energy scale shift.
\item The uncertainty on the ratio of $\langle L^{-2} \rangle$ for the upper and lower positions is taken to be 2.0\%. This introduces correlations between the spectral ratios. 
\end{itemize}

The following chi-squared function is employed in these analyses:
\begin{align}
\chi^2_{\rm DANSS} & = \left( \vec{S}_{\rm exp} - \vec{S}_{\rm pred} \right)^T \cdot \left( V_{\rm D} \right)^{-1} \cdot \left( \vec{S}_{\rm exp} - \vec{S}_{\rm pred} \right) \nn \\
& + \frac{\xi_D^2}{\sigma_D^2},
\end{align}
where $\vec{S}_{\rm exp}$ is the measured bottom/top ratio, $\vec{S}_{\rm pred}$ is the prediction of the same from \globes \, and $V_D$ is the covariance matrix describing these data. The nuisance parameter associated with the energy scale is $\xi_D$ and the corresponding uncertainty is $\sigma_D$. In our analysis, we minimize over $\xi_D$ for each point in the $\sin^2 2\theta_{ee}$--$\Delta m_{41}^2$ plane.

\begin{figure}[!t]
\includegraphics[width=\linewidth]{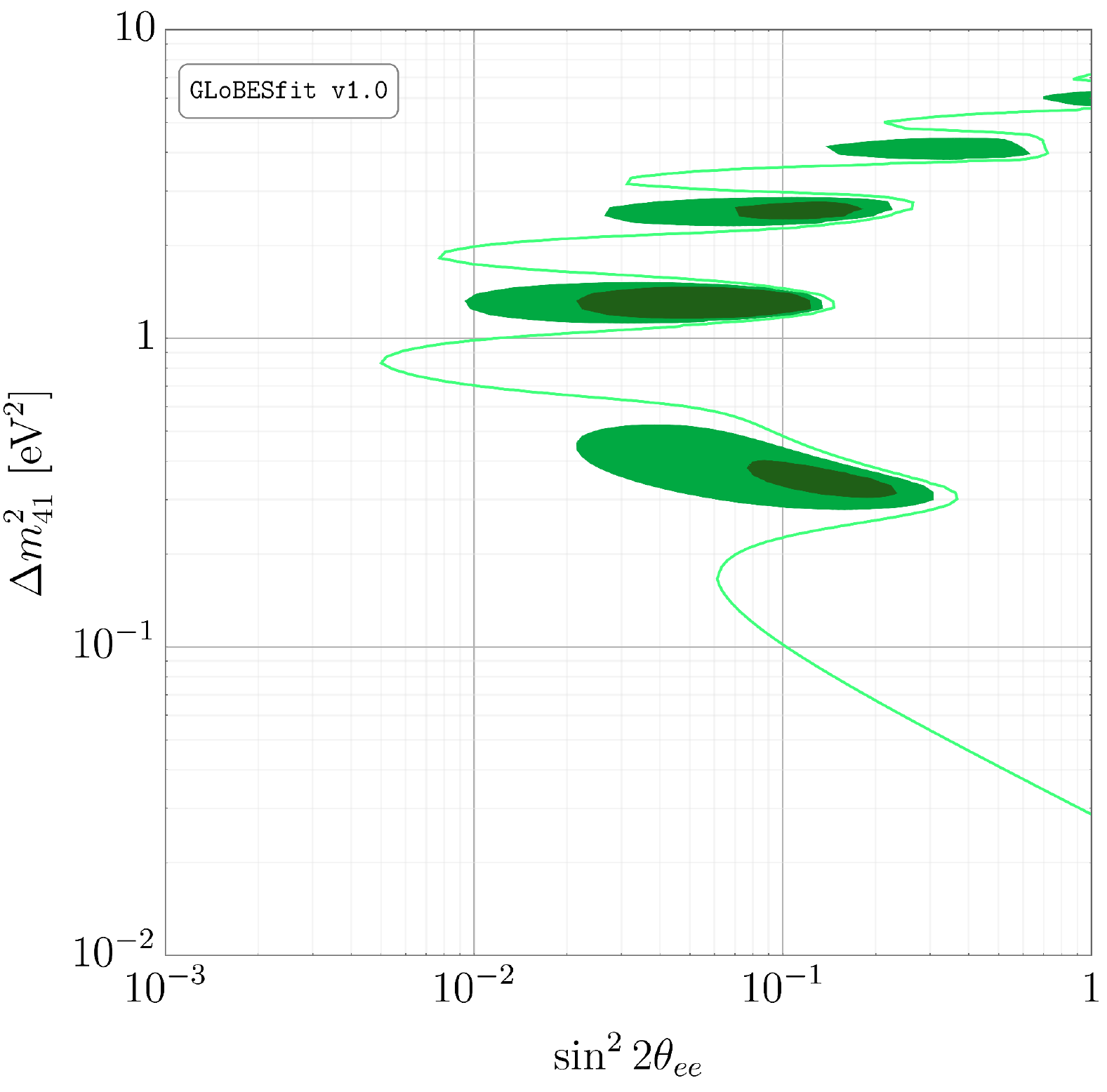}
\caption{The 95\% (dark green), 99\% (green) and 99.9\% (light green) C.L. curves for our sterile neutrino analysis of DANSS.}
\label{fig:DANSS}
\end{figure}

The results of our analysis are shown in Fig.~\ref{fig:DANSS}. The dark green, green and dark green contours represent the 95\%, 99\% and 99.9\% C.L. contours, respectively. Relevant statistics are summarized in Table \ref{tab:SpecSummary}. DANSS presents the most compelling single piece of evidence for a sterile neutrino of the experiments that we have considered -- the evidence rises to the level of $3.0\sigma$.


\subsection{Daya Bay}
\label{subsec:DayaBaySpec}


\subsubsection*{Implementation in \globes}

Daya Bay consists of an array of eight antineutrino detectors (ADs) and six nuclear reactors. Our analysis of the antineutrino spectrum at Daya Day is based on the 1958-day data release in Ref.~\cite{Adey:2018zwh}. The data have been taken in three phases, corresponding to the active number of ADs: sequentially six (6AD; 217 days), eight (8AD; 1524 days) and seven (7AD; 217 days).

In their 1230-day data release \cite{An:2016ses}, the collaboration reports the average power of each reactor core during the 6AD period and in the first 1013 days of the 8AD period; similar figures have not been published for the 1958-day data. We assume that (1) the average powers during the entire 8AD period are given by the average powers during the first 1013 days, and (2) that the average powers of during the 7AD period are equal to those from the 6AD period. We do not expect these assumptions to dramatically affect the outcome of this analysis, but we look forward to updating this in the future. The average powers during the 6AD and 8AD periods from the 1230-day data are tabulated in Table \ref{tab:DBpowers}.

The distances between pairs of reactors and detectors are tabulated in Table I of Ref.~\cite{An:2016ses}; we tabulate these in Table \ref{tab:DBlengths}. We treat the reactor cores as being point-like, but the ADs are treated as 3-meter wide targets, the stated size of the acrylic vessel that contains the Gd-doped liquid scintillator \cite{An:2016ses}. The function $F_d(q)$ (see Eq.~\eqref{eq:define_F}) for each detector $d$ is written as follows:
\begin{equation}
F_d(q) = \frac{\sum_{r=1}^8 \sum_s t_d^s P_r^s \sin^2 \left(qL_{rd}\right) / L_{rd}^2}{\sum_{r=1}^8 \sum_s t_d^s P_r^s/ L_{rd}^2},
\end{equation}
where $r$ indexes the reactor cores, $s$ indexes the three data-taking periods, $t_d^s$ is the live-time of detector $d$ during period $s$, $P_r^s$ is the power of core $r$ during period $s$ and $L_{dr}$ is the distance between detector $d$ and core $r$. The detector live-times are given as follows:
\begin{itemize}
\item ADs 1-6 have $t_d^{\rm 6AD}$ = 217 days; ADs 7 and 8 have $t_d^{\rm 6AD}$ = 0 days.
\item ADs 1-8 all have $t_d^{\rm 8AD}$ = 1013 days.
\item ADs 2-8 have $t_d^{\rm 7AD}$ = 217 days; AD 1 has $t_1^{\rm 7AD}$ = 0 days.
\end{itemize}
The oscillation probabilities are precomputed using a look-up table for each detector, for values of $\log_{10} q \in [-1, 3]$ with a grid spacing of 0.001.

There are five factors that differentiate each detector:
\begin{enumerate}
\item For each detector, we define \texttt{@time = 1.0} but, since not all detectors have been operational for the same amount of time, we fold these differences into the definition of \texttt{@power}. Namely, this is defined to be
\begin{equation}
\sum_{r=1}^8 \sum_s t_d^s P_r^s
\end{equation}
\item The target masses have subpercent-level differences; these are tabulated in the first row of Table \ref{tab:DBfracs} and are accounted for using \texttt{\$target\_mass}.
\item The total efficiencies of the ADs also vary; these are given in the second row of Table \ref{tab:DBfracs}. We set \texttt{@norm} to be the stated value of $\varepsilon_{\rm tot}$ for each detector.
\item The effective baseline of each detector, $L_d$, is defined via
\begin{equation}
\langle L_d^{-2} \rangle \equiv \frac{\sum_{r=1}^8 \sum_s t_d^s P_r^s/ L_{rd}^2}{\sum_{r=1}^8 \sum_s t_d^s P_r^s}.
\end{equation}
These are tabulated in the third row of Table \ref{tab:DBfracs}.
\item Lastly, the effective fuel composition visible to each detector varies; these are tabulated in the last four rows of Table \ref{tab:DBfracs}.
\end{enumerate}

The energy resolution of each detector is taken to be identical. In the supplementary material to Ref.~\cite{An:2016ses}, the Daya Bay collaboration publishes the response matrix that they use in their analysis. In principle, we could use the same response matrix here; to cut down on computation time, we parametrize the response using function of the form $\dfrac{\sigma}{\rm GeV} = a \left( \dfrac{E}{\rm GeV} \right)+ b \sqrt{\dfrac{E}{\rm GeV}} + c$. We find
\begin{align*}
a & = 0.0068, \\
b & = 0.0022, \\
c & = 0.000011.
\end{align*}


\subsubsection*{Statistical Analysis}

The nominal Daya Bay analysis ranges from prompt energies $[0.7, 12.0]$ MeV with irregular binning; the first bin is $[0.7, 1.3]$ MeV, the last bin is $[7.3, 12.0]$ MeV and everything in between is regularly spaced in 0.2-MeV bins. The spectrum at each AD is calculated over $[1.5, 12.8]$ MeV in \emph{true antineutrino energy} in 226 bins. This results in a 0.05-MeV spacing over the visible spectrum; this fine a spacing allows us to recycle our spectrum calculations for our analysis of NEOS, explained in Sec.~\ref{subsec:NEOS}. These bins are combined at the analysis level to reproduce the Daya Bay binning.

We are ultimately interested in the ratios of total events in EH2 (i.e., AD3 \& AD8) and EH3 (AD4-7) to EH1 (AD1, 2), which we call EH2/EH1 and EH3/EH1. The spectrum observed at each EH is determined by simply summing the number of spectrum for each detector in the hall, accounting for the appropriate effective fuel fractions. Taking the ratios of these spectra, which we call $\vec{S}^{12}_{\rm pred}$ and $\vec{S}^{13}_{\rm pred}$ for EH2/EH1 and EH3/EH1, respectively, is then trivial.

We benefit from the extensive data release from the Daya Bay collaboration in the form of supplementary data to Ref.~\cite{Adey:2018zwh}. In particular, the total numbers of observed IBD candidates (signal+background) for each bin in each experimental hall reside in the files \texttt{DayaBay\_IBDPromptSpectrum\_EH<N>\_1958days.txt}, where \texttt{<N> = 1, 2, 3}; the background spectra have also been published and reside in the files \texttt{DayaBay\_BackgroundSpectrum\_EH<N>\_1958days.txt}. We extract the observed ratios $\vec{S}^{12}_{\rm exp}$ and $\vec{S}^{13}_{\rm exp}$; we show these data in Figs.~\ref{fig:DBSpecEH2} and \ref{fig:DBSpecEH3}, respectively. These are the basis of our chi-squared, defined as
\begin{equation}
\chi^2_{\rm DB} = (\vec{S}_{\rm exp}  - \vec{S}_{\rm pred})^T \cdot (V_{\rm DB})^{-1} \cdot (\vec{S}_{\rm exp}  - \vec{S}_{\rm pred}),
\end{equation}
where $\vec{S}^{12}$ and $\vec{S}^{13}$ have been combined into one vector $\vec{S}$ and $V_{\rm DB}$ is the covariance matrix, which accounts for both statistical and systematic errors including correlations. We describe our estimation of the covariance matrix $V_{\rm DB}$ in what follows. 

The number of signal events in any of EH1, EH2 or EH3 (from data), which we call $\vec{s}^A$, is given by the difference between the total number of events $\vec{t}^A$ and the expected number of background events $\vec{b}^A$; here, $A$ (= 1, 2, 3) indexes the experimental hall. The uncertainty on $\vec{t}^A$ is statistical; the event rates are large enough where they can assumed to Gaussian-distributed. The uncertainty on $\vec{b}^A$, however, is considered a systematic uncertainty; we discuss it further below. Therefore, the statistical uncertainty on a given element of $\vec{s}^A$, which we call $s^A_i$, is $\sigma_{s_i^A} = \sqrt{t^A_i}$. The statistical component of the covariance matrix is determined by randomly varying the $s_i^A$, calculating the ratios $S_i^{12} = s^2_i/s_i^1$ and $S_i^{13} = s^3_i/s_i^1$ separately for each bin $i$ and determining at the covariance of the resulting pseudodata. The correlation between $S^{12}_i$ and $S^{13}_i$ is also recorded.

The collaboration provides a list of systematic uncertainties in Table VIII of Ref.~\cite{An:2016ses}; in Ref.~\cite{Adey:2018zwh}, it was reported that the uncertainty on the $^{9}$Li background has been reduced from 43\% to 27\% in EH1 and EH2. We have considered the contribution of each of these components, accounting for the stated level of correlations between detectors, reactors and other systematics, using Monte Carlo methods: the fractional change in the event rate is calculated by varying each systematic and the covariance matrix of the resulting pseudodata is recorded. In practice, we find that the largest contributions to the covariance matrix come from (1) variation in the backgrounds, and (2) varying the detector response. Unsurprisingly, the uncertainty on the HM flux predictions is found to be negligible -- this is precisely the reason why the ratios of event rates have been used in this analysis.

As previously mentioned, the absence of specific information regarding the true average reactor powers and live-times for each detector during each operation period means that our predicted ratios of spectra will not agree with the true ratios, even in the absence of a sterile neutrino. To compensate for this, we introduce two calibration factors, one for each spectral ratio, which are \emph{ad hoc} attempts to reproduce Daya Bay data; these are introduced via the replacements $\vec{S}^{12} \to \zeta_{12} \vec{S}^{12}$ and $\vec{S}^{13} \to \zeta_{13} \vec{S}^{13}$. These factors are determined by minimizing the above chi-squared function assuming only three-neutrino oscillations with the best-fit three-neutrino parameters in Ref.~\cite{Esteban:2018azc}, and are found to be $\zeta_{12} = 0.9933$ and $\zeta_{13} = 0.9973$. A more accurate estimate of the sensitivity can be obtained using more accurate information regarding the cores and detectors.

\begin{figure}[!t]
\includegraphics[width=\linewidth]{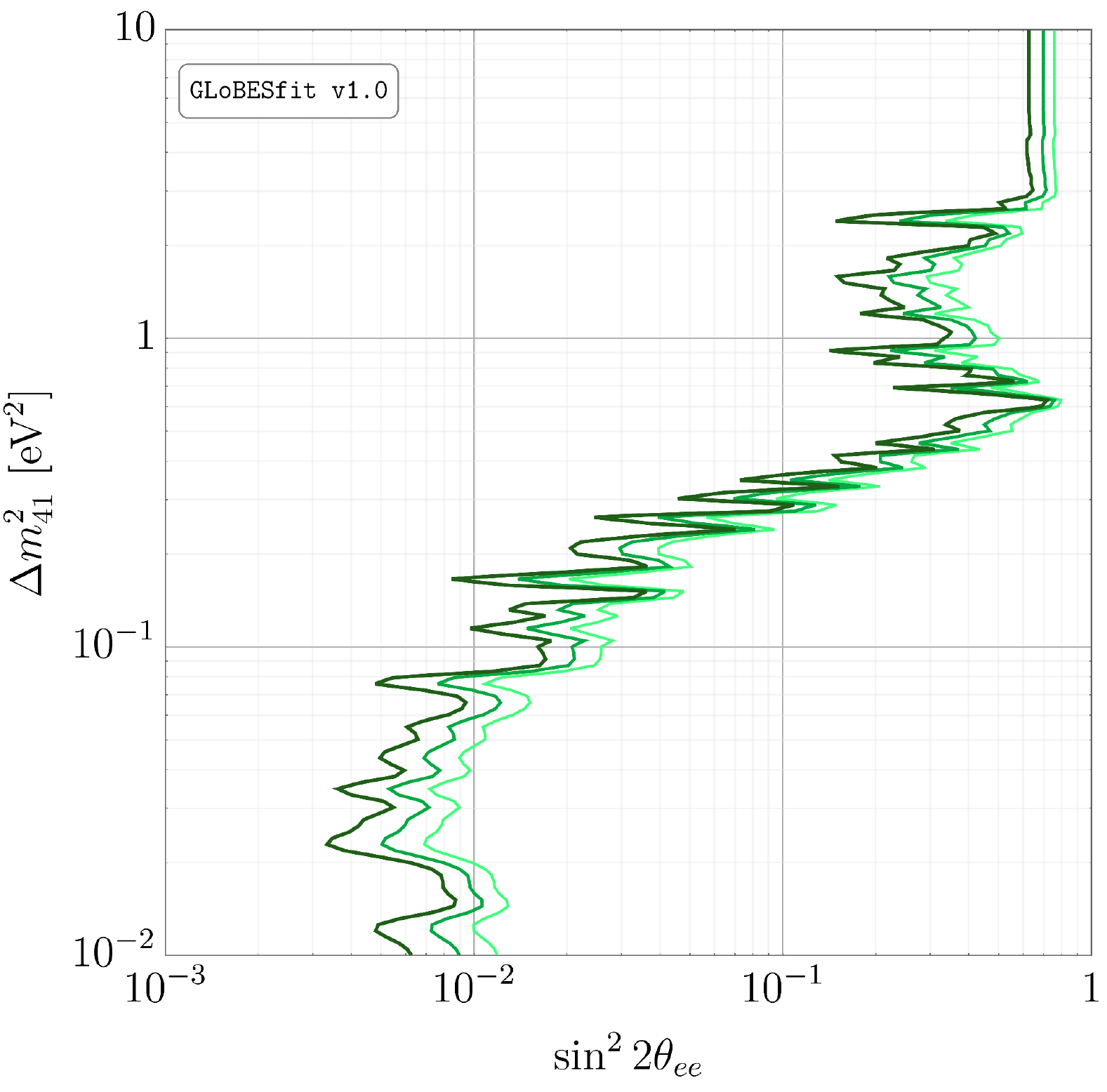}
\caption{The 95\% (dark green), 99\% (green) and 99.9\% (light green) C.L. curves for our sterile neutrino analysis of Daya Bay.}
\label{fig:DB}
\end{figure}

The results of this analysis are shown in Fig.~\ref{fig:DB}. The dark green, green and dark green contours represent the 95\%, 99\% and 99.9\% C.L. contours, respectively. Relevant statistics are summarized in Table \ref{tab:SpecSummary}. Clearly, the evidence for a fourth neutrino from Daya Bay spectra is rather weak.


\subsection{Double Chooz}
\label{subsec:DoubleChooz}

\subsubsection*{Implementation in \globes}

Double Chooz consists of two monolithic, Gd-doped liquid scintillator detectors located in the vicinity of two commercial power reactors. The separations between the reactors and the detectors are taken from Fig.~1 of Ref.~\cite{DoubleChooz:2019qbj}; the distances are tabulated in Table \ref{tab:DCinfo}. Given the $\mathcal{O}(100-1000)$ m baselines, we safely assume that both reactors are point-like. However, the detector is taken to have a width of 3 m; this helps to ensure that fast oscillations average out in our calculations, and is consistent with the physical width of the detector.

The collaboration has not published the (relative) average powers of the two reactors during the data collection periods for either the near or far detector; we assume that the average powers are the same, and comment on this further below. We again calculate the quantity $F(q)$ as in Eq.~\eqref{eq:defineF2}, accounting for both reactors at each detector. This quantity is precomputed in steps of 0.01 in $\log_{10} q$; we calculate this for $q \in [10^{-3}, 10^1]$ for the near detector and $q \in [10^{-2.5}, 10^1]$ for the far detector.

Each detector is separately defined within \globes, using the information contained in Table \ref{tab:DCinfo}. We set \texttt{@time} to the operating time of the detector (in days), \texttt{@power} to 8.5 GW (assumed to be the same for the neat and far detectors) and \texttt{@norm} to be the efficiency. We set \texttt{lengthtab} to the corresponding value of $1/\sqrt{\langle L^{-2} \rangle}$, given by 0.4003 km and 1.052 km for the near and far detectors, respectively. The collaboration reports \cite{DoubleChooz:2019qbj} that the near detector has $1.0042 \pm 0.0010$ times as many protons in its Gd-doped target as the far detector does, and $1.0045 \pm 0.0067$ times as many protons in its gamma catcher. Given the relative volumes of these two regions, we estimate that the near detector has $1.0044$ times as many protons as the far detector. Consequently, we set \texttt{\$targetmass} to be 1.0 for the far detector and 1.0044 for the near detector.

The energy spectrum in each detector is calculated in 26 bins between [1.8, 8.3] MeV in antineutrino energy, corresponding to [1.0, 7.5] MeV in prompt energy. The effective fuel fractions are the same as for our rate analysis,
\[ \left( f_{235}, \, f_{238}, \, f_{239}, \, f_{241} \right) = \left( 0.520, \, 0.087, \, 0.333, \, 0.060 \right). \]
and we again take the energy resolution to be $8\%/\sqrt{E\rm{ [MeV]}}$.

\subsubsection*{Statistical Analysis}

Figure 4 of Ref.~\cite{DoubleChooz:2019qbj} shows the measured ratio of spectra at the Double Chooz near and far detectors; we reproduce these data in Fig.~\ref{fig:DoubleChoozSpectrum} in Appendix~\ref{app:SuppData}. For our analyses, we have digitized these data and the corresponding statistical uncertainties. We also consider the following sources of systematic uncertainty:
\begin{itemize}
\item \emph{Detection Uncertainties}. These are enumerated in the third column of Table 2 in Ref.~\cite{DoubleChooz:2019qbj} and amount to a 0.475\% uncertainty, assumed to be correlated between all energy bins.
\item \emph{Flux Uncertainties}. A 0.47\% reactor-uncorrelated thermal power uncertainty has also been included. This results in a 0.4\% uncertainty on the spectral ratio, assumed to be correlated between all energy bins.
\item \emph{Backgrounds.} Background rates at the near and far detector are given in Table 3 of Ref.~\cite{DoubleChooz:2019qbj} and the corresponding spectra are shown in Fig.~3 for both the near and far detectors. From these figures, we estimate the contribution of each background process to the background-subtracted event rate and the corresponding uncertainty using a simple Monte Carlo routine. This procedure yields the background-related uncertainty on the spectrum, including nontrivial correlations.
\end{itemize}

These are all included into a chi-squared function of the following form:
\begin{equation}
\chi^2_{\rm DC} = (\vec{S}_{\rm exp}  - \vec{S}_{\rm pred})^T \cdot (V_{\rm DC})^{-1} \cdot (\vec{S}_{\rm exp}  - \vec{S}_{\rm pred}),
\end{equation}
where $\vec{S}_{\rm exp}$ and $\vec{S}_{\rm pred}$ are, respectively, the measured and predicted spectral ratios at Double Chooz. We note that we have reweighted the measured data by a factor
\[ \varphi = \left( \frac{t_{\rm far}}{t_{\rm near}} \right) \left( \frac{\langle L^{-2}_{\rm far} \rangle}{\langle L^{-2}_{\rm near} \rangle} \right)  \left( \frac{M_{\rm far}}{M_{\rm near}} \right)  \left( \frac{\varepsilon_{\rm far}}{\varepsilon_{\rm near}}, \right)\]
i.e., by the ratio of the predicted numbers of events in either detector. The covariance matrix $V_{\rm DC}$ contains the statistical and systematic uncertainties described above.

In the absence of published information regarding the relative powers of the two reactors over the operation of each detector, comparing $\vec{S}_{\rm exp}$ and $\vec{S}_{\rm pred}$ can be problematic -- it's unclear what the appropriate three-neutrino-oscillation baseline should be. To address this, we reweight our predicted spectrum by a factor $\zeta_{\rm DC}$, determined as follows. We calculate $\vec{S}_{\rm pred}$ assuming only three-neutrino oscillations using the value of $\sin^2 \theta_{13}$ measured by Double Chooz in Ref.~\cite{DoubleChooz:2019qbj}, i.e, $\sin^2 2\theta_{13} = 0.105$. We then replace $S_{\rm pred} \to \zeta_{\rm DC} S_{\rm pred}$ and minimize over $\zeta_{\rm DC}$; we find $\zeta_{\rm DC} = 1.0026$. We then include this factor in all subsequent calculations.

\begin{figure}[!t]
\includegraphics[width=\linewidth]{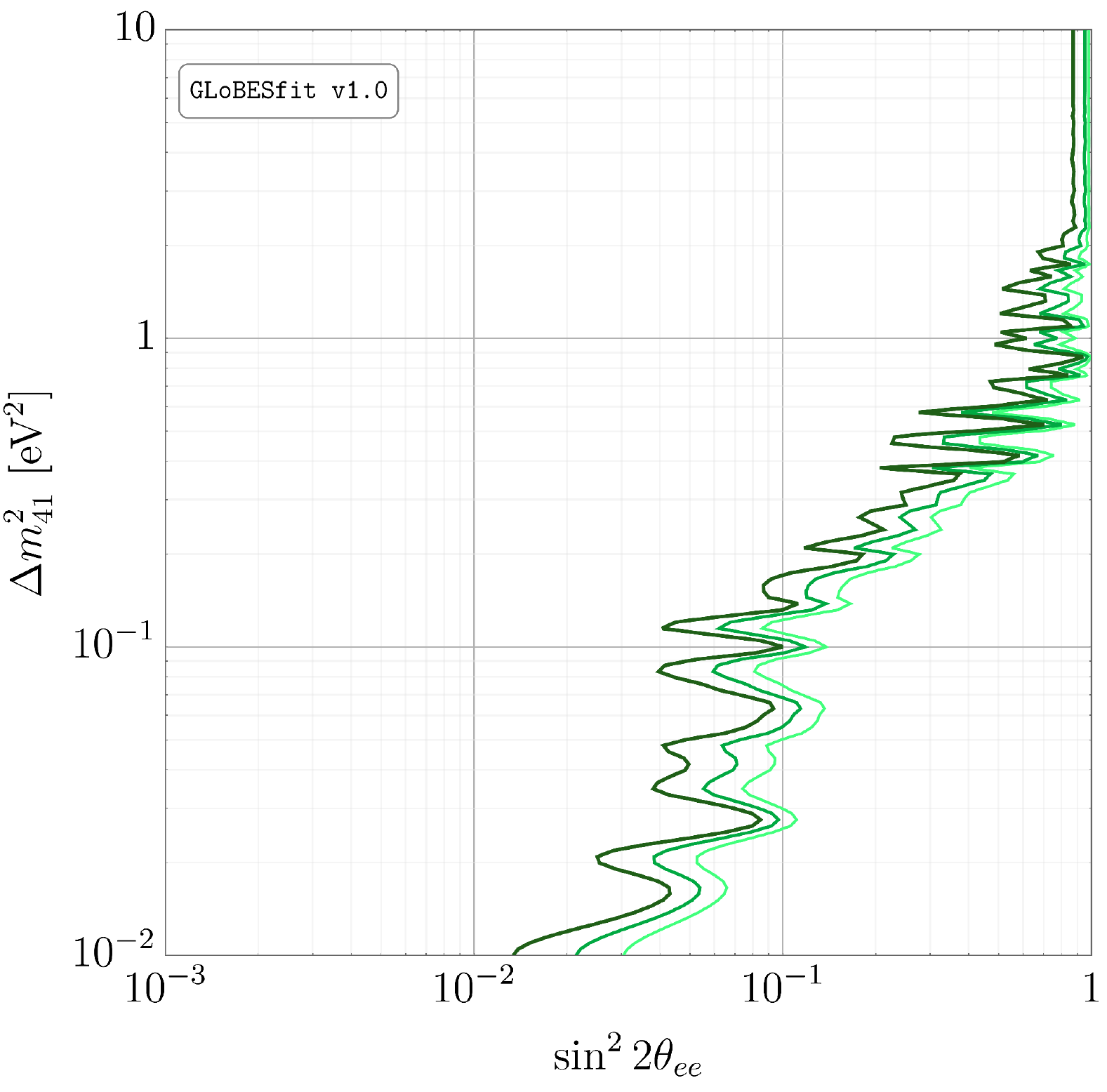}
\caption{The 95\% (dark green), 99\% (green) and 99.9\% (light green) C.L. curves for our sterile neutrino analysis of Double Chooz.}
\label{fig:DoubleChooz}
\end{figure}

The results of our calculation are shown in Fig.~\ref{fig:DoubleChooz}.\footnote{We have resumed calculating with $\sin^2 2\theta_{13} = 0.084$, from Ref.~\cite{Esteban:2018azc}.} The dark green, green and dark green contours represent the 95\%, 99\% and 99.9\% C.L. contours, respectively. Relevant statistics are compiled in Table \ref{tab:SpecSummary}. The evidence for a sterile neutrino at Double Chooz is weak -- while the data show a clear deficit stemming from nonzero $\theta_{13}$, the data are sufficiently imprecise that any fine structure beyond the expected patter is washed out.


\subsection{NEOS}
\label{subsec:NEOS}


\subsubsection*{Implementation in \globes}

Details of the geometry of NEOS are taken from Ref.~\cite{Ko:2016owz}. The reactor core at unit 5 of the Hanbit Nuclear Power Complex, where NEOS is housed, is 3.1 m in diameter and 3.8 m in height. The detector is located in the plane 10.0 m beneath the center of reactor; labeling the center of core as the origin of our coordinate system and call the axis of the cylinder the $z$ axis, the $z$ coordinate of the detector is $z=-10.0$ m. The center-to-center distance is 23.7 m; the remaining coordinates of the center of the detector are taken to be $x = +21.487$ m and $y = 0$ m. The detector is 1.21-meter-long cylinder with diameter 1.03 m and is oriented so that its axis is in the $xy$ plane, i.e., the axes of the detector and the core are orthogonal. Ref.~\cite{Ko:2016owz} does not further specify the orientation of the detector axis in the $xy$ plane, but the analysis is not particularly sensitive to this detail, given the size of the detector. For concreteness, we orient the detector along the $x$ axis.

We estimate the distribution of baselines (more concretely, the distribution of $1/L^2$ values between points in the core and the detector) using a simple Monte Carlo routine. The effective length to be $L_{\rm eff} \equiv 1/\sqrt{\langle L^{-2} \rangle} = 23.69$ m. These distances are also used to calculate the function $F(q)$ (Eq.~\ref{eq:define_F}) over the range $q = \left[10^1, \, 10^4 \right]$ in steps of 0.01 in $\log_{10} q$.

NEOS, itself, requires one experiment file within \globesfit. The NEOS collaboration reports a prompt energy spectrum on $[1.0, \, 10.0]$ MeV. We choose to ignore the last bin, covering $[7.0, \, 10.0]$ MeV, in our analysis; there are very few events in this bin, so we do not expect to lose much information. The NEOS collaboration presents their data in terms of a ratio relative to the antineutrino spectrum measured at the Daya Bay near detectors in Ref.~\cite{An:2016srz}; see Fig.~3(c) of Ref.~\cite{Ko:2016owz}.\footnote{The collaboration also publishes the ratio of their data relative to the HM fluxes; see Fig.~3(b) of Ref.~\cite{Ko:2016owz}. Given the size of the theoretical uncertainties on the flux predictions and the unexplained spectral features at 1 MeV and 5 MeV, we consider the ratio of measured spectra, even though these have been measured by two different experiments.} The spectrum that we use to normalize the NEOS result is the sum of the antineutrino spectra in AD1 and, AD2, i.e., the detectors in EH1. The ratio of NEOS data relative to the Daya Bay spectrum is shown in Fig.~\ref{fig:NEOSSpectrum} in Appendix~\ref{app:SuppData}.

Since we are again interested in a ratio of spectra, \texttt{@time}, \texttt{@power}, \texttt{@norm} and \texttt{\$target\_mass} are all fixed to be 1.0. We use the same parametrization of the energy resolution at NEOS as Ref.~\cite{Huber:2016xis}; however, since \globes \, assumes all energies are in GeV, we write the resolution as
\begin{equation}
\frac{\sigma}{\text{GeV}} = 0.00012 + 0.00158 \sqrt{\frac{E}{\text{GeV}}}.
\end{equation}
Lastly, we fix \texttt{\$lengthtab} = 0.02369 km, as described above. The spectrum is calculated for each of the four main fissile isotopes, and these are added together, weighted by the relevant fuel fractions \cite{Huber:2016xis,Ko:2016owz}:
\[
\left( f_{235}, \, f_{238}, \, f_{239}, \, f_{241} \right) = \left( 0.655, \, 0.072, \, 0.235, \, 0.038  \right).
\]


\subsubsection*{Statistical Analysis}

The object in which we are ultimately interested is the double ratio of NEOS and Daya Bay spectra, to wit,
\begin{equation}
\label{eq:NEOSratio}
S_i \equiv \frac{S^{\rm NEOS}_{4\nu, \,i}}{S^{\rm NEOS}_{3\nu, \,i}}  \frac{S^{\rm DB, EH1}_{3\nu, \,i}}{S^{\rm DB, EH1}_{4\nu, \,i}},
\end{equation}
where $S^{A}_{n\nu, \,i}$ is the number of antineutrino events calculated using \globes \, for experiment $A$ (= NEOS; DB, EH1) under the assumption of the existence of $n$ neutrinos in bin $i$, defined using NEOS's 0.1-MeV binning. The justification for this is as follows. 

As mentioned, the NEOS collaboration normalizes their spectrum relative to the Daya Bay measured flux. The latter has been inferred from the Daya Bay 1230-day data after unfolding three-neutrino oscillations. To properly account for the presence of a fourth neutrino in this denominator, one must first refold the best-fit three-neutrino oscillations before unfolding four-neutrino oscillations, with some assumptions about the new mass and mixing. This amounts to multiplying by the ratio of the expected numbers of events in a given bin at the Daya Bay near detectors with and without a sterile neutrino. Additionally, because the Daya Bay prediction has been rescaled to the NEOS data, there should also appear a factor of the ratio of oscillation probabilities at NEOS with and without oscillations. We are thus led to consider Eq.~\eqref{eq:NEOSratio}, which has been previously considered in Refs.~\cite{Dentler:2017tkw,Dentler:2018sju}.

We use EH1 from Daya Bay to provide the rescaling of the measured flux in Eq.~\eqref{eq:NEOSratio} in our calculations. The oscillation probabilities at NEOS are calculated assuming that only oscillations into the sterile state are relevant, but the full four-neutrino machinery is applied to Daya Bay. The following chi-squared is used in order to study the extent to which NEOS can probe the existence of a sterile neutrino:
\begin{align}
\chi^2_{\rm NEOS} & = \left( \vec{S}_{\rm exp} - \vec{S}_{\rm pred} \right)^T \cdot \left( V_{\rm N} \right)^{-1} \cdot \left( \vec{S}_{\rm exp} - \vec{S}_{\rm pred} \right) \nn \\
& + \frac{\xi_N^2}{\sigma_N^2},
\end{align}
where $\vec{S}^{\rm exp}$ are the NEOS data, $\vec{S}_{\rm pred}$ is the ratio calculated using \globes \, and $V_N$ is the covariance matrix of these data. As with DANSS, we introduce an explicit nuisance parameter $\xi_N$ for the energy scale uncertainty, whose uncertainty is nominally 0.5\% \cite{Ko:2016owz}. This is studied using the \globes \, function \texttt{glbShiftEnergyScale}; in our fits, this parameter is minimized for every point in the sterile neutrino parameter space.

We turn now to the covariance matrix $V_N$. There are two contributions that we include in this matrix. The first is the statistical uncertainties on the ratio $\vec{S}^{\rm exp}$, which are simply read off of Fig.~3(c) in Ref.~\cite{Ko:2016owz}. The second is the covariance matrix published with the antineutrino spectrum in Ref.~\cite{An:2016srz}. Daya Bay publishes their results using a different binning than NEOS; we reformulate the covariance matrix using a simple Monte Carlo routine. Random antineutrino spectra are generated using the Daya Bay spectrum and covariance matrix, including correlations. The antineutrino spectrum that NEOS would see is determined by interpolating these generated spectra to the stated NEOS binning. The contents of these bins using the above-stated energy resolution, and the resulting covariance matrix is calculated.

\begin{figure}[!t]
\includegraphics[width=\linewidth]{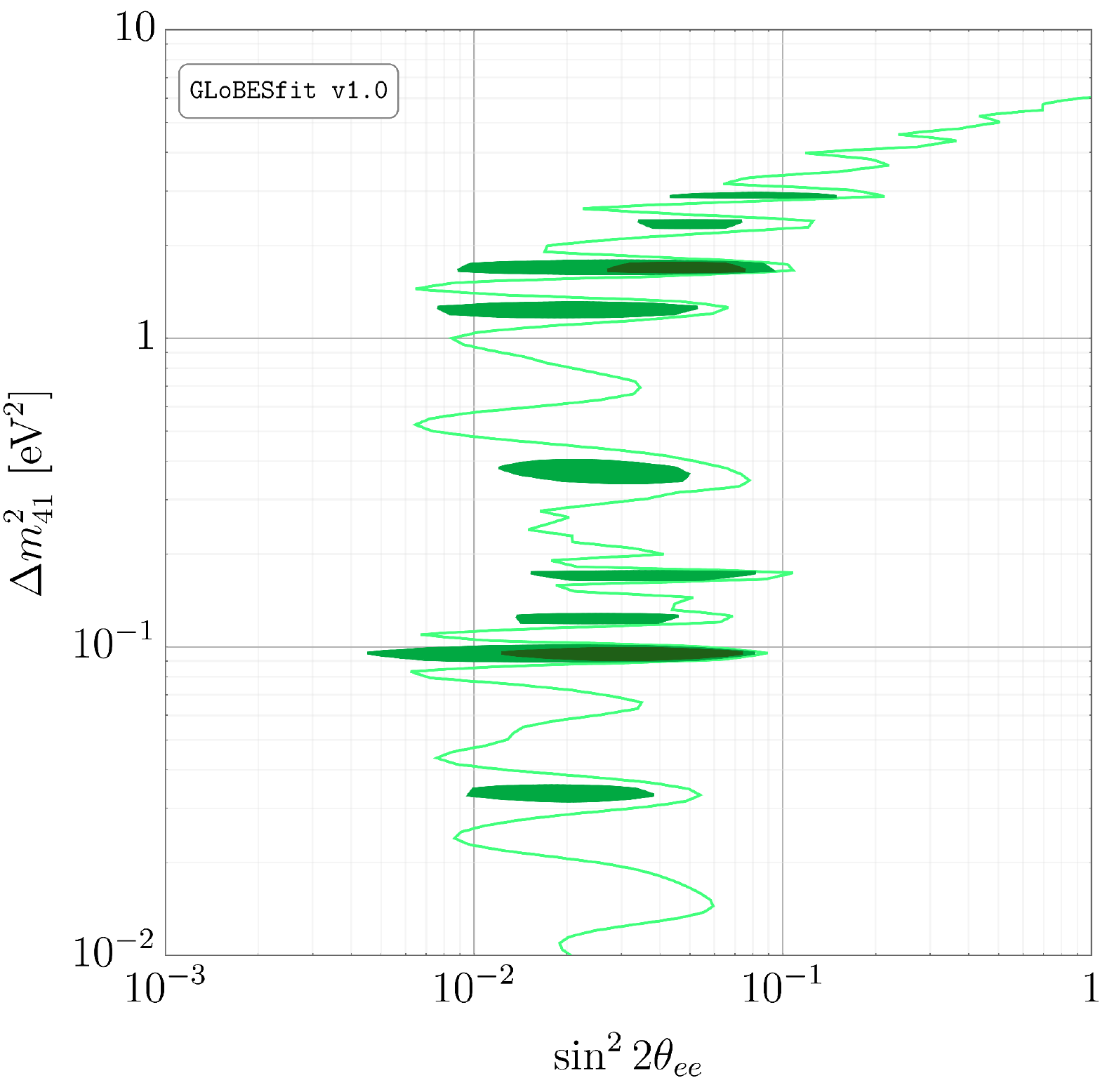}
\caption{The 95\% (dark green), 99\% (green) and 99.9\% (light green) C.L. curves for our sterile neutrino analysis of NEOS.}
\label{fig:NEOS}
\end{figure}

Formally, we need to consider the predicted spectrum given the stated NEOS fuel fraction, not the Daya Bay fuel fraction from which this result is derived. This introduces model dependence on the HM fluxes -- and with it, dependence on the uncertainties on these fluxes. These contributions have been included with a similar Monte Carlo routine, but their contribution to the final result is small~\cite{Huber:2016xis}.

It bears mentioning that the use of the Daya Bay antineutrino spectrum to normalize the NEOS spectrum implies that these experiments are formally correlated with one another. However, we ignore the correlations between these experiments for the following reason. The Daya Bay antineutrino spectrum used at NEOS is based on the 1230-day data; in contrast, the Daya Bay analysis described in Sec.~\ref{subsec:DayaBaySpec} is based on the 1958-day data. Consequently, it's difficult to accurately assess the degree of correlation between these data sets. We hope to improve this aspect of our analysis with the collection and dissemination of more data.

The results of this analysis are shown in Fig.~\ref{fig:NEOS} The dark green, green and dark green contours represent the 95\%, 99\% and 99.9\% C.L. contours, respectively. Relevant statistics are compiled in Table \ref{tab:SpecSummary}. The sensitivity at low $\Delta m_{41}^2$ ($\lesssim 0.3$ eV$^2$) is driven by the use of the Daya Bay spectrum for normalization; NEOS itself drives the sensitivity in the $\sim1$ eV$^2$ region. One could, in principle, develop an analysis based on the ratio of NEOS data to the HM flux prediction, the data for which are shown in Fig.~3(b) of Ref.~\cite{An:2016srz}; this is precisely what is done in, for instance, Ref.~\cite{Gariazzo:2018mwd}. The exclusions derived there only have support in the region around $\sim1$ eV$^2$, but, as mentioned above, are more dependent on the uncertainties of the HM fluxes.


\subsection{RENO}

\subsubsection*{Implementation in \globes}

The geometry of the RENO experiment \cite{Bak:2018ydk} has been previously discussed in the context of our rate analysis. We continue to assume that each of the six reactors is point-like, but that both the near and far detector have a width of 3.0 m, to facilitate fast oscillations in averaging out. The distances between either detector and each of the six reactors are shown in Table \ref{tab:RENObaselines}; the average powers of each reactor during the operating period of either detector are shown in Table \ref{tab:RENOpowers}. These data have been obtained from private communications with the collaboration \cite{SBKim}. As with previous experiments, we precalculate $F(q)$ (Eq.~\eqref{eq:defineF2}) on a grid with spacing 0.01 in $\log_{10} q$ over the range $q \in [10^{-2.5}, 10^1]$ for both detectors.

The near and far detectors are defined as separate experiments within \globes. For each, we set \texttt{@time} to the operating time of each detector (in days), \texttt{@power} to the total thermal power from all six reactors and \texttt{@norm} to be the efficiency of each detector. The relevant quantities are tabulated in Table \ref{tab:RENOpowers}. In the absence of any information to the contrary, we assume the detectors to be of equal mass; consequently, we set \texttt{\$target\_mass} to be 1.0 for each. We set \texttt{\$lengthtab} equal to the appropriate value of $1/\sqrt{ \langle L^{-2} \rangle}$ for the near and far detectors; these values are, respectively, 433.1 m and 1446.9 m.

The antineutrino spectrum is calculated in 29 bins of equal width on [2.1, 7.8] MeV in antineutrino energy, corresponding to [1.3, 7.0] MeV of prompt energy. The RENO data, however, are presented with nonuniform binning; the third- and second-to-last bins have width 0.2 MeV and the last has width 0.3 MeV, whereas all the others have width 0.1 MeV. In our calculations, we manually combine events in our 0.1-MeV wide bins as appropriate to achieve the correct final binning. The effective fission fractions of the four main fissile isotopes for each detector have been provided by the collaboration \cite{SBKim}; for the near detector, we use
\[
\left( f_{235}, \, f_{238}, \, f_{239}, \, f_{241} \right) = \left( 0.573, \, 0.073, \, 0.299, \, 0.055 \right),
\]
whereas we use
\[
\left( f_{235}, \, f_{238}, \, f_{239}, \, f_{241} \right) = \left( 0.574,\, 0.073, \, 0.297, \, 0.055 \right)
\]
for the far detector. As in the rate analysis, we assume a constant energy resolution of 0.4 MeV.

\subsubsection*{Statistical Analysis}

Our analysis centers around the spectral ratio presented in the bottom panel of Fig.~2 of Ref.~\cite{Bak:2018ydk}. These data and the corresponding statistical uncertainties have been digitized for use in our analyses; we reproduce these data in Fig.~\ref{fig:RENOSpectrum} in Appendix~\ref{app:SuppData}. Aside from these statistical uncertainties, we consider the following sources of systematic uncertainty:
\begin{itemize}
\item \emph{Energy Scale.} The collaboration claims a 0.15\% uncertainty on the experimental energy scale \cite{Bak:2018ydk}. We include this in our analyses via a nuisance parameter.
\item \emph{Flux Uncertainty.} A 0.9\% uncertainty on the thermal power of each reactor has been included. Assuming that this is totally uncorrelated between cores, we estimate that this implies a $\sim0.3\%$ uncertainty on the far-to-near ratio.
\item \emph{Backgrounds.} Table 1 of Ref.~\cite{Bak:2018ydk} gives the estimated total background rates at the near and far detectors, and the insets to Fig.~1 of the same reference show the background spectra. We have digitized these event spectra in order to estimate the contributions of the backgrounds to the spectral ratio via Monte Carlo. The resulting uncertainties and correlations are recorded for use in our analyses.
\end{itemize}

\begin{figure}[!t]
\includegraphics[width=\linewidth]{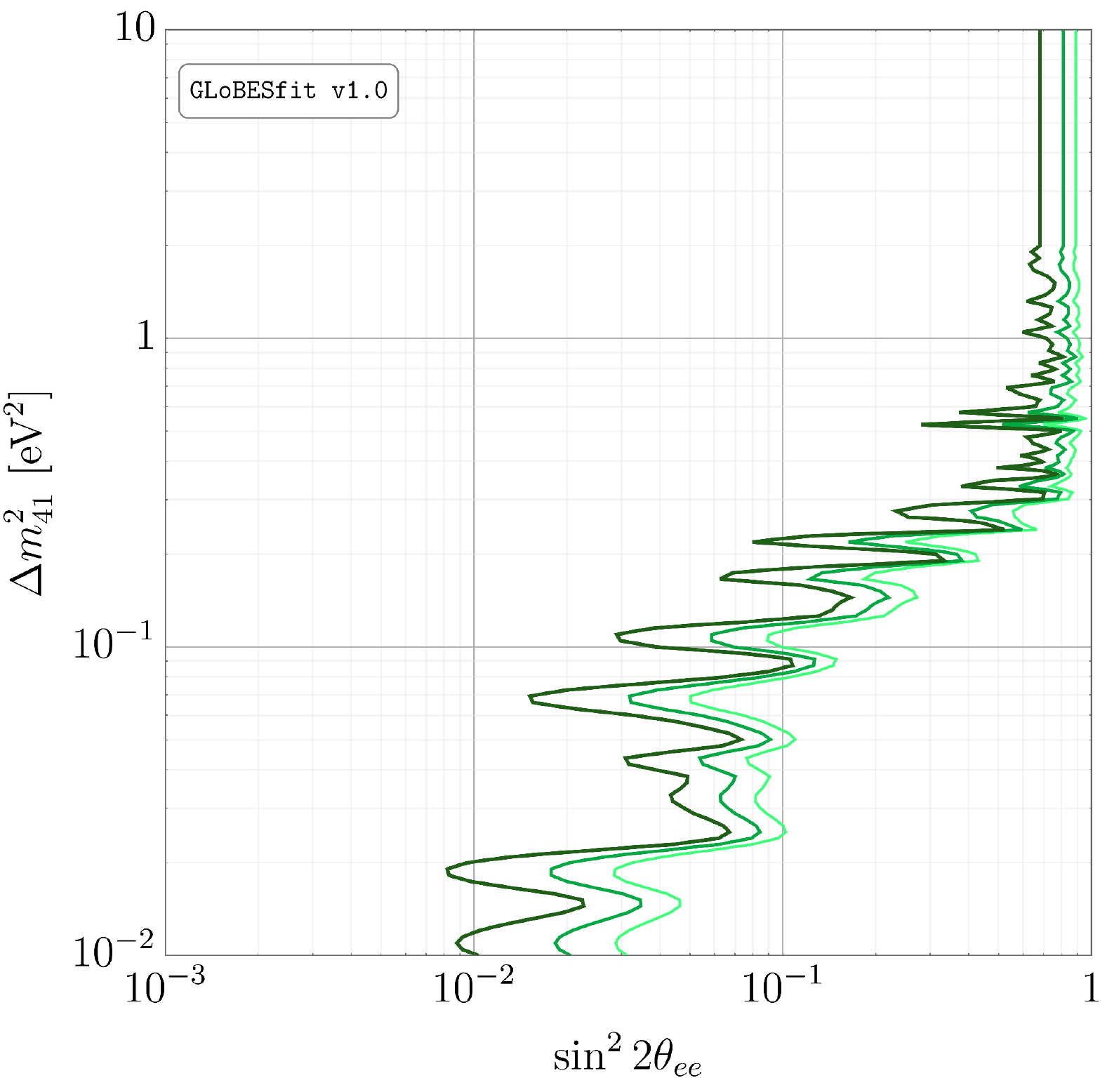}
\caption{The 95\% (dark green), 99\% (green) and 99.9\% (light green) C.L. curves for our sterile neutrino analysis of RENO.}
\label{fig:RENO}
\end{figure}

Once more, we employ a chi-squared function of the form
\begin{align}
\chi^2_{\rm RENO} & = \left( \vec{S}_{\rm exp} - \vec{S}_{\rm pred} \right)^T \cdot \left( V_{\rm R} \right)^{-1} \cdot \left( \vec{S}_{\rm exp} - \vec{S}_{\rm pred} \right) \nn \\
& + \frac{\xi_R^2}{\sigma_R^2},
\end{align}
where $\vec{S}_{\rm exp}$ and $\vec{S}_{\rm pred}$ are the experimental and predicted IBD spectra, respectively. We reweight the measured data by a factor of
\[ \varphi = \left( \frac{t_{\rm far}}{t_{\rm near}} \right) \left( \frac{\langle L^{-2}_{\rm far} \rangle}{\langle L^{-2}_{\rm near} \rangle} \right)  \left( \frac{\langle P \rangle_{\rm far}}{\langle P \rangle_{\rm near}} \right)  \left( \frac{\varepsilon_{\rm far}}{\varepsilon_{\rm near}} \right)\]
to ensure consistency in our \globes \, analysis. The covariance matrix $V_{\rm R}$ includes the statistical and systematic uncertainties described above, neglecting the energy scale uncertainty, including correlations.

The nuisance parameter $\xi_R$ describes adjustments to the energy scale at RENO, accounting for its uncertainty $\sigma_R = 0.15\%$. The effect of the energy scale shift is included through the \globes \, function \texttt{glbShiftEnergyScale}. In our calculations, we minimize $\chi^2_{\rm RENO}$ with respect to this nuisance parameter for every point in the $\sin^2 2\theta_{ee}$--$\Delta m_{41}^2$ plane.

In an attempt to faithfully reproduce this experiment in \globes, we reweight our predicted spectrum by an \emph{ad hoc} factor $\zeta_R$, determined as follows. We have simulated the IBD spectra at RENO assuming the best-fit value of $\sin^2 2\theta_{13}$ determined in Ref.~\cite{Bak:2018ydk}, i.e., $\sin^2 2\theta_{13}$ = 0.0896. We then replace $\vec{S}_{\rm pred} \to \zeta_R \vec{S}_{\rm pred}$, and minimize over $\zeta_R$. We find the minimum at $\zeta_R = 1.0088$. We then use this value of $\zeta_R$ in all subsequent calculations.

The results of our calculation are shown in Fig.~\ref{fig:RENO}.\footnote{We have again resumed calculating with $\sin^2 2\theta_{13} = 0.084$, from Ref.~\cite{Esteban:2018azc}.} Relevant statistics are compiled in Table \ref{tab:SpecSummary}. The evidence for a sterile neutrino from RENO is relatively weak.

\section{Spectral Analyses}
\label{sec:SpecAnalysis}
\setcounter{equation}{0}

In this section, we present and discuss our analyses of the experiments presented in the previous section. In particular, we consider several groupings of these experiments and assess the significance of the evidence for the existence of a sterile neutrino.

\subsection{Aggregating Results}

In Table \ref{tab:SpecSummary}, we combine relevant statistical quantities from our sterile-neutrino analyses of the experiments described in Sec.~\ref{sec:SpectrumExp}. We also tabulate the locations of the best-fit points in the $\sin^2 2\theta_{ee}$--$\Delta m_{41}^2$ plane for each experiment; these suppressed in Figs.~\ref{fig:Bugey}-\ref{fig:RENO} for clarity.

\begin{table*}[]
\centering\begin{tabular}{|c||c|c|c|c|c|c|}\hline
Experiment(s) & $\chi^2_{3\nu}$ & $n_{\rm data}$ & $\chi^2_{\rm min}$ & $p$ & $n\sigma$ & Best-Fit Point, $\left( \sin^2 2\theta_{ee}, \, \Delta m_{41}^2/\text{eV}^2 \right)$ \\ \hline \hline
Bugey-3 & 12.6 & 25 & 9.2 & 0.17 & 01.4 & ($2.00 \times 10^{-1}$, $2.88$) \\ \hline
DANSS & 33.5 & 24 & 21.3 & $2.3 \times 10^{-3}$ & 3.0 & ($7.24 \times 10^{-2}$, $1.31$)\\ \hline
Daya Bay & 45.0 & 52 & 42.7 & 0.33 & 0.98 & ($4.22 \times 10^{-1}$, $6.31 \times 10^{-1}$) \\ \hline
Double Chooz & 8.0 & 26 & 5.7 & 0.31 & 1.0 & ($3.31 \times 10^{-2}$, $2.75 \times 10^{-2}$) \\ \hline
NEOS & 65.4 & 60 & 54.1 & $3.4 \times 10^{-3}$ & 2.9 & ($4.37 \times 10^{-2}$, $9.55 \times 10^{-2}$) \\ \hline
RENO & 23.6 & 25 & 19.7 & 0.15 & 1.5 & ($4.57 \times 10^{-1}$, $ 3.63 \times 10^{-1}$) \\ \hline \hline
DANSS + NEOS & 98.9 & 84 & 84.9 & $9.2 \times 10^{-4}$ & 3.3 & ($4.37 \times 10^{-2}$, $1.26$) \\ \hline
Daya Bay + NEOS & 110.4 & 112 & 103.6 & $3.3 \times 10^{-2}$ & 2.1 & ($4.37 \times 10^{-2}$, $6.31\times10^{-1}$) \\ \hline \hline
Total & 188.2 & 212 & 175.2 & $1.6 \times 10^{-3}$ & 3.2 & ($3.80 \times 10^{-2}$, $1.26$) \\ \hline
Modern & 175.5 & 187 & 161.7 & $9.9 \times 10^{-4}$ & 3.3 & ($4.17 \times 10^{-2}$, $1.26$) \\ \hline
\end{tabular}
\caption{A summary of all results for spectral analyses.}
\label{tab:SpecSummary}
\end{table*}

\begin{figure}[!t]
\includegraphics[width=\linewidth]{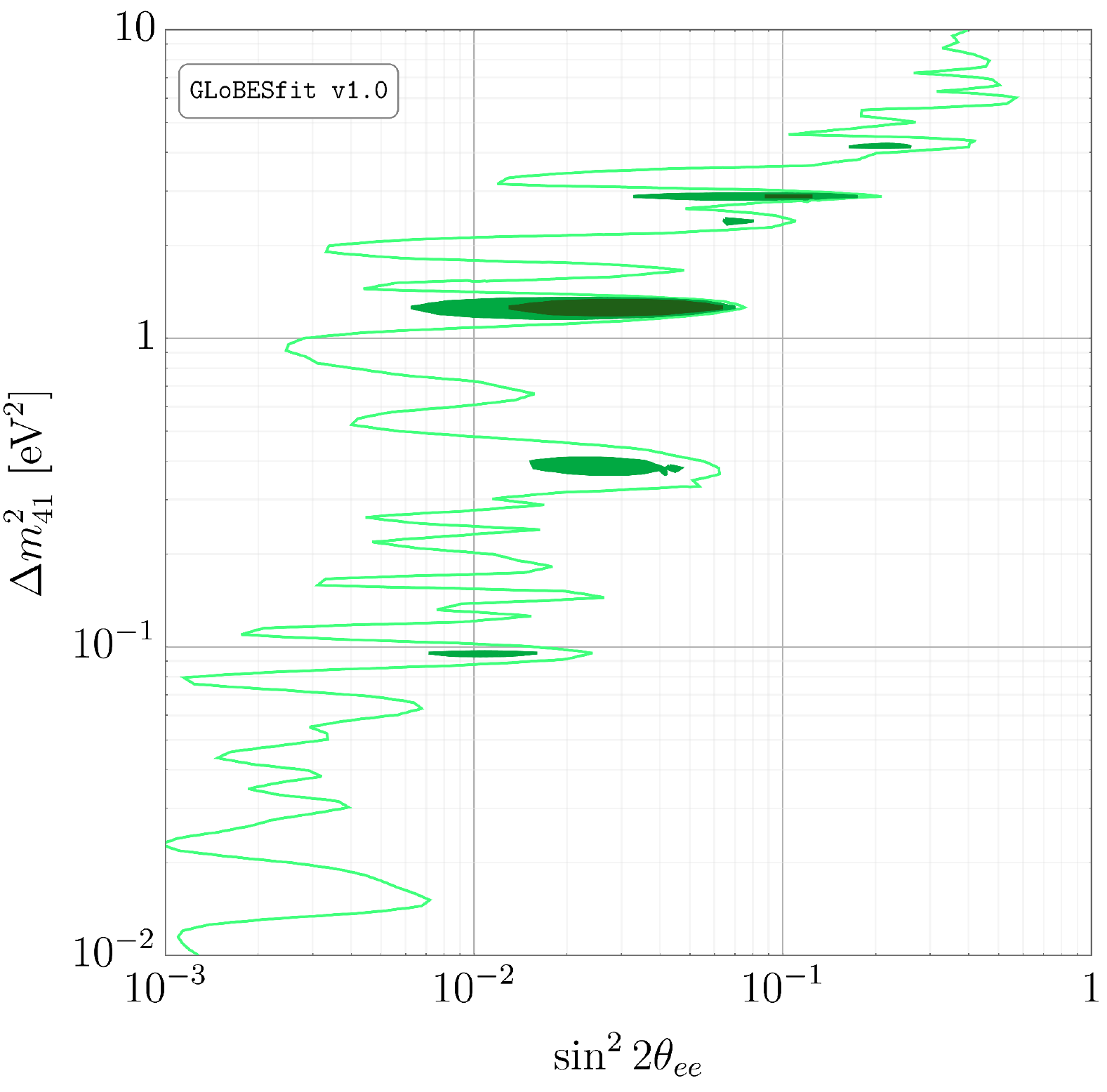}
\caption{The 95\% (dark green), 99\% (green) and 99.9\% (light green) C.L. contours derived from our combined sterile-neutrino analysis of all spectrum experiments.}
\label{fig:Total}
\end{figure}

\subsection{Combined Sterile Neutrino Analysis}

The exclusion contours from a combined analysis of these spectral measurements is shown in Fig.~\ref{fig:Total}. Since we have assumed correlations between these experiments are small, this figure is equivalent to the naive additional of the $\chi^2$ maps in Figs.~\ref{fig:DANSS}-\ref{fig:NEOS}. Relevant statistics are compiled in the penultimate line of Table \ref{tab:SpecSummary}. At the best-fit point, we find $\chi^2_{\rm min} = 175.2$; given the three-neutrino value $\chi^2_{3\nu} = 188.2$, this corresponds to $p=1.6\times10^{-3}$, or roughly $3.2\sigma$.

While introducing a sterile neutrino has significantly improved the fit, the three-neutrino hypothesis is not obviously insufficient --- for 212 degrees of freedom, the above value of $\chi^2_{3\nu}$ corresponds to $p=0.88$. All of Bugey-3, Daya Bay, Double Chooz and RENO have values of $\chi^2_{3\nu}$ that are less than the number of data points used in each analysis; simply put, these experiments do not convey a need to introduce a sterile neutrino. On the other hand, DANSS and NEOS show a strong preference for a sterile neutrino, as previously discussed (particularly the former); these experiments drive the global preference for a sterile neutrino, as has been previously observed \cite{Dentler:2017tkw,Dentler:2018sju,Gariazzo:2018mwd}. The best-fit point in Fig.~\ref{fig:Total} is at $(\sin^2 2\theta_{ee}, \, \Delta m_{41}^2) = (3.80 \times 10^{-2}, \, 1.26 \text{ eV}^2)$, which is roughly consistent with the best-fit point from the spectral analyses in Refs.~\cite{Gariazzo:2018mwd,Dentler:2018sju}.

\subsection{Additional Analyses}

In this subsection, we discuss particularly interesting subsets of the spectral-ratio data and the extent to which these indicate the possible existence of a sterile neutrino.

\subsubsection*{Calibration: Measuring $\sin^2 2\theta_{13}$ and $\Delta m_{31}^2$}

\begin{figure}[!t]
\includegraphics[width=\linewidth]{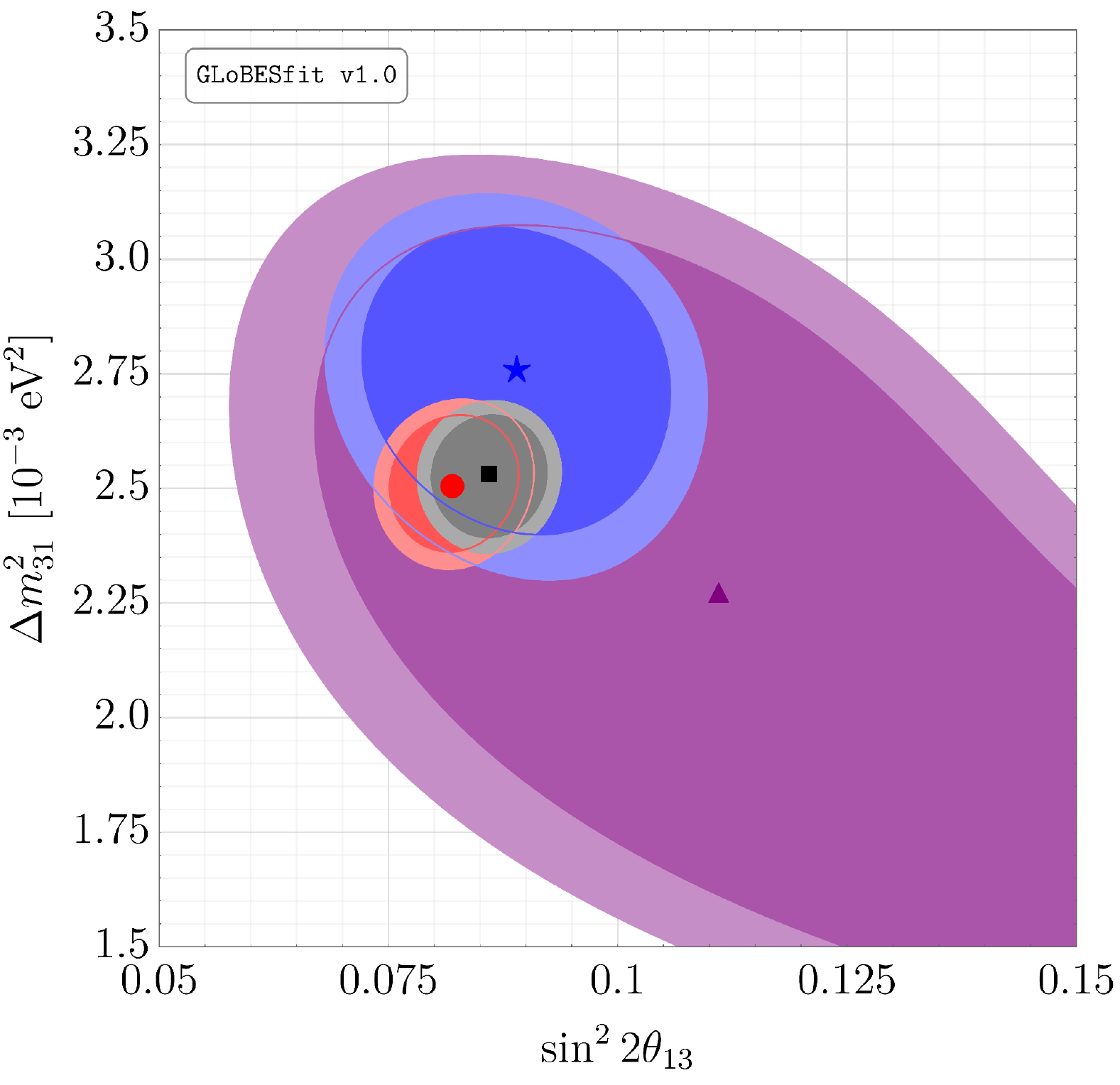}
\caption{Our replication of the measurements of $\sin^2 2\theta_{13}$ and $\Delta m_{31}^2$ assuming three-neutrino oscillations only. Shown are results for Daya Bay (red), Double Chooz (purple) and RENO (blue), as well as the combination of the three of these (gray). Dark (light) regions represent 95\% (99\%) C.L. The best-fit points for each analysis are represented by shapes of the corresponding color.}
\label{fig:Calibration}
\end{figure}

We begin this section by discussing how we verify that our \globes \, analyses of Daya Bay, Double Chooz and RENO reproduce the experimental measurements of the three-neutrino parameters $\sin^2 2\theta_{13}$ and $\Delta m_{31}^2$ in the absence of a sterile neutrino. We scan over these parameters and calculate the $\chi^2$ precisely as described above; the results are shown in Fig.~\ref{fig:Calibration}. The dark (light) regions represent the 95\% (99\%) C.L. contours derived for each of Daya Bay (red), Double Chooz (purple) and RENO (blue). The gray curves represent a combined analysis of all three of these.

These regions are broadly consistent with the best-fit values found in the analyses of the respective collaborations \cite{Adey:2018zwh,Bak:2018ydk,DoubleChooz:2019qbj}.\footnote{We elect to frame this analysis in terms of $\Delta m_{31}^2$ instead of $\Delta m_{ee}^2$. The precise relationship between these two quantities has been debated in the literature \cite{Parke:2019mhz,Adey:2019ucn}, but this is not relevant for our purposes here.} This figure can also be compared against Fig.~5 of Ref.~\cite{Esteban:2018azc}, where a combined analysis of these medium-baseline experiments has also been performed. We again find general agreement between the calculated regions. The correspondence is not exact -- for instance, the region preferred by Double Chooz is slightly wider here than the corresponding region in Ref.~\cite{Esteban:2018azc} -- but given that these fits have been performed with often incompletely reported information, we consider this a modest success.

One may worry that introducing a sterile neutrino may cause a significant mismeasurement of $\sin^2 2\theta_{13}$ at reactor experiments. We have investigated how sensitivity to a sterile neutrino changes if $\sin^2 2\theta_{13}$ is unfixed from its best-fit value, and how introducing a sterile neutrino may shift the preferred value of $\sin^2 2\theta_{13}$.\footnote{Given that accelerator neutrino experiments also provide a measurement of $\Delta m_{31}^2$, and that these measurements are consistent with the value measured at reactors, we leave this parameter fixed at its best-fit value for this study.} We find, however, that this is not the case. The measurement of $\sin^2 2\theta_{13}$ is driven primarily by Daya Bay (see Fig.~\ref{fig:Calibration}), whereas the measurement of $\sin^2 2\theta_{ee}$ is driven primarily by DANSS, and that these measurements are largely uncoupled when performed simultaneously. This validates our fixing $\sin^2 2\theta_{13}$ in our analyses above.

\subsubsection*{DANSS and NEOS}

\begin{figure}[!t]
\includegraphics[width=\linewidth]{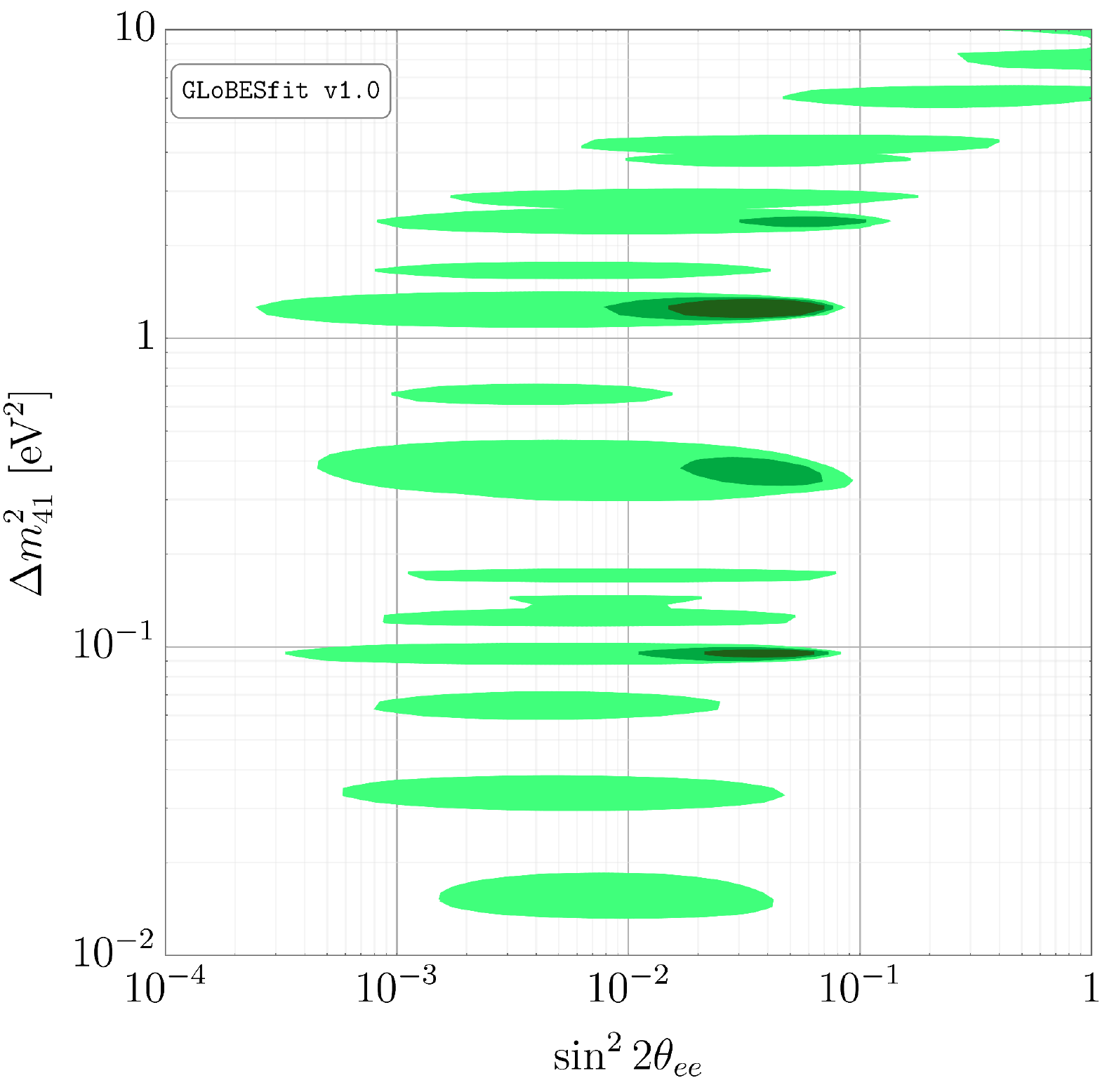}
\caption{The 95\% (dark green), 99\% (green) and 99.9\% (light green) C.L. contours derived from our combined analysis of DANSS and NEOS.}
\label{fig:DANSS_NEOS}
\end{figure}

Given that DANSS and NEOS drive the sensitivity to a sterile neutrino in our global fit, it seems pertinent to ask how the combination of just these two experiments constrains the $\sin^2 2\theta_{ee}$--$\Delta m_{41}^2$ plane. The resulting 95\%, 99\% and 99.9\% C.L. contours are shown in dark green, green and light green, respectively, in Fig.~\ref{fig:DANSS_NEOS}. Relevant statistics are compiled in Table \ref{tab:SpecSummary} in the row headed ``DANSS + NEOS.''

Restricting to these two experiments has resulted in $p = 9.2 \times 10^{-4}$, corresponding to $3.3\sigma$. Notice, however, that the value of the chi-squared at the best-fit point, $\chi^2_{\rm min}$, is 84.9. For 82 degrees of freedom, this corresponds to $p = 0.39$; this is an acceptable fit, but hardly compelling evidence for the existence of a sterile neutrino. Moreover, the best-fit points from our analyses of DANSS and NEOS do not coincide, and the best-fit point from their combination lies close to that from DANSS alone. Consequently, we have investigated the compatibility of these data sets using a parameter goodness-of-fit test \cite{Maltoni:2003cu}. We define
\begin{equation}
\chi^2_{\rm PG} \equiv \chi^2_{\rm DANSS+NEOS} - \chi^2_{\rm DANSS} - \chi^2_{\rm NEOS},
\end{equation}
where $\chi^2_{A}$ is the minimum value of the chi-squared for experiment $A$. We find $\chi^2_{\rm PG} = 6.4$; assuming this is chi-squared distributed implies $p = 4.0 \times 10^{-2}$. These data sets seem to be in mild tension. While parameter goodness-of-fit values should be viewed with some skepticism (see, for instance, Ref.~\cite{Diaz:2019fwt}), it bears mentioning that the evidence for a sterile neutrino from this experiments, while modestly strong, is not entirely overwhelming.

\subsubsection*{Daya Bay and NEOS}

\begin{figure}[!t]
\includegraphics[width=\linewidth]{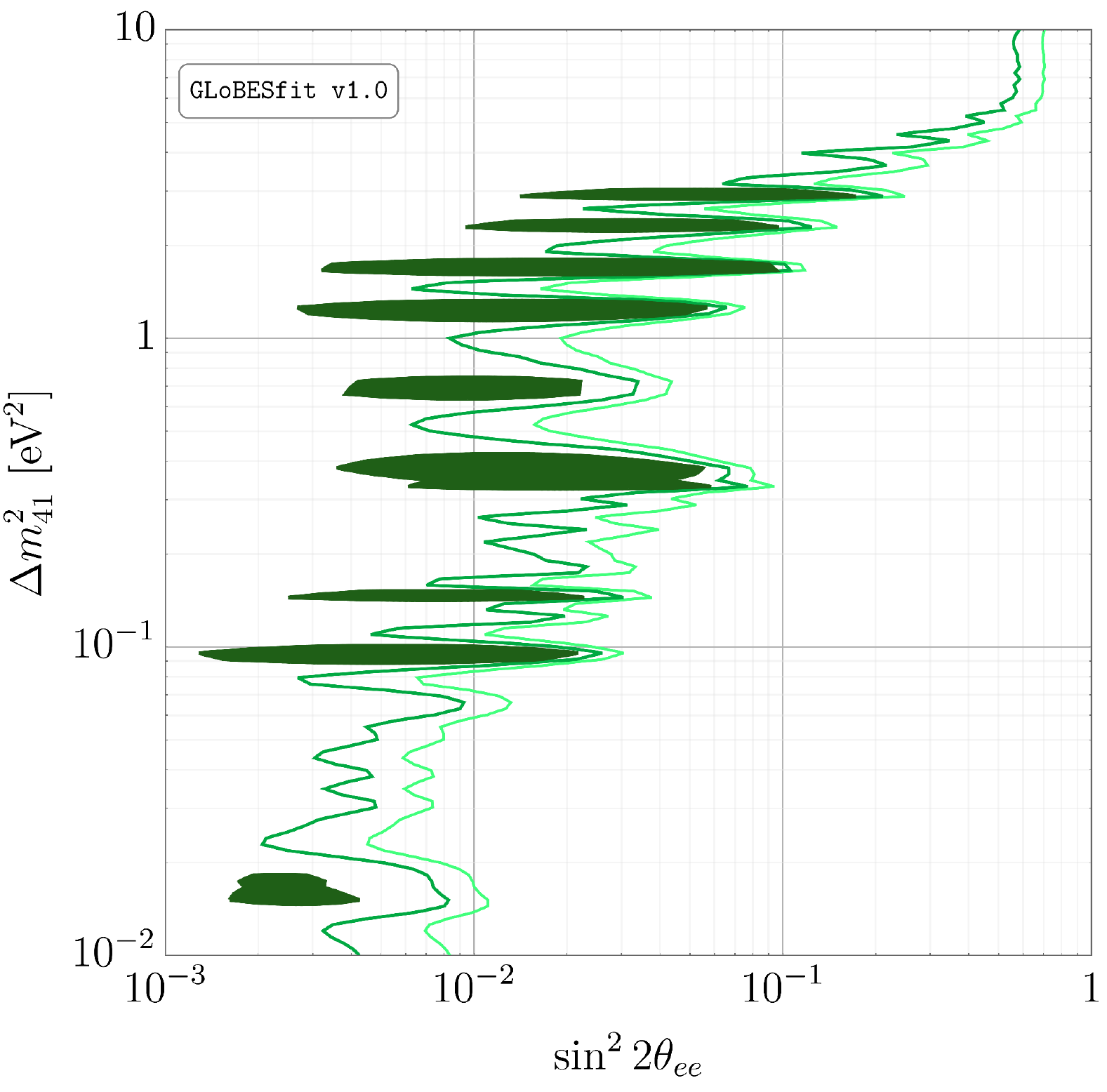}
\caption{The 95\% (dark green), 99\% (green) and 99.9\% (light green) C.L. contours derived from our combined analysis of Daya Bay and NEOS.}
\label{fig:DB_NEOS}
\end{figure}

Given the manner in which we have been forced to include NEOS -- namely, as a ratio with respect to Daya Bay's EH1 -- one could reasonably object that showing results from NEOS on its own has stripped this analysis of valuable context. To this end, we have performed a combined analysis of Daya Bay and NEOS, the results of which we show in Fig.~\ref{fig:DB_NEOS}. The resulting 95\%, 99\% and 99.9\% C.L. contours are shown in dark green, green and light green, respectively, and relevant statistics are compiled in Table \ref{tab:SpecSummary} in the row headed ``Daya Bay + NEOS.''

While the best-fit point to the NEOS data unto itself lies in the region around $\Delta m_{41}^2 \sim 10^{-1}$ eV$^2$, this part of the parameter space is disfavored by the medium-baseline experiments, including Daya Bay. Combining Daya Bay and NEOS makes this apparent; while this part of the parameter space is still allowed, much more of the sterile neutrino parameter space is included in the at 95\% C.L. region.

As mentioned in Sec.~\ref{sec:SpectrumExp}, we have ignored possible correlations between Daya Bay and NEOS in our analysis because they are hard to quantify exactly. Consequently, this analysis is tantamount to adding together the individual $\chi^2$ maps over the sterile neutrino parameter space. We note, however, that accounting for these correlations is one way in which this analysis can be strengthened in the futurel e look forward to improving this treatment in future versions of \globesfit.

\subsubsection*{Analyzing Modern Experiments}

In our analyses of rate experiments, we considered how considering only experiments from the 2010s alters the evidence for a sterile neutrino. In the interest of parity, we perform a similar analysis here. For the experiments that we consider, this is tantamount to removing Bugey-3 from the fit.

\begin{figure}[!t]
\includegraphics[width=\linewidth]{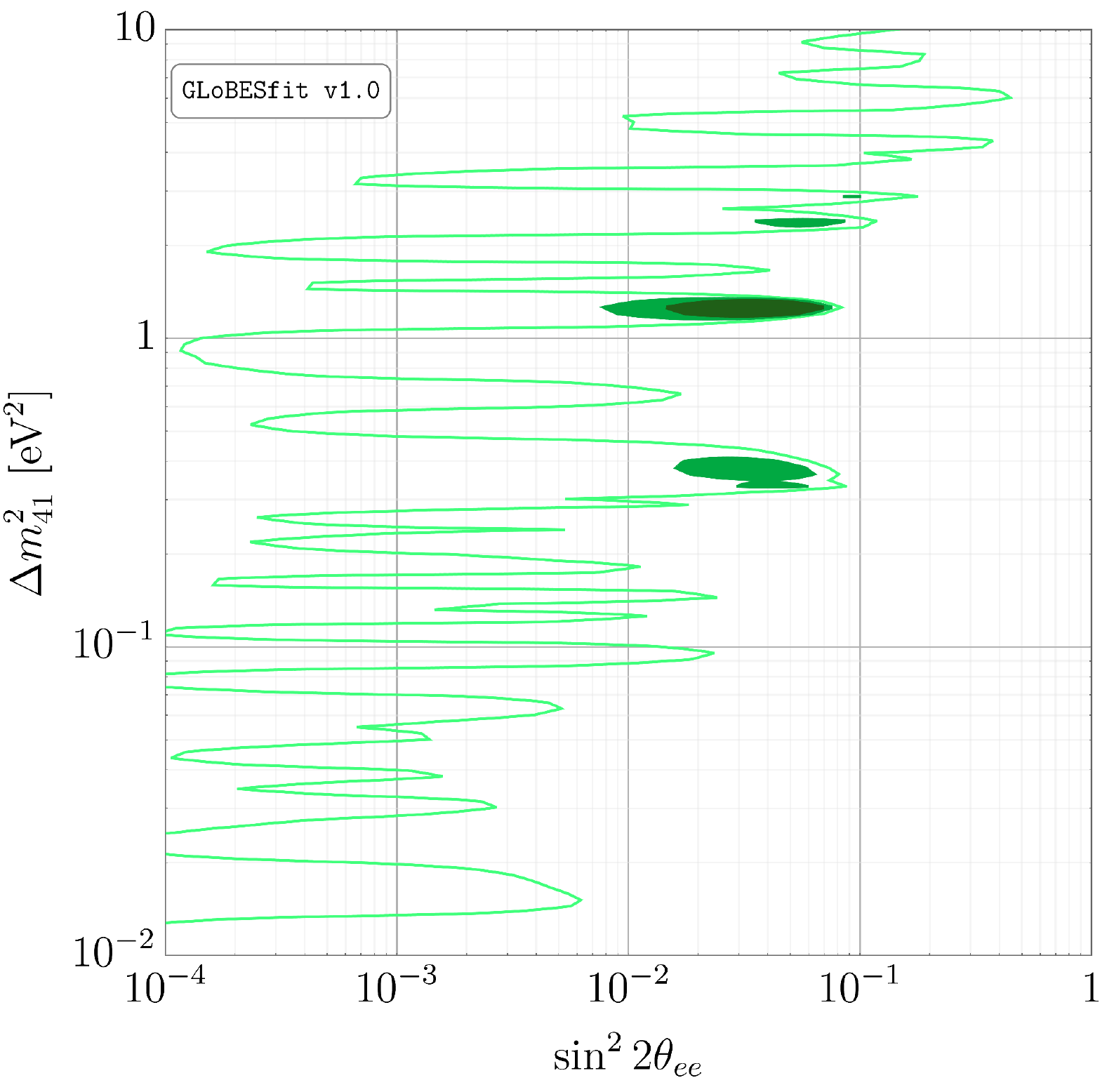}
\caption{The 95\% (dark green), 99\% (green) and 99.9\% (light green) C.L. contours derived from our combined sterile-neutrino analysis of spectrum experiments from the 2010s.}
\label{fig:ModernSpec}
\end{figure}

We show the results of this analysis in Fig.~\ref{fig:ModernSpec}. As before, the 95\%, 99\% and 99.9\% C.L. contours are shown in dark green, green and light green, respectively. Relevant statistics are compiled in the last line of Table~\ref{tab:SpecSummary}. It is clear that removing Bugey-3 from the fit has not dramatically changed the best-fit region (cf. the dark green region in Fig.~\ref{fig:Total}). In fact, the fit is slightly stronger for Bugey-3's omission --- the preference for a sterile neutrino rises to $3.3\sigma$ which, while not much higher than $3.1\sigma$, is an increase relative to the analysis of all experiments. The reason for this is clear from Fig.~\ref{fig:Bugey}: Bugey-3 indicates no preference for a sterile neutrino at $\Delta m_{41}^2 = 1.26$ eV$^2$. In fact, our analysis implies that the best-fit point from our analysis of modern experiments is less favored than zero mixing by the Bugey-3 data.

We emphasize that we know of no concrete reason why any experiment performed before 2010 (in this case, Bugey-3) should be disregarded in any of the fits we have performed. We mean merely to demonstrate that if one were suspicious of these data, for whatever reason, then their inclusion has not dramatically enhanced the preference for a sterile neutrino -- the evidence is, in fact, largely driven by experiments from the past decade.

\subsubsection*{Energy Scale Uncertainties at Bugey-3, DANSS and NEOS}

An important source of systematic uncertainty in these experiments is the energy scale of the detector. While the energy scale of liquid scintillator detectors is nonlinear, we assume in our analyses that the energy response is linear, and that the only uncertainty is on the slope of this energy response. This is precisely the meaning of the nuisance parameters $\xi_{\rm 15, 40, D, N}$ in Sec.~\ref{sec:SpectrumExp}.

Loosely speaking, the uncertainty on the energy scale translates into uncertainty on the inferred value of $\Delta m_{41}^2$; a systematic shift in reconstructed values of $E_\nu$ induces a corresponding offset in the measured value of $\Delta m_{41}^2$. If the energy scales were changed at these experiments, then the fringe-like structures in the resulting confidence level contours would similarly be shifted up and down in Figs.~\ref{fig:Bugey}-\ref{fig:RENO}.

As we have discussed, the evidence for a sterile neutrino is driven by the combination of DANSS and NEOS. This can be seen from Figs.~\ref{fig:DANSS} and \ref{fig:NEOS} by looking at the fringes in the confidence level contours: the best-fit point resides in the region where the patterns of fringes line up and leave a gap in the sterile neutrino parameter space.  One could then ask if the patterns of fringes can be moved relative to one another to produce an allowed region elsewhere. Conversely, we have also seen that Bugey-3 presents a mild challenge to the sterile-neutrino interpretation of modern spectral experiments. One could also ask if allowing the energy scale to vary at Bugey-3 allows for this tension to be mitigated.

Unsurprisingly, we find that shifting the energy scales within their stated error budgets does negligibly little to address either of these concerns -- clearly, this would have already appeared in our analyses, wherein the energy scales are allowed to vary as nuisance parameters. One could, however, ask how increasing these uncertainties modifies our analyses. We have repeated these analyses with the energy scale uncertainties all increased to 20\%.\footnote{We do not claim that any of these experiments has underreported their energy scale uncertainty. We mean merely to isolate the effect of this particular contribution to the overall error budget.} Even here, we find that the resulting patterns of fringes have not appreciably moved in the $\Delta m_{41}^2$ direction relative to the nominal analyses. Therefore, we are led to conclude that energy scale uncertainties do not contribute meaningfully to the preference for a sterile neutrino in these experiments.


\section{Sterile Neutrinos and the 5 MeV Bump}
\label{sec:Bump}
\setcounter{equation}{0}

The presence of an unexplained spectral feature at 5.0 MeV in the absolute prompt energy spectrum, independently observed by several experiments~\cite{RENO:2015ksa,An:2015nua,Abe:2015rcp}, has caused significant interest in recent years. In a sense, the so-called 5 MeV bump is a microcosm of the current situation regarding reactor antineutrino anomalies: there is a lingering suspicion that some unaccounted-for nuclear physics effect is operative, but precisely how this filters into the hunt for sterile neutrinos is unclear. In Ref.~\cite{Berryman:2018jxt}, it was proposed that the bump could arise from the process $^{13}$C$(\overline{\nu}, \overline{\nu}^\prime n)^{12}$C in the presence of some new interaction. While this interaction could, in principle, explain the excess, it was found that concrete models of potential new physics that could provide such an interaction are already strongly constrained. This all suggests that nuclear physics effects are a much more likely explanation of the bump.

Still, if the bump is indeed the result of a misunderstanding of the antineutrino flux, then this does not imply that a sterile neutrino does not exist. However, it is important to understand the effect that the bump has on searches for sterile neutrinos; this is the subject of this section. In the first subsection, we study a subset of the global dataset indicating the existence of the bump to determine how strong the evidence for this bump is. In the second, we present sterile neutrino analyses of the global reactor dataset with a phenomenological parametrization of the bump, and discuss the results.

\subsection{The Evidence for the Bump}

We begin by considering the most relevant subset of the evidence for the existence of the bump. In particular, we consider the following three measurements:
\begin{enumerate}
\item The measurements of the isotopic prompt energy spectra for $^{235}$U and $^{239}$Pu inferred from Daya Bay burn-up measurements in Ref.~\cite{Adey:2019ywk}. We ignore the first three and the last two bins for each of these spectra, where experimental systematic uncertainties can be large.
\item The fractional excess of events in the region [3.8, 7.0] MeV (prompt energy) relative to the HM flux predictions, published by RENO in Fig.~5 of Ref.~\cite{RENO:2018pwo}.
\item The $^{235}$U spectrum measured by PROSPECT in Ref.~\cite{Ashenfelter:2018jrx}.\footnote{While we have included PROSPECT as a part of this analysis, we reiterate that we do not include this experiment as a part of our sterile neutrino exclusion.}
\end{enumerate}
We make extensive use of the supplementary material published by the Daya Bay and PROSPECT collaborations, as pertains to these measurements. For the RENO data, we have digitized Fig.~5 of Ref.~\cite{RENO:2018pwo} for our analysis. More specifically, Daya Bay and PROSPECT have published covariance matrices for their measured spectra, including nontrivial correlations; for RENO, we assume that the data are uncorrelated in the absence of concrete information.

We ask how well these data are described by the HM fluxes and determine the extent to which introducing a bump affects the quality of this fit. To these ends, we use the Markov Chain Monte Carlo (MCMC) package {\sc emcee} \cite{ForemanMackey:2012ig} to perform two types of analyses. In the first, we allow only the magnitudes of the isotopic fluxes of $^{235}$U and $^{239}$Pu to vary -- multiplied by quantities $r_{235}$ and $r_{239}$, respectively -- while maintaining the shapes given by the HM fluxes. In the second, we add a separate Gaussian feature to each of these isotopic fluxes. The Gaussians themselves are normalized to unit area, though we multiply them by $n_{235}$ and $n_{239}$, respectively; we simplify our analysis by fixing the means of these Gaussians to 5.8 MeV antineutrino energy (corresponding to 5.0 MeV prompt energy), and we assume they share a common width $\sigma_{bump}$. We scan over the appropriate parameter spaces and determine the posterior probability densities for each case, and compare the maximum likelihoods.

We have considered two combinations of the data mentioned above. For the first, we simultaneously analyze Daya Bay and RENO; for the second, we add PROSPECT into the analysis. The PROSPECT spectrum of Ref.~\cite{Ashenfelter:2018jrx} has been presented with arbitrary normalization. To account for this, we introduce a free-floating parameter $\eta$ to rescale the predicted $^{235}$U spectrum at PROSPECT, independent of $r_{235}$ above. Table \ref{tab:MCMC1} shows relevant statistics from our analysis performed in the absence of a bump. We present the maximum value of the likelihood $\mathcal{L}$ in terms of the minimum value of $-2 \log \mathcal{L}$. Shown also are the number of degrees of freedom $n_{\rm dof}$, and the $p$-value derived assuming $(-2 \log \mathcal{L})_{\rm min}$ is Gaussian-distributed. Moreover, Table \ref{tab:MCMC2} shows the same quantities for our analyses in which bumps in the isotopic fluxes are introduced. Clearly, the current data greatly prefer this sort of bump over the pure HM fluxes.

\begin{table}
\centering
\begin{tabular}{|c||c|c|c|} \hline
Analysis & $(-2 \log \mathcal{L})_{\rm min}$ & $n_{\rm dof}$ & $p$ \\ \hline \hline
Daya Bay + RENO & 172.3 & 40 & $2.4\times10^{-18}$ \\ \hline
'' + PROSPECT & 212.2 & 71 & $5.5\times10^{-16}$ \\ \hline
\end{tabular}
\caption{The results of our MCMC scans over Daya Bay and RENO (and PROSPECT) in the absence of the 5 MeV bump. See text for details.}
\label{tab:MCMC1}
\vspace{2mm}
\begin{tabular}{|c||c|c|c|} \hline
Analysis & $(-2 \log \mathcal{L})_{\rm min}$ & $n_{\rm dof}$ & $p$ \\ \hline \hline
Daya Bay + RENO & 32.0 & 37 & 0.70 \\ \hline
'' + PROSPECT & 91.8 & 68 & $2.9 \times 10^{-2}$ \\ \hline
\end{tabular}
\caption{The results of our MCMC scans over Daya Bay and RENO (and PROSPECT) in the presence of the 5 MeV bump. See text for details.}
\label{tab:MCMC2}
\end{table}

More of the structure of the fits in which a bump is invoked is shown in Figs.~\ref{fig:MCMC1} and \ref{fig:MCMC2} in Appendix \ref{app:SuppData} for our analyses without and with PROSPECT, respectively. In the two-dimensional planes, the blue, orange and red contours represent the 68.3\%, 95\% and 99\% credible regions (C.R.), respectively. Additionally, the one-dimensional plots also show the 90\% C.R. curve in green. Firstly, both analyses prefer the HM fluxes to be rescaled downward in magnitude, particularly for $r_{235}$. Secondly, while the data are largely consistent with the absence of a bump in $^{239}$Pu, nonzero $n_{235}$ is preferred at $\gtrsim 7 \sigma$ --- the data unequivocally prefer a spectral distortion.

The fit to Daya Bay and RENO in the presence of a bump is fairly high quality ($p = 0.70$) but degrades quite steeply ($p = 2.9 \times 10^{-2}$) when PROSPECT is introduced. The reason is straightforward. The Daya Bay and RENO analyses prefer for the absolute IBD yields from $^{235}$U to be scaled down by $\sim10$\%.\footnote{The best-fit rescaling of the Daya Bay $^{235}$U flux that we calculate here ($0.867\pm0.027$) is smaller than that published by the collaboration (0.92) \cite{Adey:2019ywk}. Daya Bay has determined this value by fitting to the total number of events, whereas we have fit based on the measured spectrum; the systematics associated with these measurements differ, so one would not necessarily expect these procedures to yield the same result.} This primarily stems from the deficit of events observed at low energies at Daya Bay, whereas the spectrum above $E_\nu \sim 5$ MeV is largely consistent with the HM prediction (see Fig.~\ref{fig:fig_a}). On the other hand, the PROSPECT spectrum is reasonably consistent with the shape of the HM prediction for the $^{235}$U flux, to within modest-sized uncertainties, over the entire energy range, particularly in the low-energy region. Clearly, the fit cannot successfully reconcile these differing low-energy behaviors. The issue of the reactor bump may be a commentary on the low-energy spectrum as much as the high-energy spectrum.

For context, the PROSPECT data, unto themselves, yield $(-2 \log \mathcal{L})_{\rm min} \approx 54$; for $n_{\rm dof} = 29$, this corresponds to $p = 3.2 \times 10^{-2}$. It bears mentioning that PROSPECT is still collecting data -- the discrepancy between the PROSPECT and Daya Bay measured $^{235}$U spectra may well be resolved with increased statistics and improved analysis techniques.

\subsection{Sterile Neutrino Analyses with the Bump}

We turn now to the issue of how the inclusion of a bump in the isotopic flux of $^{235}$U alters the evidence for the existence of a sterile neutrino. We will consider both rate and spectral analyses in turn.

We begin by considering the effect of the 5 MeV bump on our rate analyses in Secs.~\ref{sec:RateExps} and \ref{sec:RateResults}. Antineutrinos in the range [5.3, 6.3] MeV are predicted to constitute 13.5\% of IBD events at these experiments. If the $^{235}$U  flux is augmented with a Gaussian bump of the sort indicated by our Daya Bay/RENO/PROSPECT analysis, i.e, a Gaussian bump with height $n_{235} = 5.915 \times 10^{-3}$ (cf. Fig.~\ref{fig:MCMC2}), then this factor becomes 14.4\% --- a 0.9\% increase. Moreover, since many of the experiments considered here have fission fractions of the order $f_{235} \approx 0.5$, we see that adding this sort of Gaussian bump increases the \emph{predicted} IBD rate by $\lesssim 0.5$\%. This will not be enough to explain the entire deficit with respect to the HM prediction, as we will see.

On the other hand, we have the measurements of ratios of spectra in Secs.~\ref{sec:SpectrumExp} and \ref{sec:SpecAnalysis}. As argued previously, we expect these measurements to be largely insensitive to the details of the antineutrinos flux. Therefore, we do not expect the 5 MeV bump to significantly affect searches for sterile neutrinos in this channel, either.

We have confirmed this suspicion by calculating the distribution of $\chi^2$ over the $\sin^2 2\theta_{ee}$--$\Delta m_{41}^2$ plane -- precisely as has been done previously -- with the additional contribution from a Gaussian bump in $^{235}$U. We vary the height of the bump between $0.0$ and $10^{-2}$ in the units of Figs.~\ref{fig:MCMC1} and \ref{fig:MCMC2}, and have determined that the minimum value of the $\chi^2$ changes by less than one unit over this range. Specifically, we find $\chi^2_{\rm min}(n_{235} = 0) = 174.8$\footnote{This does not agree with the equivalent value in Table \ref{tab:SpecSummary} because we are using the HM fluxes as our baseline here, whereas before we had been using the \emph{ab initio} fluxes.} and $\chi^2_{\rm min}(n_{235} = 10^{-2}) = 175.5$. The bulk of this difference is attributable to NEOS, given their unconventional approach to normalizing their spectrum. Moreover, we find that the statistical significance, in terms of $n\sigma$, does not vary meaningfully from $3.1\sigma$.

We are led to conclude that the presence of the 5 MeV bump does not substantively impact searches for sterile neutrinos at reactor experiments, because (1) the contribution of events contained within the bump is at the sub-percent level, and (2) spectral ratios are, by design, largely insensitive to the particulars of the flux model employed.


\section{Regarding Neutrino-4}
\label{sec:Neutrino4}
\setcounter{equation}{0}

Neutrino-4 \cite{Serebrov:2018vdw,Neutrino4talk} is a liquid-scintillator antineutrino detector located at the SM-3 reactor in Dimitrovgrad, Russia. The moveable detector covers a range of distances -- between 6 and 12 m -- from a 100 MW core. Recently, the collaboration has reported \cite{Serebrov:2018vdw} results that imply 3.5$\sigma$ evidence for the existence of a sterile neutrino; preliminary results presented in December 2019 \cite{Neutrino4talk} suggest that this feature persists with the collection of more data.

\begin{figure}[t]
\includegraphics[width=\linewidth]{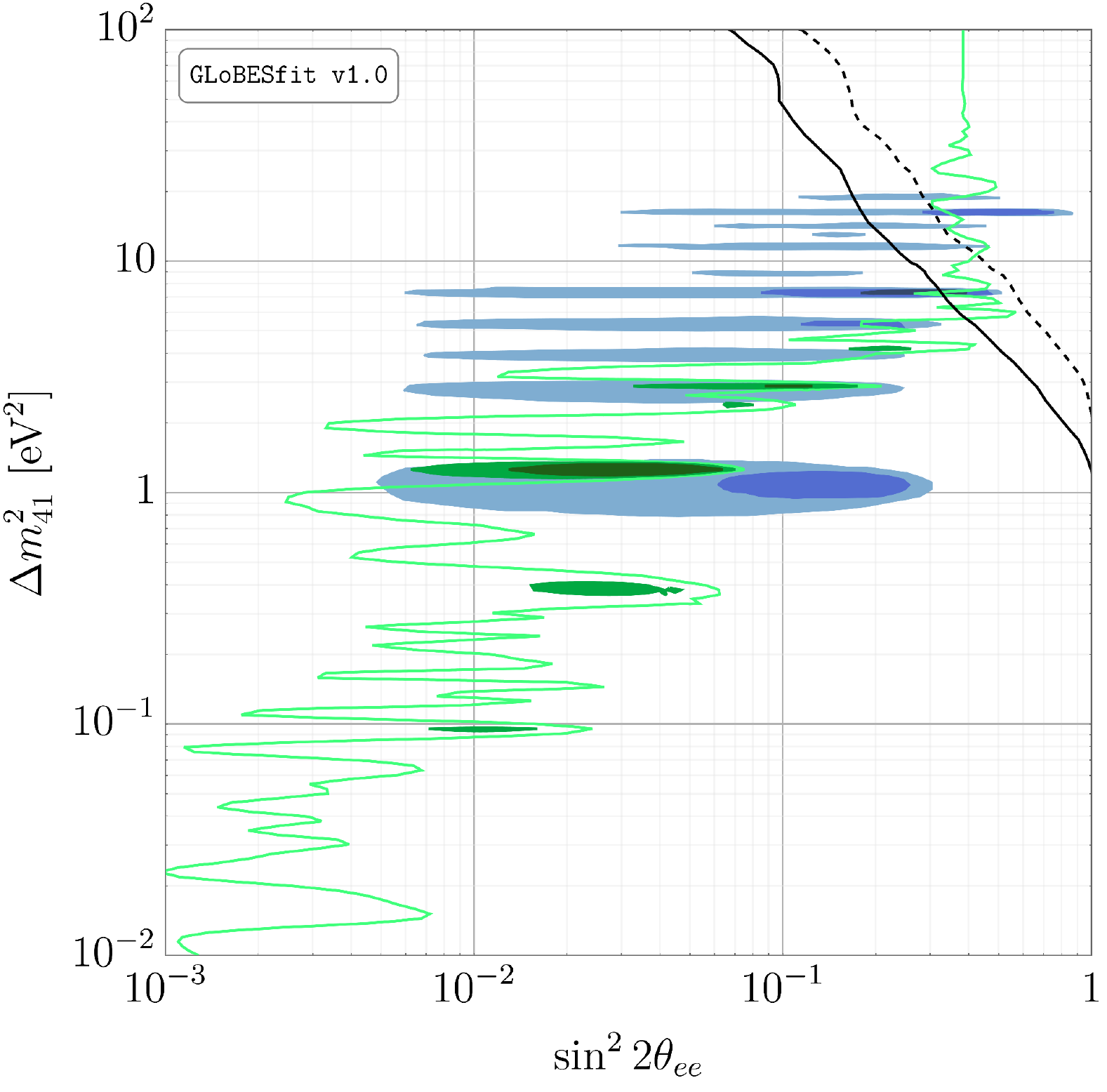}
\caption{A comparison between our spectral results, recent results from Neutrino-4 \cite{Neutrino4talk} and a global analysis of tritium-endpoint experiments \cite{Giunti:2019fcj}. We show our 95\%, 99\% and 99.9\% C.L. contours in dark green, green and light green; the 1$\sigma$, 2$\sigma$ and $3\sigma$ C.L. contours from Neutrino-4 in dark blue, blue and light blue; and the 90\% and 99\% C.L. contours for tritium experiments in solid and dashed black.}
\label{fig:Neutrino4Comp}
\end{figure}

The data currently available to the public are insufficient to allow us to attempt to replicate the collaboration's analysis. However, we can get a sense for how compatible Neutrino-4 is with the other experiments that we have considered by overlaying their respective allowed contours. We show precisely this in Fig.~\ref{fig:Neutrino4Comp}. The dark green, green and light green curves are, respectively the 95\%, 99\% and 99.9\% C.L. contours from our combined analysis of measured spectral ratios from Bugey-3, DANSS, Daya Bay, Double Chooz, NEOS and RENO. The dark blue, blue and light blue curves are the $1\sigma$, $2\sigma$ and $3\sigma$ allowed contours shown on slide 30 of Ref.~\cite{Neutrino4talk}; we caution that these results are preliminary, but proceed regardless. We also show the 90\% (solid black) and 99\% (dashed black) C.L. contours from the combined analysis of the Mainz, Troitsk and KATRIN tritium-endpoint experiments presented in Ref.~\cite{Giunti:2019fcj}, where a similar comparison has been performed on the basis of Neutrino-4 results from Ref.~\cite{Serebrov:2018vdw}.

As observed in Ref.~\cite{Giunti:2019fcj}, the best-fit point from Neutrino-4 ($\sin^2 2\theta_{ee}$ = 0.31, $\Delta m_{41}^2$ = 7.32 eV$^{2}$) is disfavored, to a moderate degree, by the tritium-endpoint experiments. Similarly, the best-fit point to Neutrino-4 data is inconsistent with the region preferred by reactor spectral ratios. Moreover, while not shown in Fig.~\ref{fig:Neutrino4Comp}, the Neutrino-4 best-fit point is also disfavored by the integrated rate measurements presented in Sec.~\ref{sec:SpecAnalysis} -- in particular, see Fig.~\ref{fig:RateCombo}. The degree of (in)compatibility has been tabulated in Table \ref{tab:Neutrino4Comp}. For each of our three rate analyses and for our spectral analysis, we show the difference in $\chi^2$ between our best-fit point and that of Neutrino-4 ($\Delta \chi^2_{\rm N4}$), as well as the corresponding $p$-value and the equivalent number of $\sigma$. 

\begin{table}
\begin{tabular}{|c||c|c|c|}\hline
Analysis & $\Delta \chi^2_{\rm N4}$ & $p$ & $n\sigma$ \\ \hline \hline
Rates + HM & 25.9 & $2.3 \times 10^{-6} $ & 4.7 \\ \hline
Rates + AI & 63.5 & $1.6 \times 10^{-14} $ & 7.7 \\ \hline
Rates + HKSS & 17.9 & $1.3 \times 10^{-4} $ & 3.8 \\ \hline
Spectra & 13.9 & $9.5 \times 10^{-4} $ & 3.3 \\ \hline
\end{tabular}
\caption{The (in)compatibility between the best-fit point from the Neutrino-4 analysis presented in Ref.~\cite{Neutrino4talk} and the analyses we've presented here.}
\label{tab:Neutrino4Comp}
\end{table}

Clearly, none of the four analyses are particularly tolerant of this result from Neutrino-4. Our spectral analysis disfavors the best-fit point at the $3.3\sigma$ level, but all three rate analyses disfavor this point at $\gtrsim4\sigma$. It is perhaps imprudent for us to focus solely on the best-fit point from Neutrino-4: significant portions of the $3\sigma$-preferred parameter space cannot be firmly refuted by the tritium-endpoint experiments or by our analyses. However, in the absence of enough information to perform our own analysis of Neutrino-4, it is difficult to precisely discuss the global compatibility of these results. Interestingly, the updated analysis of Ref.~\cite{Neutrino4talk} contains a $3\sigma$-preferred region around $\Delta m_{41}^2 \sim 1$ eV$^2$ that was not present in Ref.~\cite{Serebrov:2018vdw}. If confirmed in an official publication, then this would be broadly consistent with the best-fit region from our analysis of spectral ratios. It is our hope that an extensive data release will allow us to study Neutrino-4 in more depth in the future.

\section{Conclusions}
\label{sec:Conclusions}
\setcounter{equation}{0}

In this work, we have studied the evidence for the existence of sterile neutrinos contained in the global reactor antineutrino dataset using a new open-source, \globes-based toolkit that we call \globesfit. We have demonstrated that inferences from IBD rate measurements are confounded by the necessary dependence on a particular flux model. In particular, while the evidence obtained using the traditional HM fluxes is at the level of $2.5\sigma$, updated \emph{ab initio} fluxes imply $<1\sigma$ significance, while updated spectral-conversion fluxes yield $2.6\sigma$ significance. This discrepancy can only be resolved with continued improvements to predictions of reactor antineutrino fluxes, which in turn rely on the collection of improved data. Ratios of measured IBD spectral, on the other hand, seem to paint a more optimistic picture: taken together, the current dataset suggests that the evidence rises to the level of $3.2\sigma$. This is certainly tantalizing, but is hardly ironclad. Therefore, we are led to conclude that that issue of the existence of additional species of neutrinos cannot be resolved one way or another with current data. We have also addressed the 5 MeV bump, and have found that the presence of this feature does not dramatically alter inferences about sterile neutrinos.

In addition to presenting our results, this manuscript is also intended to document \globesfit. This software is available for download from either GitHub \cite{OurGitHub} or from our dedicated website \cite{OurWebsite}. The function of making \globesfit \, publicly available is twofold:
\begin{enumerate}
\item Publishing our code allows for members of the community to levy informed commentary on our techniques. Global fitting in an inherently tricky business; reproducing an experiment's results can be challenging for those not involved in the collaboration. Invariably, there are aspects of this program that can be improved. Our hope is that public input, over time, will reinforce this work.
\item Though we intend to revisit and expand on this work in the future, it is not possible for us to test every interesting physical hypothesis that could explain current neutrino anomalies. We have focused on the 3+1 scenario because of its simplicity, but Nature could certainly be far more interesting than this. If a user is interested in testing a particular model, then the tools we have developed ought to allow them to at least start that process. Because \globesfit \, is developed using pre-existing \globes \, architecture, it should lend itself to much a wider set of applications than the relatively simple scenario we have considered here.
\end{enumerate}
Our hope is that making these tools available contributes meaningfully to the wider neutrino physics community.

\subsubsection*{Regarding The Use of Statistics}

A common sin in global analyses is assuming that $\Delta \chi^2$ is chi-squared distributed. This issue has been discussed in detail in the literature; see, for instance, Refs.~\cite{Feldman:1997qc,Agostini:2019jup, Silaeva:2020yot}. Analyses of spectral ratios are particularly susceptible to this statistical idiosyncrasy, as is demonstrated nicely in Fig.~5 of Ref.~\cite{Agostini:2019jup}; we direct the interested reader to this work for more details. The primary issue is that while the expected ratio of spectra is flat in the absence of a sterile neutrino, statistical noise invariably results in a preference for oscillations with some nonzero amplitude. Failing to take this into account yields final results that are overconfident. In this work, we have been conservative in talking about the statistical significance of our findings, particularly those in Sec.~\ref{sec:SpecAnalysis}. While the significance we have presented here ostensibly rises to the $\sim3\sigma$ level, the true significance is certainly less than this. The actual preference for a sterile neutrino in the global reactor dataset, accounting for this effect, remains to be determined.

The difficulty is that the precise relationship between $\Delta \chi^2$ and a $p$-value depends strongly on the experiment(s) under consideration. How $\Delta \chi^2$ is distributed -- and thus the $p$-value associated with a particular measurement -- can be determined numerically as a function of $\sin^2 2\theta_{ee}$ and $\Delta m_{41}^2$ using Bayesian methods; this is the basis of the Feldman-Cousins technique \cite{Feldman:1997qc}. However, this requires the ability to reliably simulate the experiment(s) under consideration; the capacity to do so is limited for those not involved in a given collaboration. Even with perfect simulation capabilities, however, this sort of analysis can be prohibitively computationally demanding, particularly if considering multiple experiments simultaneously. We hope to address these concerns in future releases of \globesfit, but for the time being, we are left wanting for a more sophisticated statistical treatment.\footnote{As this work neared completion, Ref.~\cite{Giunti:2020uhv} appeared on the preprint arXiv. In this work, the author performs precisely the sort of analysis advocated above. They find that such a Feldman-Cousins correction diminishes the overall significance from $2.4\sigma$ to $1.8\sigma$. We note that the author has included data from PROSPECT \cite{Ashenfelter:2018iov}, as well as preliminary, updated data from DANSS \cite{Danilov:2019aef}, but not from Daya Bay, Double Chooz nor RENO; this explains the difference in significance relative to our findings here.}

While we believe that the results we have presented here are merited in their own right, our hope is that presenting the results as we have -- and acknowledging the ways in which the underlying treatment is deficient -- is sufficient to promote dialogue between experimentalists and theorists on how to best employ the available data. It is incumbent on theorists to ensure that experimental data is used correctly in their analyses; it is incumbent on experimentalists to follow good data preservation practices. Decisions for how to best commit the neutrino community's finite resources ultimately rely on understanding what is (and is not) contained in current data.

\textbf{Acknowledgments:} We thank Pedro Ochoa-Ricoux and Liang Zhan (Daya Bay); Giorgio Gratta (Palo Verde); Karsten Heeger, Bryce Littlejohn and Pranava Teja Surukuchi (PROSPECT); Soo-Bong Kim (RENO); and David Lhuillier (STEREO) for providing data and useful discussions. We further thank Muriel Fallot for providing the \emph{ab initio} fluxes in machine-readable format and Leendert Hayen for information on the HKSS model, as well as Carlo Giunti and Marco Laveder for discussions regarding Neutrino-4. JMB thanks the Fermilab Neutrino Physics Center for their hospitality during the completion of this work, as well as Andr\'{e} de Gouv\^{e}a, Peter Denton, Kevin Kelly and Yue Zhang for helpful conversations. We again thank Bryce Littlejohn for pointing out bugs in a previous version of this analysis. This work is supported by DOE Office of Science awards~DE-SC0018327 and~DE-SC0020262. The work of JMB is also supported by NSF Grant~PHY-1630782 and by Heising-Simons Foundation Grant~2017-228.



\appendix

\section{Sterile Neutrinos in \globes}
\label{app:OscEngines}
\setcounter{equation}{0}

In this appendix, we outline how oscillations involving sterile neutrinos are calculated using existing \globes \, machinery. We begin by reviewing the basics of oscillations with sterile neutrinos before turning to our implementation thereof. We stress that this machinery can be modified to implement any new-physics scenario that the user wishes to probe; our discussion of how to create custom oscillation engines and probability matrices is completely generalizable.

\subsection{Oscillations with a Sterile Neutrino}

The generic expression for the vacuum oscillation probability $P_{\alpha \beta}$ -- the probability for a neutrino produced in the flavor eigenstate $\nu_\alpha$ to be observed in the flavor eigenstate $\nu_\beta$ ($\alpha, \, \beta = e, \, \mu, \, \tau, \, s$) -- is given as follows:
\begin{align}
\label{eq:DefineProbs}
P_{\alpha \beta} & = \left| \delta_{\alpha \beta} \right. \\
& \left. - 4 \sum_{i=2}^4 \sum_{j=1}^{i-1} \Re \left[ U_{\alpha i}^* U_{\alpha j} U_{\beta i} U_{\beta j}^*\right] \sin^2 \left( \frac{\Delta m_{ij}^2 L}{4E_\nu} \right) \right. \nonumber \\
& \left.+ 2 \sum_{i=2}^4 \sum_{j=1}^{i-1} \Im \left[ U_{\alpha i}^* U_{\alpha j} U_{\beta i} U_{\beta j}^* \right] \sin \left( \frac{\Delta m_{ij}^2 L}{2E_\nu} \right) \right| \nonumber,
\end{align}
where $U_{\alpha i}$ ($\alpha = e, \, \mu, \, \tau, \, s$; $i = 1, \, 2, \, 3, \, 4$) are the elements of the leptonic mixing matrix, $L$ is the distance propagated (i.e., the baseline), $E_\nu$ is the neutrino energy and we define $\Delta m_{ij}^2 \equiv m_i^2 - m_j^2$, with $m_i$ the neutrino masses. For oscillations involving antineutrinos, we make the replacement $U \to U^*$; this is equivalent to changing the sign of the last term in Eq.~\eqref{eq:DefineProbs}.

There are only four relevant elements of the leptonic mixing matrix, for our purposes here. We parametrize them as follows:
\begin{align}
|U_{e1}|^2 & = \cos^2 \theta_{12} \cos^2 \theta_{13} \cos^2 \theta_{14}, \\
|U_{e2}|^2 & = \sin^2 \theta_{12} \cos^2 \theta_{13} \cos^2 \theta_{14}, \\
|U_{e3}|^2 & = \sin^2 \theta_{13} \cos^2 \theta_{14}, \\
|U_{e4}|^2 & = \sin^2 \theta_{14};
\end{align}
one can easily verify the unitarity constraint $\sum_{i=1}^4 |U_{ei}|^2 = 1$. In our analyses, we have assumed that the values of $\theta_{12}$ and $\theta_{13}$ determined from global fits to oscillation data assuming three-neutrino oscillations \cite{Esteban:2018azc} can be applied without alteration. Of course, what experiments actually measure is (some combinations of) the elements of the leptonic mixing matrix. Consequently, a determination of, say, $|U_{e2}|^2$ is absolute, whereas the inferred value of $\theta_{12}$ will depend on values of $\theta_{13}$ and $\theta_{14}$ inferred elsewhere. However, as long as $\theta_{14}$ is small -- which we find to be the case in our fits -- the implied modifications to the three-neutrino mixing angles are also small.

When (anti)neutrinos propagate through matter, the background potential from ambient protons, neutrons and electrons modifies their propagation. The effects of this matter potential can be included via modifications to the mixing angles and mass-squared differences. Moreover, \globes \, does contain the functionality to calculate neutrino propagation in a matter background. Given the low energies under consideration here, however, we assume that the vacuum-oscillation formalism is sufficient.

For short-baseline experiments -- experiments with baselines $\lesssim \mathcal{O}(100)$ m -- with point-like sources and a point-like detectors, we use the two-flavor approximation for the electron-type survival probability, $P_{ee}$:
\begin{equation}
P_{ee} = 1 - \sin^2 2\theta_{ee} \sin^2 \left( \frac{\Delta m_{41}^2 L}{4E_\nu}\right),
\end{equation}
where $\sin^2 2\theta_{ee} = 4 |U_{e4}|^2 (1-|U_{e4}|^2) = \sin^2 2\theta_{14}$ is the effective active-sterile mixing angle. We work in terms of $\sin^2 2\theta_{ee}$ even though this is equal to $\sin^2 2\theta_{14}$ in our parametrization; the effective angle $\sin^2 2\theta_{ee}$ is parameterization independent and thus more appropriate for comparison with the broader literature.

For medium-baseline experiments, this two-flavor formalism is insufficient --- we consider the full, four-flavor formalism. The electron-type survival probability is instead written as
\begin{equation}
\label{eq:DefinePee}
P_{ee} = 1 - 4 \sum_{i=2}^4 \sum_{j=1}^{i-1} |U_{ei}|^2 |U_{ej}|^2 \sin^2 \left( \frac{\Delta m_{ij}^2 L}{4E_\nu} \right);
\end{equation}
this can be immediately derived from Eq.~\eqref{eq:DefineProbs} by setting $\alpha = \beta = e$.

For the experiments we consider, we often cannot consider oscillations with one well-defined baseline: either (1) the experiment is sourced by multiple reactors, or (2) the physical extent of the core or detector (or both) is not much smaller than the distance between them. In either case, one must generalize the term $\sin^2 \left( \frac{\Delta m^2 L}{4E_\nu} \right)$ to a flux-weighted average over baselines. The appropriate quantity is 
\begin{equation}
\label{eq:define_F}
\sin^2 \left( \frac{\Delta m^2 L}{4E_\nu} \right) \to F(q) \equiv \dfrac{\int \frac{f(L) \sin^2 \left(q L\right)}{L^2} \, dL}{\int \frac{f(L)}{L^2} \, dL},
\end{equation}
where we define $q \equiv \frac{\Delta m^2}{4E_\nu}$ and introduce $f(L)$ as the distribution of baselines in the experiment. The denominator is the mean-inverse-squared baseline, which we denote $\langle L^{-2} \rangle$. We discuss our implementation of these integrals below.

\subsection{Implementation in \globesfit}

The survival probability $P_{ee}$ is calculated in \globes \, with custom oscillation engines using the function \texttt{glbDefineOscEngine}, which is called in the main program file. A sample call of this has the following form:

\texttt{glbDefineOscEngine(NUMP, }

\texttt{\&<exp>\_probability\_matrix,}

\texttt{\&glf\_get\_oscillation\_parameters,}

\texttt{\&glf\_set\_oscillation\_parameters,}

\texttt{``<engine\_name>'', user\_data);}

We explain the arguments of this function below.

\vspace{2mm}

\underline{\texttt{NUMP}}: The maximum number of oscillation parameters used. For four-neutrino oscillations, this number should be \texttt{12}; practically, we take it to be \texttt{6}.

\underline{\texttt{<exp>\_probability\_matrix}}: This is the function in which the oscillation probabilities are calculated for experiment \texttt{<exp>}. This function produces a three-by-three matrix \texttt{P} whose elements are the active-active oscillation probabilities (i.e., \texttt{P[0][0]} corresponds to $P_{ee}$, \texttt{P[0][1]} to $P_{e\mu}$, etc.). Currently, these functions only calculate $P_{ee}$; all other probabilities are fixed to \texttt{0}. We discuss the evaluation of the probabilities in the next subsection.

\underline{\texttt{glf\_get\_oscillation\_parameters}}: This function retrieves the current values of the oscillation parameters.

\underline{\texttt{glf\_set\_oscillation\_parameters}}: This function sets the values of the oscillation parameters.

\underline{\texttt{``<engine\_name>''}}: This is the internal name of the oscillation engine, which is called in the corresponding AEDL file via \texttt{\$oscillation\_engine=``<engine\_name>''}.

\underline{\texttt{user\_data}}: A void pointer to user-specified data that are relevant for the evaluation of the probability. We discuss the use of this apparatus below.

\subsection{Custom Probability Matrices}

For most of the experiments we consider, it is sufficient to take the core to be point-like but to give the detector finite extent. We usually make the approximation that the detector constitutes a (portion of a) spherical shell centered around the core. The function $f(L)$ is then flat over the extent of the detector, which we consider to be $L_a < L < L_b$; this leads to a relatively simple expression for 
$\langle L^{-2} \rangle$,
\begin{equation}
\langle L^{-2} \rangle = \frac{1}{L_a L_b},
\end{equation}
as well as the function $F(q)$ in Eq.~\eqref{eq:define_F}:
\begin{align}
\label{eq:ShellF}
F(q) & = \frac{1}{(L_b-L_a)} \times \left\{ L_b \sin^2 (q L_a) - L_a \sin^2 (q L_b) \right. \\ \nn
& \left. + L_a L_b q \times \left[ {\rm Si}(2qL_b) -  {\rm Si}(2qL_a) \right] \right\},
\end{align}
where ${\rm Si}(x) = \int_0^x \frac{\sin t}{t} \, dt$ is the sine integral.

For each experiment of this sort, the width of the detector is specified in \globesfit \, as the sole element of an array typically named \texttt{<exp>\_wide}. This is passed into the probability engine by setting ``\texttt{user\_data}" to be ``\texttt{(void *) <exp>\_wide}" in \texttt{glbDefineOscEngine}, above. The function in Eq.~\eqref{eq:ShellF} is evaluated, using this input, with a helper function called \texttt{lsin(q, La, Lb)}. Here, \texttt{La} and \texttt{Lb} are the nearest and further baseline at the experiment, determined from the width and the (average) baseline thereof. Note that all lengths in \globes \, are assumed to be in km. Moreover, since \globes \, also assumes all energies are in GeV, $q$ in this function is numerically given by
\begin{equation}
q = 1.267 \left(\frac{\Delta m^2}{\text{eV}^2}\right) \left( \frac{\text{GeV}}{E_\nu} \right).
\end{equation}

Occasionally, the experimental geometry results in a more complicated $f(L)$ whose structure we must take into account. This distribution is estimated using a simple Monte Carlo routine. Pairs of points are randomly drawn from the core and detector, whose geometry we must specify by hand. One can then numerically integrate over this distribution to determine $\langle L^{-2} \rangle$ and $F(q)$; this numerical integral is performed as a sum over these pairs.

To save computation time, $F(q)$ is precomputed on a grid and is fed into \globesfit \, as a look-up table, whose elements are of the form \texttt{\{ log\_10 q, F(q)\}}, that resides in auxiliary header files. Linear interpolation in \texttt{log\_10 q} is used to get intermediate values. If $q$ is less than some minimum specified value $q_{\rm min}$, then it is assumed that the integral scales quadratically and $F(q)$ is determined by extrapolation. If $q$ is instead greater than some maximum specified value $q_{\rm max}$, then it is assumed that $F(q)$ has averaged out to 0.5. The values of $q_{\rm min}$ and $q_{\rm max}$ depend on the experiment under consideration. We adopt this method for Daya Bay and RENO for both the rate and spectral analyses; moreover, we do this for Bugey-3,\footnote{We do not use this technique for Bugey-3 or Bugey-4 for the rate-only analysis because we expect the antineutrino spectrum to be more sensitive to the specific experimental geometry than the total event rate.} DANSS, Double Chooz and NEOS for our spectral analysis.

The precomputed $F(q)$ is fed into the custom oscillation engine via the ``\texttt{user\_data}'' option. This is handled by a custom data structure called ``\texttt{glf\_distance\_data}'', whose elements are the length of the precomputed array and a pointer to the array itself. The structure for experiment \texttt{<exp>} is named ``\texttt{<exp>\_s}''; this is handled by by setting ``\texttt{user\_data}" to be ``\texttt{(void *) \&<exp>\_s}'' in the definition of the oscillation engine.

For medium-baseline experiments, each detector and reactor can be considered point-like, but these experiments are sourced by multiple reactors. The reactors generically have different average powers over the duration of the experiment; these weight their contributions. The integrals over baselines in $\langle L^{-2} \rangle$ and $F(q)$ are replaced by sums:
\begin{align}
\langle L^{-2} \rangle & = \left(\sum_r^{N_r} \frac{P_r}{L_r^2} \right) / \left( \sum_r^{N_r} P_r \right), \\
F(q) & = \left( \sum_r^{N_r} \frac{P_r \sin^2 (q L_r)}{L_r^2} \right) / \left( \sum_r^{N_r} \frac{P_r}{L_r^2} \right), \label{eq:defineF2}
\end{align}
where $r$ indexes the reactors, of which there are $N_r$, and the average power of each is $P_r$.

For rate-only analyses, the multiple reactors at Chooz, Double Chooz and Palo Verde are handled using different AEDL experiment definitions; the resulting event rates are added at the analysis level. Meanwhile, for rate-only analyses of Daya Bay and RENO, as well as the spectral analyses at Daya Bay, Double Chooz and RENO, the contributions of multiple reactors are combined at the level of the AEDL file. This is purely a function of the development history of \globesfit \, and does not reflect differing physics in these cases.


\section{Structure of the Code}
\label{app:TheCode}
\setcounter{equation}{0}

In this appendix, we document the component files of \globesfit \, and briefly describe the function and contents of each.

\subsection{Common Files}

We begin with files that are common to both the rate and spectral modules.

\vspace{2mm}

\underline{\texttt{Pure\_<isotope>.dat}}: The antineutrino fluxes for isotope \texttt{<isotope> = U235, U238, Pu239, Pu241}, written in the standard \globes \, format. These fluxes have been calculated from the Huber-Mueller (HM) predictions, linearly interpolated/extrapolated on a logarithmic scale (i.e., the logarithm of the flux has been linearly interpolated/extrapolated).

\underline{\texttt{Pure\_<isotope>\_SM.dat}}: Antineutrino fluxes similar to \texttt{Pure\_<isotope>.dat}, except that the HM fluxes have been replaced in favor of the \emph{ab initio} fluxes. (Note that ``\texttt{SM}'' in the file name stands for ``summation method,'' which is an alternate name for the \emph{ab initio} method.)

\underline{\texttt{Pure\_<isotope>\_HKSS.dat}}: Antineutrino fluxes similar to \texttt{Pure\_<isotope>.dat}, except that the HM fluxes have been replaced in favor of the HKSS fluxes.

\underline{\texttt{IBDnew.dat}}: The IBD cross section calculated from the $\mathcal{O}(1/M)$ results of Ref.~\cite{Vogel:1999zy} (i.e., Eqs.~(14) and (15) of that reference). This is the cross section calculation used in the analyses presented here.

\underline{\texttt{glf\_precomputed\_probabilities.h}}: An auxiliary file that contains the look-up tables used to compute $F(q)$ for Daya Bay, RENO, DANSS, Double Chooz, NEOS and Bugey. This is called by both \texttt{glf\_rate.c} and \texttt{glf\_spectrum.c} (see below).

\underline{\texttt{glf\_probability.c}}: A file that contains helper functions and probability engines (see Appendix~\ref{app:OscEngines}) for both the rate and spectral analysis. This file contains generic oscillation engines for point-like reactor cores and detectors of finite extent for both two-flavor (\texttt{glf\_probability\_matrix}) and four-flavor (\texttt{glf\_standard\_probability\_matrix}) oscillations. Moreover, for experiments for which oscillation probabilities have been precomputed, there again exist functions for both two-flavor (\texttt{glf\_two\_state\_probability\_matrix}) and four-flavor (\texttt{glf\_four\_state\_probability\_matrix}) oscillations.

\underline{\texttt{glf\_probability.h}}: The header file associated with \texttt{glf\_probability.c}. This is called by both \texttt{glf\_rate.c} and \texttt{glf\_spectrum.c}.

\underline{\texttt{glf\_type.h}}: The header file wherein the structure ``\texttt{glf\_distance\_data}'' is defined. This is called by several files.

\underline{\texttt{Makefile}:} The file used to make the executables \texttt{$.$/glf\_rate} and \texttt{$.$/glf\_spectrum} (see below). The user may need to modify this file to link to their local \globes \, libraries -- specifically, the ``\texttt{prefix}'' option may need to be set to the location of the user's \globes \, directory.

\subsection{Rate Files}

We next consider files specific to the rate module.

\vspace{2mm}

\underline{\texttt{rate\_combo\{2,3\}.glb}}: The files containing the AEDL definitions of the experiments we will consider, described in Sec.~\ref{sec:RateExps}. These files employ the HM flux model. The order of the experiments is important in the evaluation of the chi-squared; we enumerate the experiments contained in each file below.
\begin{spacing}{0.75}
\begin{itemize}
\item Short-baseline experiments are contained in \texttt{rate\_combo.glb}:
\begin{enumerate}
\setcounter{enumi}{-1}
\item Bugey-4 \& Bugey-3, 15 m
\item Rovno 91
\item Bugey-3, 40 m
\item Bugey-3, 95 m
\item G\"osgen, 38 m
\item G\"osgen, 46 m
\item G\"osgen, 65 m
\item ILL
\item Krasnoyarsk 87, 33 m
\item Krasnoyarsk 87, 92 m
\item Krasnoyarsk 94
\item Krasnoyarsk 99
\item Savannah River, 18 m
\item Savannah River, 24 m
\item Rovno 88, All 18 m
\item Rovno 88, 25 m
\item Nucifer
\end{enumerate}

\item Medium-baseline, total-rate experiments are contained in \texttt{rate\_combo2.glb}:
\begin{enumerate}
\setcounter{enumi}{16}
\item Palo Verde, 750 m
\item Palo Verde, 890 m
\item Double Chooz, 355 m
\item Double Chooz, 469 m
\item Chooz, 998 m
\item Chooz, 1115 m
\end{enumerate}

\item Medium-baseline, fuel-evolution experiments are contained in \texttt{rate\_combo3.glb}:
\begin{enumerate}
\setcounter{enumi}{22}
\item Daya Bay, EH1 AD1
\item Daya Bay, EH1 AD2
\item Daya Bay, EH2 AD3
\item Daya Bay, EH2 AD8
\item RENO, Near Detector
\end{enumerate}
\end{itemize}
\end{spacing}
These numberings correspond to the internal \globes \, ordering. Note that some AEDL files apply to more than one experiment. The ordering of these experiments is a function of the development history of \globesfit \, and is in no way a commentary of any of these experiments. 

We define a \globes \, rule for each of the four main, fissile isotopes for each experiment. This allows the user to separate out contribution of each isotope in whatever calculation they are interested in. Moreover, it cuts down on the number of flux files that one must keep on hand; instead of defining a separate flux file for each experiment, one need only combine the fluxes from each isotope with the appropriate fuel fraction. Moreover, the theoretical uncertainties on the isotopic IBD yields for $^{235}$U, $^{238}$U, $^{239}$Pu and $^{241}$Pu are declared in the AEDL definition for Bugey-4/Bugey-3 (15 m) in the field \texttt{@sys\_on\_errors}. We have also made clones of this file that modify how these systematics are treated; see below.

Because the absolute numbers of events are not relevant for our analyses, \texttt{@time}, \texttt{@power}, \texttt{@norm} and \texttt{\$target\_mass} are all generally set to 1.0 in the AEDL definitions; we comment below on instances in which we deviate from this. The fluxes from each of the four main fissile isotopes for each experiment are considered separately and reweighted by the effective fuel fractions given by each experiment. Consequently, \texttt{@time}, \texttt{@power} and \texttt{@norm} are set to be the same for each component of the flux. The value of \texttt{\$length\_tab} is set to $1/\sqrt{\langle L^{-2} \rangle}$ for each experiment;  

Exceptions to this occur for medium-baseline experiments -- namely, Palo Verde, Double Chooz, Chooz, Daya Bay and RENO -- which are typically sourced by multiple reactors, or are comprised of multiple detectors, whose relative contributions must be taken account. \begin{itemize}
\item For Palo Verde and Chooz, \texttt{@norm} is set equal to the numerical value of the exposure, in GW$\cdot$yr, for each detector, which we have calculated in Sec.~\ref{sec:RateExps}.
\item We do something similar for Daya Bay. We set \texttt{@power} equal to the total power delivered to each detector, namely
\begin{equation}
\sum_s \sum_r t_d^s P^s_r,
\end{equation}
where $r$ indexes the eight reactor cores and $s$ indexes the periods of operation; the Daya Bay rate-evolution analysis is based on the 1230-day data set from Ref.~\cite{An:2017osx}, so we only consider the 6AD and 8 AD periods.
We set \texttt{@norm} to be equal to each detectors total efficiency, and \texttt{\$target\_mass} to be equal to each detector's mass in tons.
\item Since the relative contributions of the reactors used in Double Chooz near detector have not been specified, we assume them to be equal. This allows us to set \texttt{@norm} to 1.0 for each.
\end{itemize}

\underline{\texttt{rate\_combo\_no\_sys.glb}}: Similar to \texttt{rate\_combo.glb}, except no systematic uncertainties are associated with the HM flux predictions. In this mode, \globesfit \, recycles the AEDL definitions in \texttt{rate\_combo2.glb} and \texttt{rate\_combo3.glb}.

\underline{\texttt{rate\_combo\_unfix.glb}}: Similar to \texttt{rate\_combo.glb}, except systematic uncertainties are only applied to $^{238}$U and $^{241}$Pu. The IBD yields from $^{235}$U and $^{239}$Pu are left unfixed to allow them to be scanned in our analysis. Note that in this mode, \globesfit \, recycles the AEDL definitions in \texttt{rate\_combo2.glb} and \texttt{rate\_combo3.glb}.

\underline{\texttt{rate\_combo\_SM\{2,3\}.glb}}: Similar to \texttt{rate\_combo\{2,3\}.glb}, except that the SM fluxes are used in lieu of the HM fluxes.

\underline{\texttt{rate\_combo\_HKSS\{2,3\}.glb}}: Similar to \texttt{rate\_combo\{2,3\}.glb}, except that the HKSS fluxes are used in lieu of the HM fluxes.

\underline{\texttt{glf\_rate\_aux.h}}: Experimental inputs for our rate analyses. These include the fuel fractions and experimental IBD rates at each experiment, as well as the (inverses of the) covariance matrices that are used in the determination(s) of the chi-squared function(s). This is called in \texttt{glf\_rate\_chi.c}. 

\underline{\texttt{glf\_rate\_chi.c}}: The file containing the chi-squared functions to be used with the \globes \, files described above. There are five such functions -- \texttt{glf\_rate\_chi\_nosys}, \texttt{glf\_rate\_chi}, \texttt{glf\_rate\_chi\_SM}, \texttt{glf\_rate\_chi\_HKSS} and \texttt{glf\_rate\_chi\_unfix} -- each to be used in conjunction with a different set of \globes \, files, depending on how the antineutrino flux prediction is to be handled. The default is \texttt{glf\_rate\_chi\_nosys}; we discuss how to change this below.

\underline{\texttt{glf\_rate\_chi.h}}: The header file associated with \texttt{glf\_rate\_chi.c}. This is called in \texttt{glf\_rate.c}.

\underline{\texttt{glf\_rate.c}}: The file that gets run by the executable \texttt{$.$/glf\_rate}. The contents are as follows:
\begin{enumerate}
\item Code for parsing command-line options. Options are discussed below.
\item Oscillation engine definitions for each experiment.
\item Initialization of the chi-squared functions. The chi-squared function is fed the list of experiments to include in its evaluation though the integer array \texttt{YesNo}.
\item The calculation of the chi-squared. Here, \globesfit \, calculates the chi-squared over ranges of some parameters (described in the main text) and writes the output per the user's specification. We simulate the rate at each experiment both with (using the \globes \, function \texttt{glbGetSignalFitRatePtr}) and without (\texttt{glbGetRuleRatePtr}) oscillations involving sterile neutrinos and take the ratio of these for each such flux model.
\end{enumerate}

\subsection{Spectrum Files}

We conclude with files specific to the spectrum module.

\vspace{2mm}

\underline{\texttt{spectra\{2-6\}.glb}}: The files containing the AEDL definitions. The implementations of these experiments are described in Sec.~\ref{sec:SpectrumExp}; we provide here the internal ordering of experiments, decomposed into host \texttt{.glb} file:
\begin{spacing}{0.75}
\begin{itemize}
\item \texttt{spectra.glb}
\begin{enumerate}
\setcounter{enumi}{-1}
\item DANSS, Upper Position
\item DANSS, Lower Position
\end{enumerate}
\item \texttt{spectra2.glb}
\begin{enumerate}
\setcounter{enumi}{1}
\item Daya Bay, EH1 AD1
\item Daya Bay, EH1 AD2
\item Daya Bay, EH2 AD3
\item Daya Bay, EH2 AD8
\end{enumerate}
\item \texttt{spectra3.glb}
\begin{enumerate}
\setcounter{enumi}{5}
\item Daya Bay, EH3 AD4
\item Daya Bay, EH3 AD5
\item Daya Bay, EH3 AD6
\item Daya Bay, EH3 AD7
\item NEOS
\end{enumerate}
\item \texttt{spectra4.glb}
\begin{enumerate}
\setcounter{enumi}{10}
\item Double Chooz, Near Detector
\item Double Chooz, Far Detector
\end{enumerate}
\item \texttt{spectra5.glb}
\begin{enumerate}
\setcounter{enumi}{12}
\item Bugey-3, 15 m
\item Bugey-3, 40 m
\end{enumerate}
\item \texttt{spectra6.glb}
\begin{enumerate}
\setcounter{enumi}{14}
\item RENO, Near Detector
\item RENO, Far Detector
\end{enumerate}
\end{itemize}
\end{spacing}
The indices used here correspond to the internal \globes \, indices for each experiment. As with rate experiments, we define separate \globes \, rules for each of the four main fissile isotopes.

\underline{\texttt{glf\_spectrum\_aux\_1.h}}: Covariance matrices used in our spectral analyses. This is called in \texttt{glf\_spectrum\_chi.c}. 

\underline{\texttt{glf\_spectrum\_aux\_2.h}}: Experimental data for our spectral analyses, including the relevant measured spectral ratios and the stated fuel fractions. This is called in \texttt{glf\_spectrum\_chi.c}. 

\underline{\texttt{glf\_spectrum\_chi.c}}: The file containing the definition of the chi-squared function(s) detailed in Sec.~\ref{sec:SpectrumExp}.

\underline{\texttt{glf\_spectrum\_chi.h}}: The header file corresponding to \texttt{glf\_spectrum\_chi.c}. This is called by \texttt{glf\_spectrum.c}.

\underline{\texttt{glf\_spectrum.c}}: The main file that gets run by the executable \texttt{$.$/glf\_spectrum}. The structure of this file is similar to \texttt{main\_rate.c}, above.


\subsection{Executables and Options}

\underline{\texttt{$.$/glf\_rate}}: The main executable for the rate-only aspect of \globesfit.

\underline{\texttt{$.$/glf\_spectrum}}: The main executable for the spectrum aspect of \globesfit.

\vspace{2mm}

We summarize particularly useful command-line options common to either executable.
  \begin{itemize}
\item \underline{\texttt{-bN}}: Allows the user to turn off a particular block of experiments corresponding to the index \texttt{N}. (We call these ``blocks'' because ``experiment'' can be ambiguous in the parlance of \texttt{GLoBES}. Removing Daya Bay from the calculation, for instance, requires that we ignore several experiments in the AEDL file.) We comment on these blocks below.
\item \underline{\texttt{-rN}}: Sets the number of points used in each direction in the parameter scan to be \texttt{N}. By default, \texttt{N} = 51.
\item \underline{\texttt{-xM,N}}: Sets the range in direction \texttt{x}, corresponding to $\log_{10} \sin^2 2\theta_{ee}$ to be [\texttt{M, N}]. The default is $\texttt{M}=-4$, $\texttt{N}=0$.
\item \underline{\texttt{-yM,N}}: Sets the range in direction \texttt{y}, corresponding to $\log_{10} \left[ \frac{\Delta m_{41}^2}{\text{eV}^2} \right]$ to be [\texttt{M, N}]. The default is $\texttt{M}=-2$, $\texttt{N}=2$.
\item \underline{\texttt{-o [file\_name]} or \texttt{\ldots \, > [file\_name]}:} Sets the output file of the parameter scan to be \texttt{[file\_name]}. If left blank, then the output is set to \texttt{stdout}, i.e., it writes to screen.
\end{itemize}
Additionally, there are some options that are specific to the rate-only analysis.
\begin{itemize}
\item \underline{\texttt{-S}}: Turns on systematics stemming from the HM flux predictions. This changes the chi-squared function from \texttt{glf\_rate\_chi\_nosys} to \texttt{glf\_rate\_chi}.
\item \underline{\texttt{-M}}: Replaces the HM fluxes with the SM fluxes, using the systematic uncertainties of the former. This changes the chi-squared function from \texttt{glf\_rate\_chi\_nosys} to \texttt{glf\_rate\_chi\_SM}.
\item \underline{\texttt{-H}}: Replaces the HM fluxes with the HKSS fluxes, including their own uncertainties. This changes the chi-squared function from \texttt{glf\_rate\_chi\_nosys} to \texttt{glf\_rate\_chi\_HKSS}.
\item \underline{\texttt{-u}}: Turns on systematics stemming from $^{238}$U and $^{241}$Pu and scans over the normalizations of the $^{235}$U and $^{239}$Pu fluxes \emph{instead of} considering a sterile neutrino. This changes the chi-squared function from \texttt{glf\_rate\_chi\_nosys} to \texttt{glf\_rate\_chi\_unfix}.
\end{itemize}
Note that these options are mutually exclusive and have been listed in order of priority -- for instance, entering ``\texttt{-M -S -H}'' at the command line will be treated as if only ``\texttt{-S}'' has been entered. There is one further rate-specific option:
\begin{itemize}
\item \underline{\texttt{-w}}: In addition to the values of $\sin^2 2\theta_{ee}$, $\Delta m_{41}^2$ and $\chi^2$, this option toggles writing the best-fit values of the four nuisance parameters $\xi_{235}$, $\xi_{238}$, $\xi_{239}$ and $\xi_{241}$ to disk, in that order. This only works in conjunction with \texttt{-S}, \texttt{-M} and \texttt{-H}.
\end{itemize}

Finally, we list some options that are specific to the spectral-ratio analysis.
\begin{itemize}
\item \underline{\texttt{-T}}: In addition to scanning over $\sin^2 2\theta_{ee}$ and $\Delta m_{41}^2$, this option triggers an additional scan over $\sin^2 2\theta_{13}$ on the range $[0.0, 0.2]$.
\item \underline{\texttt{-C}}: This option eschews the scan over the sterile neutrino parameter space altogether and instead scans over $\sin^2 2\theta_{13}$ on the range $[0.05, 0.2]$ and $\Delta m_{31}^2$ on the range $[1.0, 4.0] \times 10^{-3}$ eV$^2$. (This option has been used to perform the calibration scan whose results are shown in Fig.~\ref{fig:Calibration}.)
\end{itemize}

\subsubsection*{Experimental Blocks}

Currently, \texttt{glf\_rate}\, is not set up to isolate every individual experiment. However, given that the covariance matrix decomposes into blocks, this allows us to trivially remove blocks of experiments from the analysis. Most of these blocks consist of only one experiment, while the rest include multiple experiments. Here, we tabulate which experiments correspond to which of these blocks.
\begin{spacing}{0.75}
\begin{enumerate}
\setcounter{enumi}{-1}
\item Bugey-4 + Rovno 91 \vspace{-2mm}
\item Bugey-3 (15 m+ 40 m + 95 m) \vspace{-2mm}
\item G\"osgen (38 m + 46 m + 65 m) + ILL \vspace{-2mm}
\item Krasnoyarsk 87 (33 m + 92 m) \vspace{-2mm}
\item Krasnoyarsk 94 \vspace{-2mm}
\item Krasnoyarsk 99 \vspace{-2mm}
\item Savannah River (18 m) \vspace{-2mm}
\item Savannah River (24 m) \vspace{-2mm}
\item Rovno 88 (1I + 2I + 1S + 2S + 3S) \vspace{-2mm}
\item Nucifer \vspace{-2mm}
\item Palo Verde (750 m + 890 m) \vspace{-2mm}
\item Double Chooz (355 m + 469 m) \vspace{-2mm}
\item Chooz (998 m + 1115 m) \vspace{-2mm}
\item Daya Bay (EH1 + EH2) \vspace{-2mm}
\item RENO \vspace{-2mm}
\end{enumerate}
\end{spacing}
The number associated to each block is the appropriate choice of \texttt{N} to be specified at the command line. As an example, if one wanted to remove Nucifer and RENO from the analysis, one would include ``\texttt{-b9 -b14}" at the command line.

\texttt{glf\_spectrum} uses similar options for toggling on/off experimental blocks. However, since we treat the six above-mentioned experiments as uncorrelated, each experiment resides in its own block, i.e., there is a one-to-one mapping of experiments to blocks.


\section{Supplementary Data}
\label{app:SuppData}
\setcounter{equation}{0}


\subsection{Data Tables}

\begin{table*}[t]
\centering\begin{tabular}{|c||c|c|c|c|c|c|}\hline
 & D1 & D2 & L1 & L2 & L3 & L4 \\ \hline \hline
EH1 AD1 & 362.38 & 371.76 & 903.47 & 817.16 & 1353.62 & 1265.32\\ \hline
EH1 AD2 & 357.94 & 368.41 & 903.35 & 816.90 & 1354.23 & 1265.89\\ \hline
EH2 AD3 & 1332.48 & 1358.15 & 467.57 & 489.58 & 557.58 & 499.21\\ \hline
EH2 AD8 & 1337.43 & 1362.88 & 472.97 & 495.35 & 558.71 & 501.07\\ \hline
EH3 AD4 & 1919.63 & 1894.34 & 1533.18 & 1533.63 & 1551.38 & 1524.94\\ \hline
EH3 AD5 & 1917.52 & 1891.98 & 1534.92 & 1535.03 & 1554.77 & 1528.05\\ \hline
EH3 AD6 & 1925.26 & 1899.86 & 1538.93 & 1539.47 & 1556.34 & 1530.08\\ \hline
EH3 AD7 & 1923.15 & 1897.51 & 1540.67 & 1540.87 & 1559.72 & 1533.18\\ \hline
\end{tabular}
\caption{The distances between each reactor and detector at Daya Bay~\cite{An:2016ses}. The columns are labeled by the reactor; rows are labeled by the detector. All distances are given in meters.}
\label{tab:DBlengths}
\end{table*}

\begin{table*}
\begin{tabular}{|c||c|c|c|c|c|c|}\hline
 & D1 & D2 & L1 & L2 & L3 & L4 \\ \hline \hline
$P^{\rm 6AD}_r$ [GW$_{\rm th}$] & 2.082 & 2.874 & 2.516 & 2.554 & 2.825 & 1.976 \\ \hline
$P^{\rm 8AD}_r$ [GW$_{\rm th}$] & 2.514 & 2.447 & 2.566 & 2.519 & 2.519 & 2.550 \\ \hline
\end{tabular}
\caption{The average thermal power of each reactor at Daya Bay in GW$_{\rm th}$ for the 6AD and 8AD periods~\cite{An:2016ses}.}
\label{tab:DBpowers}
\end{table*}

Tables \ref{tab:DBlengths}-\ref{tab:DCinfo} collect some supplementary data that are too cumbersome to have included in the main text of this manuscript.

\begin{table*}
\centering\begin{tabular}{|c||c|c|c|c|c|c|c|c|}\hline
 & EH1 AD1 & EH1 AD2 & EH2 AD3 & EH2 AD8 & EH3 AD4 & EH3 AD5 & EH3 AD6 & EH3 AD7 \\ \hline \hline
Target Mass [$10^3$ kg]& 19.941 & 19.967 & 19.891 & 19.944 & 19.917 & 19.989 & 19.892 & 19.931 \\ \hline \hline
$\varepsilon_{\rm tot}$ & 0.8044 & 0.8013 & 0.8365 & 0.8363 & 0.9587 & 0.9585 & 0.9581 & 0.9588 \\ \hline \hline
$L_d$ [m] & 568.2 & 563.4 & 591.3 & 595.1 & 1633.3 & 1634.6 & 1638.7 &  1639.7 \\ \hline \hline
$f_{235}$ & 0.564 & 0.564 & 0.557 & 0.552 & 0.559 & 0.559 & 0.559 & 0.552 \\ \hline
$f_{238}$ & 0.076 & 0.076 & 0.076 & 0.076 & 0.076 & 0.076 & 0.076 & 0.076  \\ \hline
$f_{239}$ & 0.303 & 0.302 & 0.312 & 0.315 & 0.310 & 0.310 & 0.310 & 0.315 \\ \hline
$f_{241}$ & 0.056 & 0.056 & 0.055 & 0.057 & 0.055 & 0.055 & 0.055 & 0.057 \\ \hline
\end{tabular}
\caption{Relevant experimental data for Daya Bay. The first row shows the total fiducial target mass of the detector in metric tons \cite{An:2016ses}. The second row shows the total efficiency $\varepsilon_{\rm tot}$ for each detector \cite{An:2016srz}. The third row shows the effective baseline of each detector in meters, calculated using the distances in Table \ref{tab:DBlengths} and the powers in Table \ref{tab:DBpowers}. The last four rows are the average effective fuel fraction at each detector at Daya Bay, assumed to be the same for the 6AD and 8AD period \cite{An:2016srz}.}
\label{tab:DBfracs}
\end{table*}

\begin{table*}
\centering\begin{tabular}{|c||c|c|c|c|c|}\hline
Bin & $f_{235}$ & $f_{238}$ & $f_{239}$ & $f_{241}$ & $\sigma$ [cm$^2$/fission] \\ \hline \hline
1 & 0.5113 & 0.0767 & 0.3445 & 0.0675 & $\left(5.745 \pm 1.53\% \right) \times 10^{-43}$ \\ \hline
2 & 0.5279 & 0.0766 & 0.3326 & 0.0629 & $\left(5.762 \pm 1.52\% \right) \times 10^{-43}$ \\ \hline
3 & 0.5418 & 0.0764 & 0.3219 & 0.0599 & $\left(5.778 \pm 1.53\% \right) \times 10^{-43}$ \\ \hline
4 & 0.5553 & 0.0762 & 0.3113 & 0.0572 & $\left(5.795 \pm 1.53\% \right) \times 10^{-43}$ \\ \hline
5 & 0.5699 & 0.0760 & 0.2992 & 0.0549 & $\left(5.832 \pm 1.52\% \right) \times 10^{-43}$ \\ \hline
6 & 0.5849 & 0.0758 & 0.2878 & 0.0515 & $\left(5.837 \pm 1.52\% \right) \times 10^{-43}$ \\ \hline
7 & 0.6033 & 0.0757 & 0.2744 & 0.0466 & $\left(5.862 \pm 1.52\% \right) \times 10^{-43}$ \\ \hline
8 & 0.6304 & 0.0754 & 0.2525 & 0.0417 & $\left(5.930 \pm 1.54\% \right) \times 10^{-43}$ \\ \hline
\end{tabular}
\caption{The fuel-fraction bins for the Daya Bay data, and the corresponding measured IBD rates~\cite{An:2017osx}. These data have been corrected for three-flavor oscillations; seen text for details.}
\label{tab:DBfracs2}
\end{table*}

\begin{table*}
\begin{tabular}{|c||c|c|c|c|c|c|c|c|}\hline
Fuel Bin, $f_{235}$ & 0.5113 & 0.5279 & 0.5418 & 0.5553 & 0.5699 & 0.5849 & 0.6033 & 0.6304 \\ \hline \hline
HM & 0.9430 & 0.9411 & 0.9398 & 0.9387 & 0.9404 & 0.9372 & 0.9361 & 0.9391 \\ \hline
\emph{Ab Initio} & 0.9795 & 0.9786 & 0.9781 & 0.9778 & 0.9804 & 0.9779 & 0.9780 & 0.9827 \\ \hline
HKSS & 0.9342 & 0.9323 & 0.9310 & 0.9300 & 0.9316 & 0.9284 & 0.9273 & 0.930 \\ \hline
\end{tabular}
\caption{Measurement-to-expectation ratios calculated for the Daya Bay data ~\cite{An:2017osx}.}
\label{tab:DB_info}
\end{table*}

\begin{table*}
\centering\begin{tabular}{|c||c|c|c|c|c|c|}\hline
 & Reactor 1 & Reactor 2 & Reactor 3 & Reactor 4 & Reactor 5 & Reactor 6 \\ \hline \hline
Near Detector & 660.064 & 444.727 & 301.559 & 339.262 & 519.969 & 746.155 \\ \hline
Far Detector & 1563.771 & 1460.826 & 1397.813 & 1380.062 & 1409.389 & 1483.001  \\ \hline
\end{tabular}
\caption{The baselines between each reactor and either detector at RENO, in m~\cite{SBKim}.}
\label{tab:RENObaselines}
\end{table*}

\begin{table*}
\begin{tabular}{|c||c|c|c|c|c|c|}\hline
 & Reactor 1 & Reactor 2 & Reactor 3 & Reactor 4 & Reactor 5 & Reactor 6 \\ \hline \hline
Near Detector & 2.381 & 2.084 & 2.224 & 2.060 & 2.315 & 2.248 \\ \hline
Far Detector & 2.363 & 2.082 & 2.144 & 2.111 & 2.387 & 2.319  \\ \hline
\end{tabular}
\caption{The average thermal power of each reactor during the operating period of either RENO detector, in GW~\cite{SBKim}.}
\label{tab:RENOpowers}
\end{table*}

\begin{table*}
\begin{tabular}{|c||c|c|c|c|}\hline
 & Operating Time [Days] & $1/\sqrt{\langle L^{-2} \rangle}$ [km] & Total Average Power, $\langle P \rangle$ [GW$_{\rm th}$]& Efficiency \\ \hline \hline
Near Detector & 1807.88 & 0.4331 & 13.312 & 60.3\% \\ \hline
Far Detector & 2193.04 & 1.4469 & 13.406 & 68.7\% \\ \hline
\end{tabular}
\caption{Additional experimental information relevant for our analyses of RENO \cite{RENO:2018pwo,SBKim}.}
\label{tab:RENOinfo}
\end{table*}

\begin{table*}
\centering\begin{tabular}{|c||c|c|c|c|c|}\hline
Bin & $f_{235}$ & $f_{238}$ & $f_{239}$ & $f_{241}$ & $\sigma$ [cm$^2$/fission]  \\ \hline \hline
1 & 0.527 & 0.074 & 0.333 & 0.066 & $\left(5.681 \pm 2.13\% \right) \times 10^{-43}$ \\ \hline 
2 & 0.546 & 0.073 & 0.320 & 0.061 & $\left(5.740 \pm 2.13\% \right) \times 10^{-43}$ \\ \hline 
3 & 0.557 & 0.075 & 0.310 & 0.058 & $\left(5.746 \pm 2.13\% \right) \times 10^{-43}$ \\ \hline 
4 & 0.568 & 0.074 & 0.302 & 0.056 & $\left(5.739 \pm 2.13\% \right) \times 10^{-43}$ \\ \hline 
5 & 0.579 & 0.074 & 0.295 & 0.052 & $\left(5.786 \pm 2.14\% \right) \times 10^{-43}$ \\ \hline 
6 & 0.588 & 0.075 & 0.286 & 0.051 & $\left(5.795 \pm 2.13\% \right) \times 10^{-43}$ \\ \hline 
7 & 0.599 & 0.073 & 0.278 & 0.050 & $\left(5.813 \pm 2.13\% \right) \times 10^{-43}$ \\ \hline 
8 & 0.62 & 0.072 & 0.262 & 0.046 & $\left(5.819 \pm 2.14\% \right) \times 10^{-43}$ \\ \hline 
\end{tabular}
\caption{The fuel-fraction bins for the RENO data, and the corresponding measured IBD rates ~\cite{RENO:2018pwo}. These data have been corrected for three-flavor oscillations; seen text for details.}
\label{tab:RENOfracs}
\end{table*}

\begin{table*}
 \begin{tabular}{|c||c|c|c|c|c|c|c|c|}\hline
Fuel Bin, $f_{235}$ & 0.527 & 0.546 & 0.557 & 0.568 & 0.579 & 0.588 & 0.599 & 0.620 \\ \hline \hline
HM & 0.9348 & 0.9401 & 0.9362 & 0.9327 & 0.9374 & 0.9352 & 0.9364 & 0.9319 \\ \hline
\emph{Ab Initio} & 0.9720 & 0.9787 & 0.9753 & 0.9723 & 0.9780 & 0.9761 & 0.9780 & 0.9746 \\ \hline
HKSS & 0.9261 & 0.9313 & 0.9274 & 0.9239 & 0.9286 & 0.9264 & 0.9276 & 0.9231 \\ \hline
\end{tabular}
\caption{Measurement-to-expectation ratios calculated for the RENO data~\cite{RENO:2018pwo}.}
\label{tab:RENO_info}
\end{table*}

\begin{table*}
\centering\begin{tabular}{|c||c|c|c|c|c|}\hline
 & Reactor 1 & Reactor 2 & Operating Time [Days] & Efficiency & Relative Proton Number \\ \hline \hline
Near Detector & 355 m & 469 m & 258.0 & 85.47\% & 1.0044 \\ \hline
Far Detector & 998 m & 1115 m & 818.0 & 86.78\% & 1 \\ \hline
\end{tabular}
\caption{Double Chooz information from Ref.~\cite{DoubleChooz:2019qbj}.}
\label{tab:DCinfo}
\end{table*}


\subsection{Experimental Measurements}

\begin{figure*}
\begin{subfigure}[t]{0.45\textwidth}
\centering
\includegraphics[width=\linewidth]{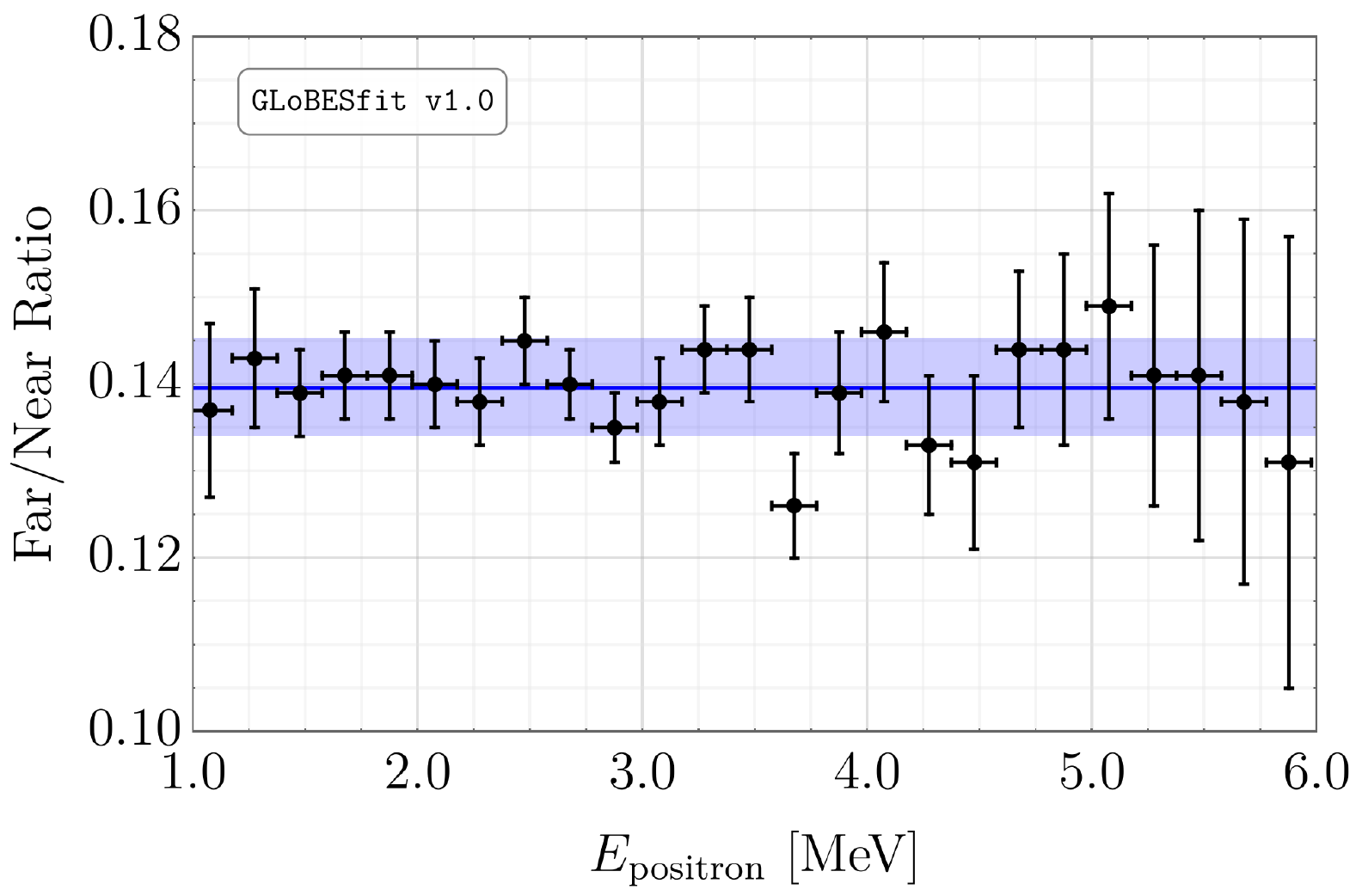}
        \caption{The far-to-near ratio measured at Bugey-3 \cite{Declais:1994su}.}
        \label{fig:Bugey3Spectrum}
\end{subfigure}
\begin{subfigure}[t]{0.45\textwidth}
\centering
\includegraphics[width=\linewidth]{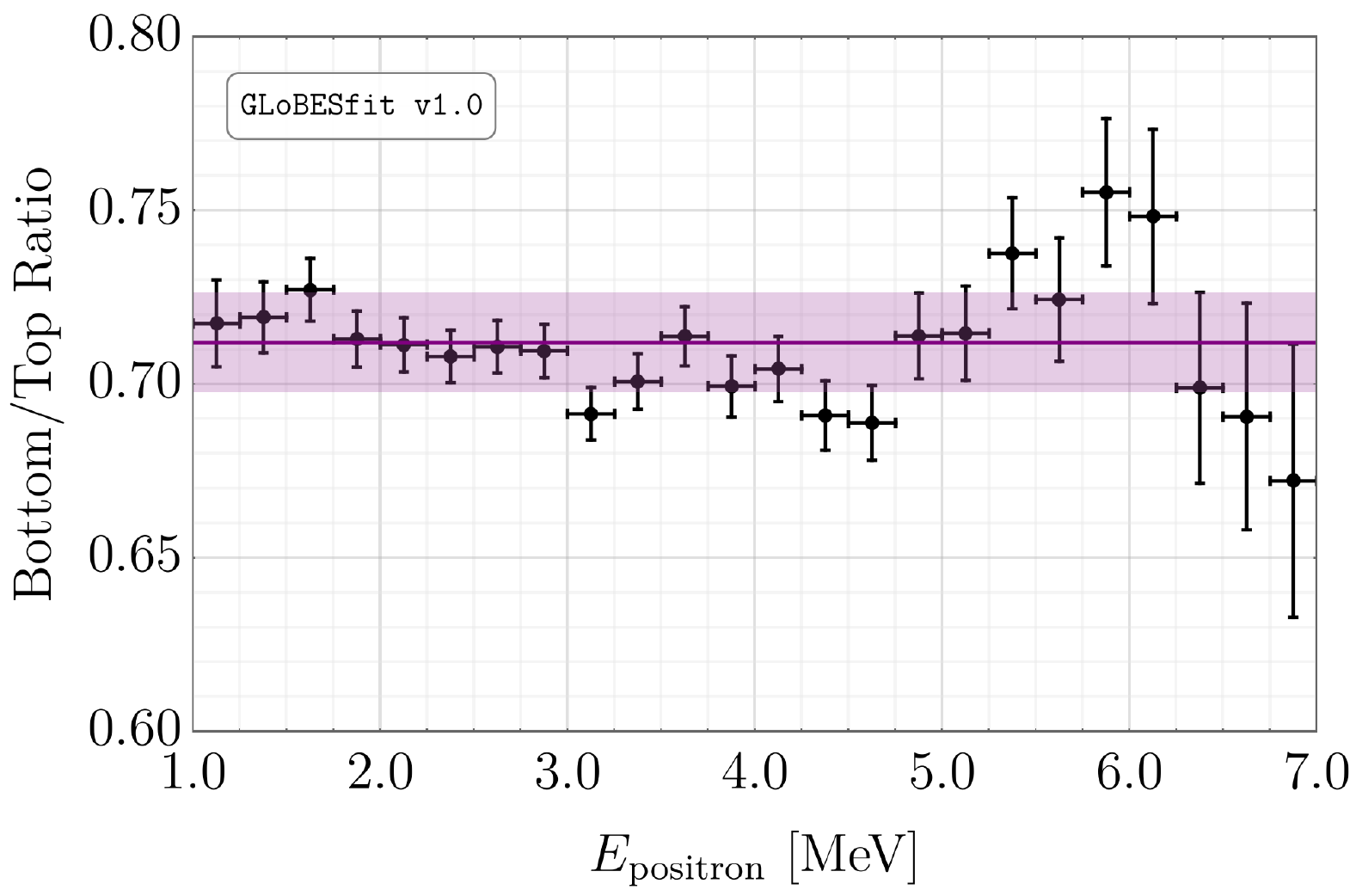}
	\caption{The bottom-to-top ratio measured at DANSS \cite{Alekseev:2018efk}.}
	\label{fig:DANSSSpectrum}
\end{subfigure}

\medskip

\begin{subfigure}[t]{0.45\textwidth}
\centering
\vspace{0pt}
\includegraphics[width=\linewidth]{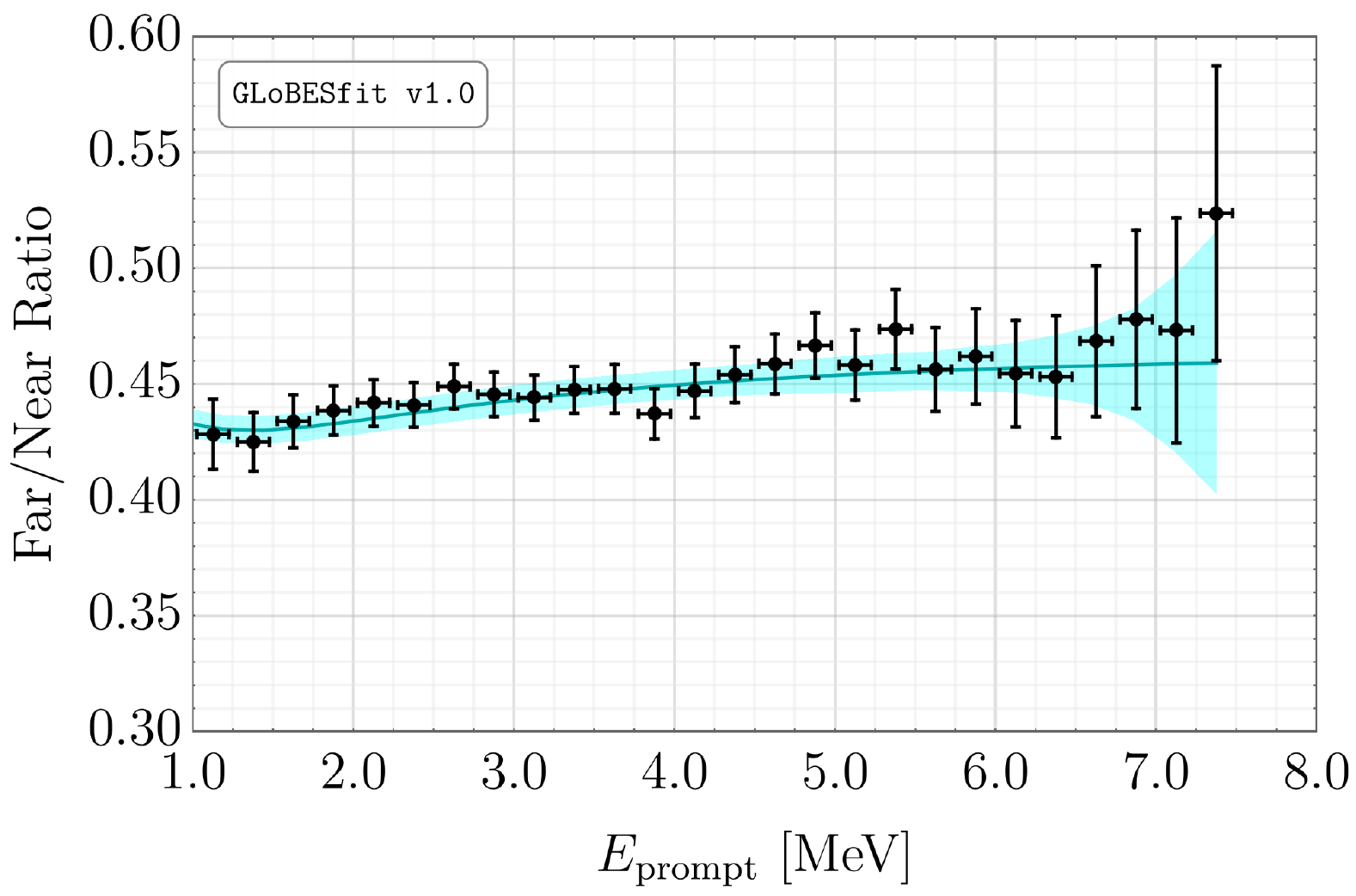}
	\caption{The far-to-near ratio measured at Double Chooz \cite{DoubleChooz:2019qbj}. The colored curve assumes $\sin^2 2\theta_{13} = 0.105$.}
	\label{fig:DoubleChoozSpectrum}
\end{subfigure}
\begin{subfigure}[t]{0.45\textwidth}
\centering
\vspace{0pt}
\includegraphics[width=\linewidth]{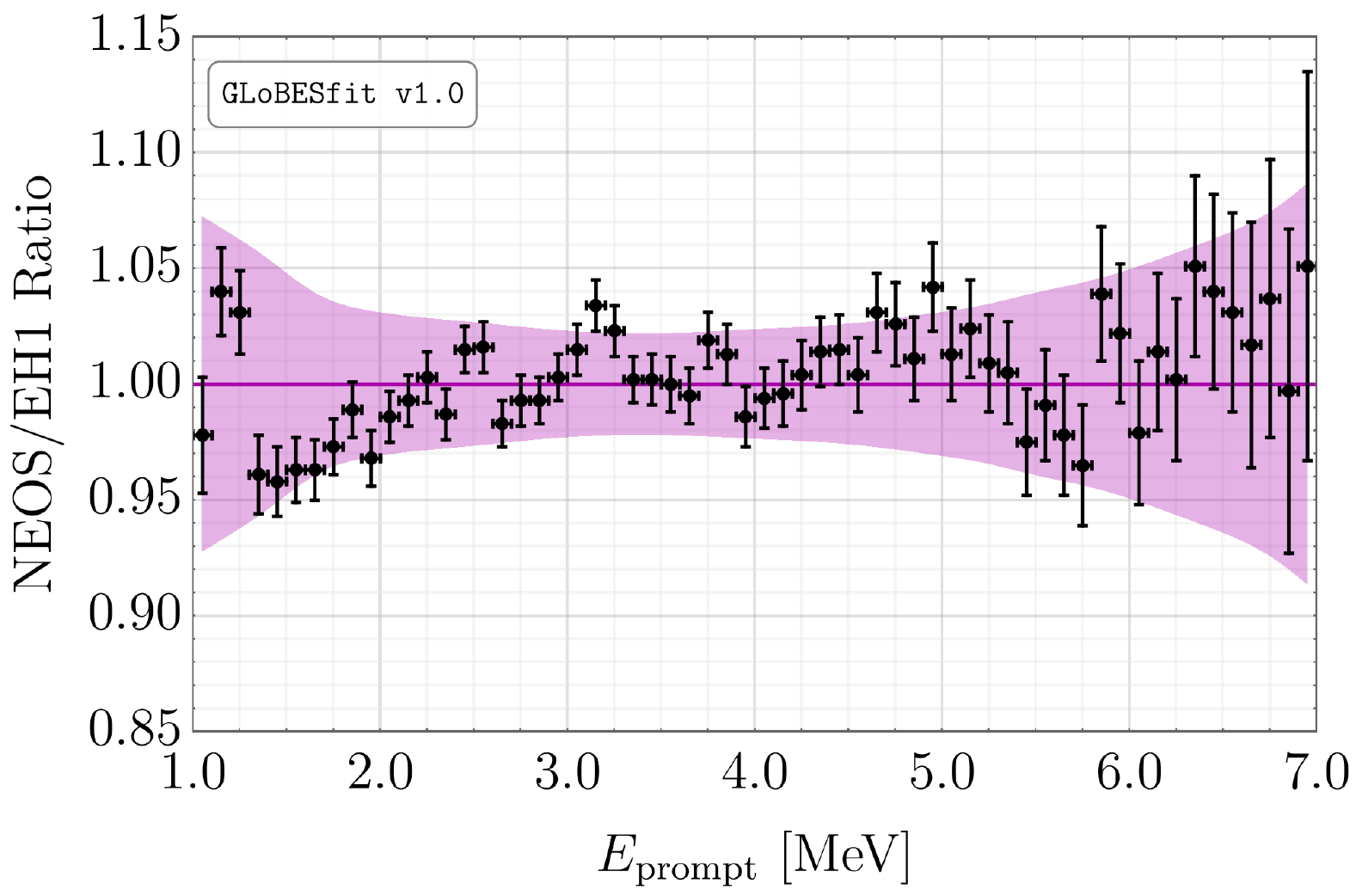}
	\caption{The spectrum measured at NEOS \cite{Ko:2016owz} relative to the Daya Bay spectrum \cite{An:2016srz}.}
	\label{fig:NEOSSpectrum}
\end{subfigure}

\medskip

\begin{subfigure}[t]{0.45\textwidth}
\centering
\vspace{0pt}
\includegraphics[width=\linewidth]{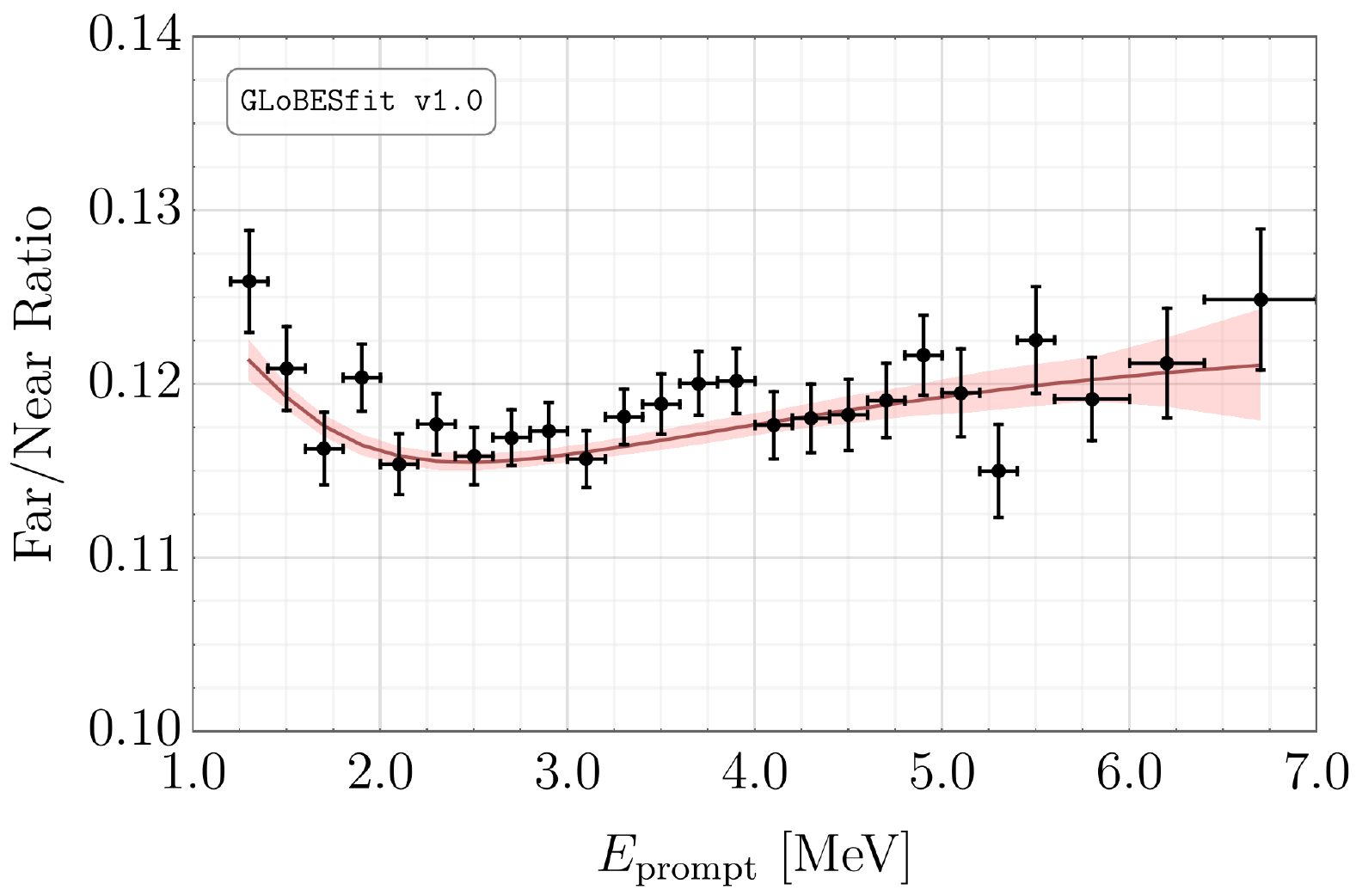}
	\caption{The far-to-near ratio measured at RENO \cite{Bak:2018ydk}. The colored curve assumes $\sin^2 2\theta_{13} = 0.0896$.}
	\label{fig:RENOSpectrum}
\end{subfigure}
	\caption{Spectral data used in our analyses, excluding Daya Bay (see Figs.~\ref{fig:DBSpecEH2} and \ref{fig:DBSpecEH3}).}
	\label{fig:spec1}
\end{figure*}

\begin{figure*}
\centering
\begin{subfigure}[t]{0.45\textwidth}
\centering
\includegraphics[width=\linewidth]{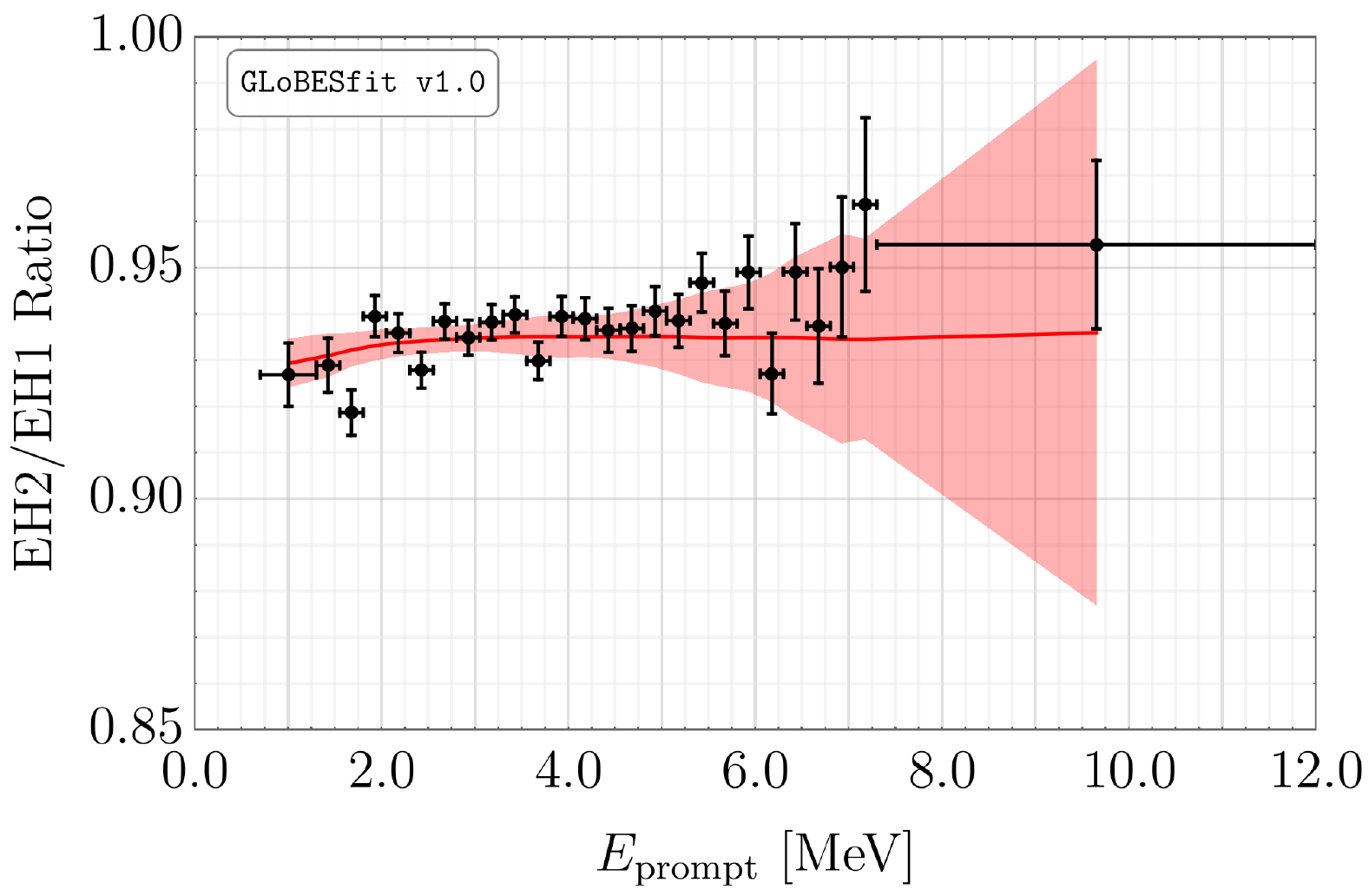}
        \caption{The EH2/EH1 ratio measured at Daya Bay \cite{Adey:2018zwh}. The red curve takes $\sin^2 2\theta_{13} = 0.0856$.}
        \label{fig:DBSpecEH2}
\end{subfigure}
\begin{subfigure}[t]{0.45\textwidth}
\centering
\includegraphics[width=\linewidth]{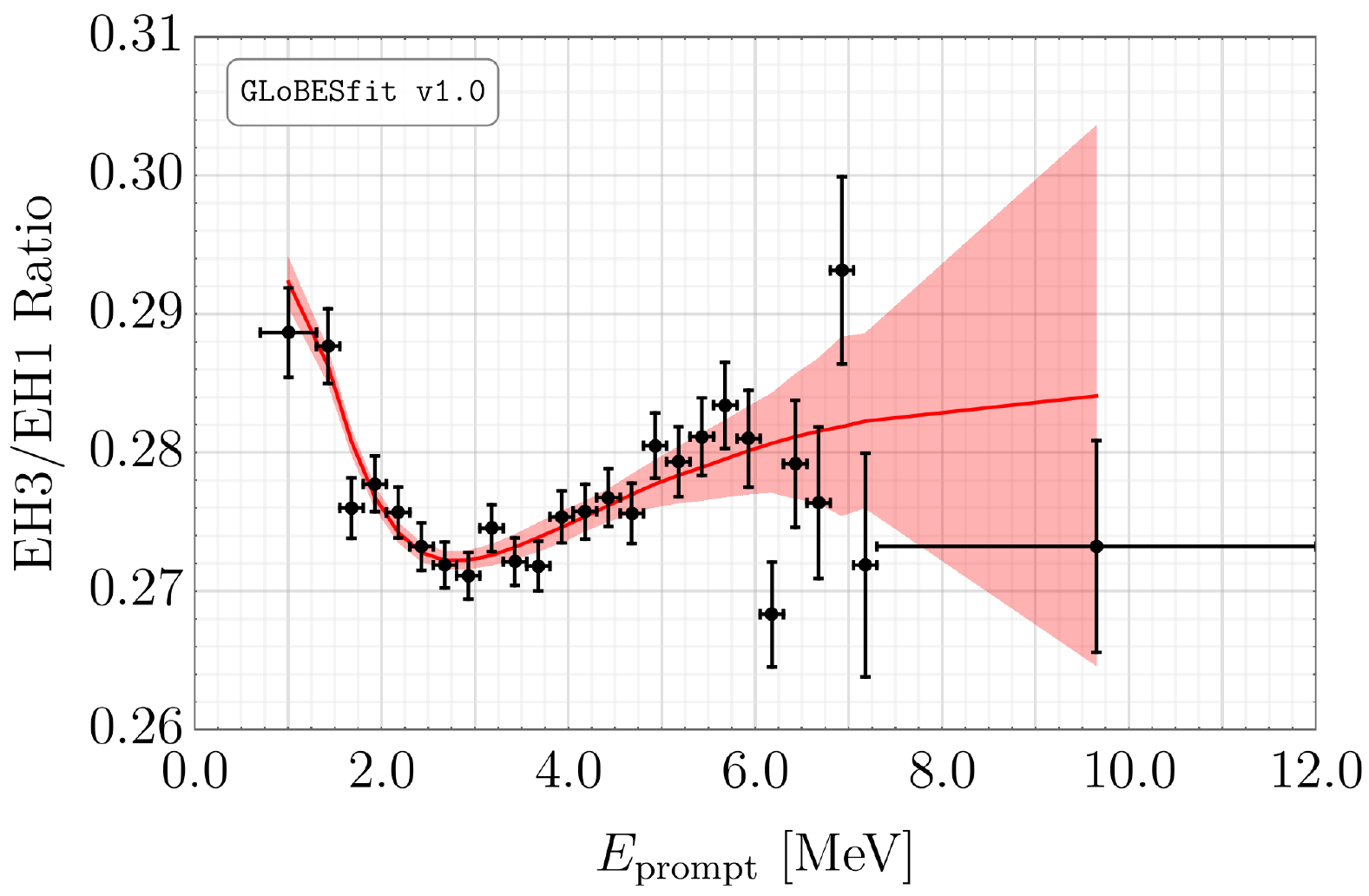}
	\caption{The EH3/EH1 ratio measured at Daya Bay \cite{Adey:2018zwh}. The red curve takes $\sin^2 2\theta_{13} = 0.0856$.}
	\label{fig:DBSpecEH3}
\end{subfigure}
	\caption{Daya Bay spectral data used in our analyses.}
	\label{fig:spec2}
\end{figure*}

We show plots of the ratios of spectra measured at Bugey-3, DANSS, Daya Bay, Double Chooz, NEOS and RENO, which underpin our analyses in Secs.~\ref{sec:SpectrumExp} and \ref{sec:SpecAnalysis}, in Figs.~\ref{fig:spec1} and \ref{fig:spec2}. The black points and error bars represent the data and their statistical uncertainties. The colored curves represent the expectations assuming three-neutrino oscillations; when relevant, the value of $\sin^2 2\theta_{13}$ used to calculate the line is the best-fit value as determined by that collaboration. The colored bands represent the total systematic uncertainties, given by the square root of the diagonal elements of the covariance matrix for each experiment.

\subsection{Markov Chain Monte Carlo Results}

\begin{figure*}
\includegraphics[width=0.9\linewidth]{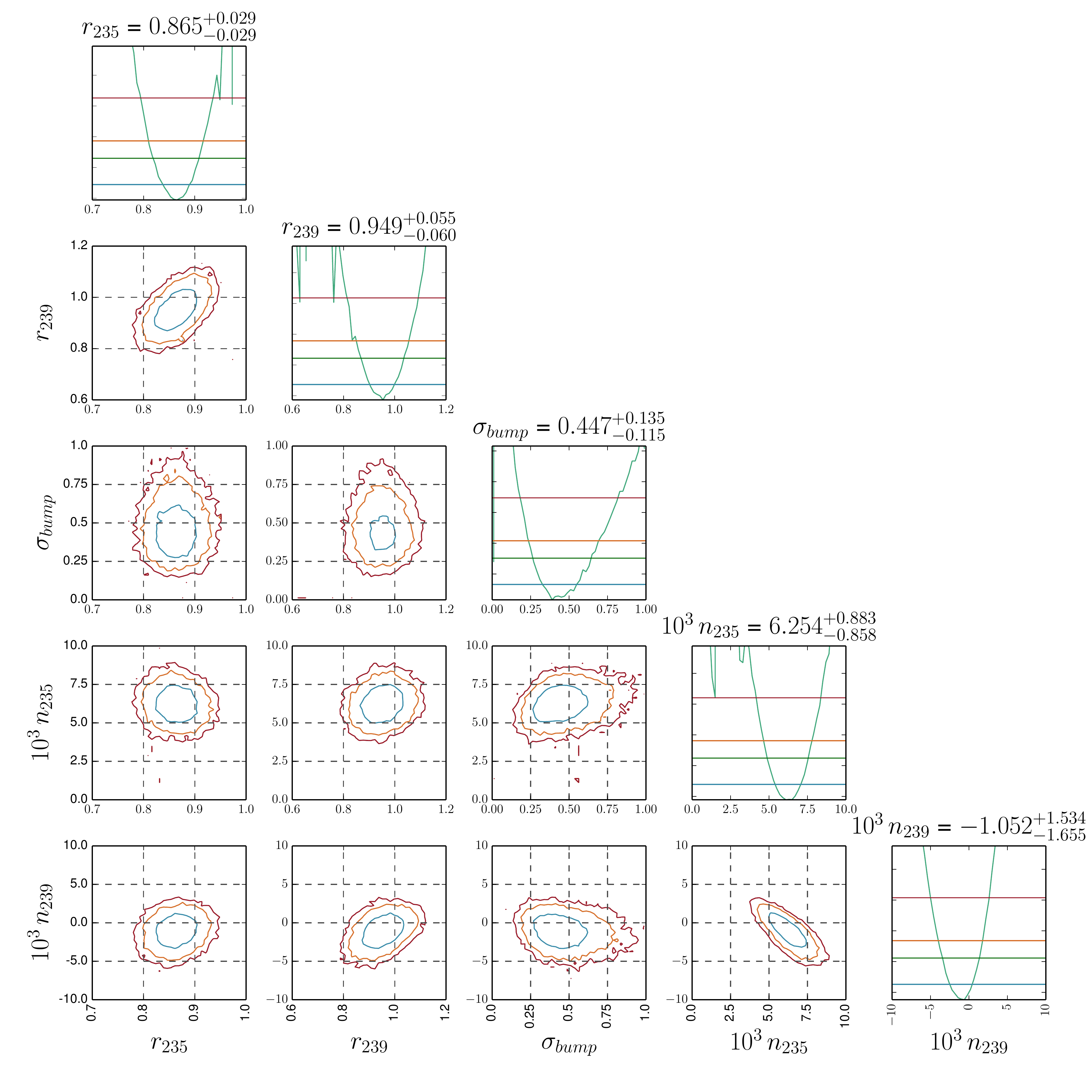}
	\caption{Results of our Markov Chain Monte Carlo analysis of Daya Bay and RENO.}
	\label{fig:MCMC1}
\end{figure*}

\begin{figure*}
\includegraphics[width=0.9\linewidth]{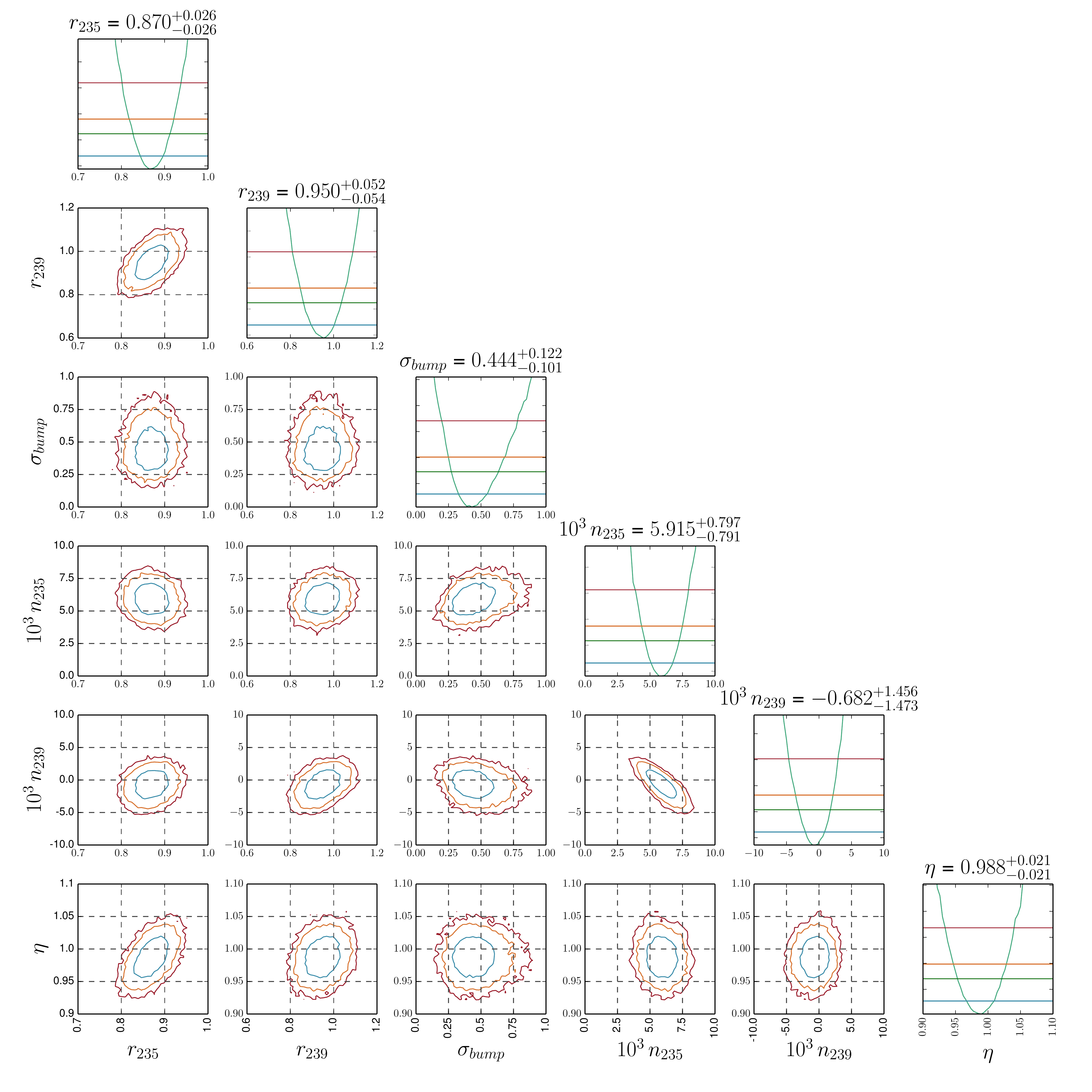}
	\caption{Results of our Markov Chain Monte Carlo analysis of Daya Bay, RENO and PROSPECT.}
	\label{fig:MCMC2}
\end{figure*}

In Figs.~\ref{fig:MCMC1} and \ref{fig:MCMC2}, we show the result of our Markov-Chain Monte Carlo analyses described in Sec.~\ref{sec:Bump}. In the two-dimensional plots, the blue, orange and red contours represent the 68.3\%, 95\% and 99\% credible regions (C.R.); the one-dimensional plots add to this list a green curve representing the 90\% C.R. Above each one-dimensional figure is shown the best-fit value of the corresponding parameter, as well as extent of the 68.3\% C.R. region. To determine the posterior probability, we use 100 chains of walkers with randomized starting points. Each walker takes 1500 steps, of which the first 500 are conservatively ignored for the purposes of burn-in. The calculations have been performed using \textsc{emcee} \cite{ForemanMackey:2012ig}.

\bibliographystyle{apsrev-title}
\bibliography{references}{}

\end{document}